\documentstyle[aps,epsfig]{revtex}
\newcommand{\idn}{\hspace{\parindent}} 
\newcommand{\snr}{\rangle \hspace{-0.05cm}{^{(0)}}}
\newcommand{\snl}{{^{(0)}}\hspace{-0.1cm}\langle}

\begin{document}

\title{The Functional Schr\"{o}dinger Picture Approach to Many-Particle Systems}
\author{Chul Koo Kim \ and \ Sang Koo You}
\address{Institute of Physics and Applied Physics, 
         Yonsei University, Seoul 120-749, Korea \\
                       and  \\
Center for Strongly Correlated Materials Research,
Seoul National University, Seoul 151-742, Korea}
\maketitle

\begin{abstract}
A self-contained pedagogical introduction to the functional Schr\"{o}dinger picture method
of many-body theory is given at a level suitable for graduate students and also for
many-body physicists who have not been exposed to the functional Schr\"{o}dinger picture
method previously. Mathematical tools necessary for the functional Schr\"{o}dinger picture
calculation are introduced. The method is first applied to various condensed matter problems
including the electron gas, the Hubbard model, the BCS superconductivity, and dilute bose gas
within the variational approximation.
It is shown that the variational approximation with the Gaussian trial functional invariably
leads into the Hartree-Fock results for both zero and finite temperature cases.

In order to go beyond the Gaussian results, concepts of variational and optimized perturbation
theories are introduced using simple quantum mechanical languages. Then, the variational perturbation
theory in the functional Schr\"{o}dinger picture is applied to the $\lambda \phi^4$ model to yield
the effective potential to the third order. The optimized perturbation theory is also applied to the
$\lambda \phi^4$ model  to yield a general expression up to the second order.
\end{abstract}

\tableofcontents
\eject

\section*{Foreword}

These notes represent the lectures given at Johannes Kepler Universit\"{a}t-Linz,
Austria in October and November, 2002 as "Special Topics in Many-Body Theory".
The intent of these lectures was to provide a self-contained tutorial introduction to the 
functional Schr\"{o}dinger picture approach to many-particle systems.
 
Condensed matter systems are basically many-particle systems.
Whether the system in consideration is a crystalline solid, an
amorphous material, or a liquid, one has to deal with
substantially large number of particle which often amounts to
$10^{23}/{\rm cm}^3$. This large number still eludes
exact numerical calculation even with ever increasing
computational
capability of modern day computers. \\
Many-body theories based on non-relativistic quantum field
theories are conceived to deal with almost infinite degrees of
freedoms of many-particles in condensed matter systems. Many-body
theory closely followed its relativistic brother, quantum field
theory in its development. In quantum field theory, there exist
basically three alternative approaches, namely operator
(Heisenberg), functional integral (path integral), and
Sch\"{o}dinger approaches. Comparing to the former two methods,
the last one Sch\"{o}dinger picture approach was early regarded
less convenient and clumsy for development of field theories.
However, in 80's and 90's, theorists have found that it offers
some distinct advantages over others when suitable schemes are
employed. Since then the Sch\"{o}dinger picture has been drawing
wide interests and field theories based on the functional
Sch\"{o}dinger picture have been
developing rapidly. \\
In condensed matter physics, so far the Sch\"{o}dinger
picture scheme has been much neglected. Only very recently,
efforts to develop the Sch\"{o}dinger picture method into a useful
many-body theoretical tool has started. In this lecture, we
present such an effort and try to give a pedagogical overview on
this method.

Since these notes are designed for beginning graduate students and
newcomers in this field, we tried to make the notes selfcontained.
In this respect, we tried to limit the number of references to those
essential to read these notes. Other relevant references can be found
in the references of the papers and books cited in these notes.
Although most of essential steps are described in detail to help the 
readers, some of the derivations are left as problems, so that
readers can gain more insights and expertise through exercise.

\section{Introduction }
Condensed matter systems are basically
many-particle systems with many particles ($\sim 10^{23}$ per $\rm
cm^3$). Now, the basic question is whether it is possible to solve
the many-particle Schr\"{o}dinger equation directly?
\begin{eqnarray}
&& H=\sum_{k=1}^{N} T(x_k)+\frac{1}{2}\sum_{k \neq l=1}^{N} V(x_k, x_l), \\
&& i\hbar \frac{\partial}{\partial t} \Psi (x_1 \cdot \cdot \cdot x_N, t)
   =H \Psi (x_1 \cdot \cdot \cdot x_N, t).
\end{eqnarray}
If we have $\Psi (x_1 \cdot \cdot \cdot x_N, t)$, then we can
calculate all the physical
quantities we need. \\
Expansion of $\Psi (x_1 \cdot \cdot \cdot x_N, t)$ can be written
in terms of the single-particle wave function $\psi_{E_k}(x_k)$
\begin{eqnarray}
\Psi (x_1 \cdot \cdot \cdot x_N, t)=\sum_{E_{1}' \cdot \cdot \cdot E_{N}'}
C(E_{1}' \cdot \cdot \cdot E_{N}', t)\psi_{E_{1}'}(x_1)
\cdot \cdot \cdot \psi_{E_{N}'}(x_N),
\end{eqnarray}
where $E_{1}' \cdot \cdot \cdot E_{N}'$ represent single particle
quantum numbers. $C(E_{1}' \cdot \cdot \cdot E_{N}', t)$ is the
coefficient for a particular configuration. In general, it is not
practical to continue to use $\Psi (x_1 \cdot \cdot \cdot x_N, t)$
with $N \sim 10^{23}$. Therefore, a suitable scheme to handle the
large degree of freedom has to be adopted. Here, we resort to the
technique of second quantization.

\subsection{Second Quantization}
There are several ways for introducing second quantization to the
many-body Schr\"{o}dinger equation. An explicit and detailed
treatise is given in Chapter 1 of Fetter \& Walecka (Gen. Ref. 3).
However, here, we follow a rather concise but less detailed path
to the second quantization commonly adopted by field theory textbooks
(Chapt. 2, Gen. Ref. 1 \& \S  12 \& 13, Gen. Ref. 4). \\ \\
{\bf The Schr\"{o}dinger Wave Field of Bose Statistics} \\ \\
(1st quantized) Schr\"{o}dinger equation is given by
\begin{eqnarray}
-\frac{\hbar^2}{2m} \nabla^2 \Psi + V(\vec{x}) \Psi = i \hbar
\dot{\Psi}.
\end{eqnarray}
The conjugate complex equation is
\begin{eqnarray}
-\frac{\hbar^2}{2m} \nabla^2 \Psi^{*} + V(\vec{x}) \Psi^{*} = -i
\hbar \dot{\Psi}^{*}.
\end{eqnarray}
Find a Lagrangian such that the corresponding Euler-Lagrange
equation leads back to Schr\"{o}dinger equation.
\begin{eqnarray}
\frac{d}{dt}\left( \frac{\partial L}{\partial \dot{q}_{i}} \right)
-\frac{\partial L}{\partial {q}_{i}}=0 \ \ \ i=1 \cdot \cdot \cdot f
\end{eqnarray}
Such a Lagrangian is given by
\begin{eqnarray}
L=\int d^3 x {\cal L} = \int \Psi^* \left\{ i\hbar
\dot{\Psi}-V(\vec{x}) \Psi + \frac{\hbar^2}{2m}\nabla^2 \Psi
\right\} d^3 x ,
\end{eqnarray}
which leads directly to
\begin{eqnarray}
\frac{d}{dt} \frac{\delta L}{\delta \dot{\Psi}^* } -\frac{\delta
L}{\delta {\Psi}^* } = -\left\{ i \hbar \dot{\Psi} - V(\vec{x})
\Psi + \frac{\hbar^2}{2m}\nabla^2 \Psi \right\} =0.
\end{eqnarray}
(Note: The above Lagrangian is not Hermitian. We can work with a Hermitian
Lagrangian (See Chap. 2, Gen. Ref. 1.) at the expense of slight complications.) \\
Define the canonically conjugate momentum in the usual way,
\begin{eqnarray}
\pi = \frac{\delta L}{\delta \dot{\Psi}} = -\frac{\hbar}{i}
\Psi^{*}.
\end{eqnarray}
The Hamiltonian is given by
\begin{eqnarray}
H &=& \int (\pi \dot{\Psi}-{\cal L}) d^3 x  \nonumber \\
&=&\int \left\{  i \hbar \Psi^* \dot{\Psi} - i \hbar \Psi^* \dot{\Psi}
-\Psi^* \frac{\hbar^2}{2m}\nabla^2 \Psi +\Psi^* V\Psi \right\} d^3 x \nonumber \\
&=&\int \Psi^* (\vec{x}) \left\{  -\frac{\hbar^2}{2m}\nabla^2 +V(\vec{x}) \right\}
 \Psi(\vec{x}) d^3 x.
\end{eqnarray}
In the above Hamiltonian, only wave nature appears. In order to
restore the particle nature, we expand $\Psi$ in terms of the 1st
quantized eigenfunctions,
\begin{eqnarray}
&&\left( -\frac{\hbar^2}{2m}\nabla^2 + V \right) \psi_\mu = i \hbar \dot{\psi}_\mu \nonumber, \\
&& \psi_{\mu}(\vec{x},t)=e^{-\frac{i}{\hbar} E_\mu t} \psi_\mu
(\vec{x}).
\end{eqnarray}
The expansions are
\begin{eqnarray}
&&\Psi(\vec{x})=\sum_{\mu} b_{\mu}(t) \psi_{\mu}(\vec{x}) \nonumber, \\
&&\Psi^{*}(\vec{x})=\sum_{\mu} b_{\mu}^{\dagger}(t)
\psi_{\mu}^{*}(\vec{x}),
\end{eqnarray}
where $b_{\mu}(t)=b_{\mu}(0) e^{-\frac{i}{\hbar} E_\mu t}$. \\
With these expressions, we obtain
\begin{eqnarray}
H= \sum_\mu E_\mu b_{\mu}^{\dagger} b_{\mu}.
\end{eqnarray}
In order to quantize the wave field $\Psi$, we require
\begin{eqnarray}
[\pi(\vec{x}), \Psi(\vec{x}')]=\frac{\hbar}{i}\delta
(\vec{x}-\vec{x}').
\end{eqnarray}
(Remember the 1st quantization postulate $[\hat{p}, \hat{x}]= \frac{\hbar}{i}$.) \\
Thus, we obtain
\begin{eqnarray}
[\Psi(\vec{x}'),\Psi^{\dagger}(\vec{x})] =\delta
(\vec{x}-\vec{x}').
\end{eqnarray}
Also we have additional commutation relations
\begin{eqnarray}
&&[\Psi(\vec{x}),\Psi(\vec{x}')]=0, \nonumber \\
&&[\Psi^{\dagger}(\vec{x}),\Psi^{\dagger}(\vec{x}')]=0.
\end{eqnarray}
We find that $b_{\mu}$ and $b_{\mu}^{\dagger}$ satisfy
\begin{eqnarray}
[b_\mu , b_{\nu}^{\dagger}]=\delta_{\mu \nu} , \ \ [b_\mu ,
b_{\nu}]=0 , \ \ [b_{\mu}^{\dagger} , b_{\nu}^{\dagger}]= 0.
\end{eqnarray}
With the new Hamiltonian operator, the `Schr\"{o}dinger equation'
is given by
\begin{eqnarray}
H \Phi &=& E \Phi , \nonumber \\
H &=& \int \Psi^{\dagger} (\vec{x}) \left\{
-\frac{\hbar^2}{2m}\nabla^2 +V(\vec{x}) \right\}
 \Psi(\vec{x}) d^3 x \nonumber \\
&\equiv& \sum_{\mu} E_{\mu} b_{\mu}^{\dagger} b_{\mu} ,
\end{eqnarray}
where
\begin{eqnarray}
&&\Phi = \prod_{\mu} \frac{1}{\sqrt{n_{\mu} !}} (b_{\mu}^{\dagger})^{n_{\mu}} \Phi_0, \\
&& E=\sum_{\mu} E_{\mu} n_{\mu} \ \ \ \ n_{\mu}=0, 1, 2, \cdot
\cdot \cdot . \nonumber
\end{eqnarray}
A series of energy levels $E_{\mu}$ are occupied by a definite number $n_{\mu}$ of quanta.
This shows clearly that field quantization guarantees the corpuscular nature of the
Schr\"{o}dinger wave field. \\ \\
{ \bf Expectation Values in the New Formalism} \\
(How do we relate the new theory to reality?)\\ \\
\hspace*{-0.2cm}$\bullet$ {\bf The particle density operator }
\begin{eqnarray}
&{\Psi^{*}(x,t) \Psi(x,t)}& \longrightarrow  \ \
{\Psi^{\dagger}(x,t) \Psi(x,t)} \\
&{\rm (1st \ quantized)}& \ \ \ \ \ \ \ {\rm (2nd \ quantized)} \nonumber
\end{eqnarray}
\begin{eqnarray}
&&\rho_{op}(x)=\Psi^{\dagger}(x) \Psi(x) \ : \ {\rm particle \ density \ operator} \nonumber \\
&&\bar{\rho}(x)=\langle \Phi | \Psi^{\dagger}(x) \Psi(x) | \Phi \rangle
\end{eqnarray}
\\
$<${\bf Problem}$>$ Show that the above $\bar{\rho}(x)$ gives an identical result to the 1st
quantized particle density $\bar{\rho}(x)= \psi^{*}(x) \psi(x)$, when only a single particle is present in state $k$. \\ \\
\hspace*{-0.2cm}$\bullet$ {\bf The position operator} \\
In the 1st quantized form, we have
\begin{eqnarray}
\bar{x}=\int \psi^{*}(x,t) x \psi(x,t) d^3 x. \nonumber
\end{eqnarray}
The second quantized form is given by
\begin{eqnarray}
x_{op}&=&\int \psi^{\dagger}(x) x \psi(x) d^3 x, \\
\bar{x}&=&\langle \Phi | x_{op}| \Phi \rangle \nonumber \\
&=&\langle \Phi |\int \psi^{\dagger}(x) x \psi(x) d^3 x  | \Phi
\rangle.
\end{eqnarray}
\hspace*{-0.2cm}$\bullet$ {\bf The potential energy operator}
\begin{eqnarray}
V_{op}=\int \psi^{\dagger}(x) V(x) \psi(x) d^3 x.
\end{eqnarray}
\hspace*{-0.2cm}$\bullet$ {\bf The interaction energy operator}
\begin{eqnarray}
V(x,x')=V(x-x') \nonumber .
\end{eqnarray}
In the first quantized form, it is given by
\begin{eqnarray}
&&\frac{1}{2} \int \rho(x,t) V(x,x') \rho(x',t) d^3 x d^3 x' \nonumber \\
&=&\frac{1}{2} \int \Psi^{*}(x,t) \Psi(x,t)
V(x-x')\Psi^{*}(x',t)\Psi(x',t) d^3 x d^3 x' .
\end{eqnarray}
In the second quantized form, the order of the operators is important. \\
When there is one or zero particle present, the interaction energy is zero.\\
Thus, we should have $\Psi$ on the right and $\Psi^{\dagger}$ on the left,\\
\begin{eqnarray}
\frac{1}{2} \int \Psi^{\dagger}(x) \Psi^{\dagger}(x')
V(x-x')\Psi(x')\Psi(x) d^3 x d^3 x'.
\end{eqnarray}
The expectation value is given by
\begin{eqnarray}
\langle \Phi |\frac{1}{2} \int \Psi^{\dagger}(x)
\Psi^{\dagger}(x') V(x-x')\Psi(x')\Psi(x) d^3 x d^3 x' | \Phi
\rangle.
\end{eqnarray}
\\
$<${\bf Problem}$>$ Calculate the expectation value of the Coulomb interaction energy $V(x-x')=e^2 /|x-x'|$ for
$\Phi = b_{\mu_1}^{\dagger} \Phi_0 $ and $\Phi =b_{\mu_1}^{\dagger} b_{\mu_2}^{\dagger} \Phi_0$ . \\ \\
{\bf Quantization of the Schr\"{o}dinger Wave Field of Fermi-Dirac Statistics; Fermions} \\ \\
In reality, electrons which are main constituents of the many-body problems in condensed matter are fermions,
which obey F-D statistics. \\ \\
We expand the wave field into
\begin{eqnarray}
&&\Psi(x)=\sum_{\mu} a_{\mu} \psi_{\mu}(x), \nonumber \\
&&\Psi^{\dagger}(x)=\sum_{\mu} a_{\mu}^{\dagger} \psi_{\mu}^{*}(x),
\end{eqnarray}
where $\psi_{\mu}(x)$ satisfies
\begin{eqnarray}
\left( -\frac{\hbar^2}{2m} \nabla^2 + V(x)  \right) \psi_{\mu}(x)
= E_{\mu} \psi_{\mu}(x). \nonumber
\end{eqnarray}
In order to satisfy the F-D statistics (Pauli exclusion
principle), they should obey the following commutation relations.
\begin{eqnarray}
&& \{a_{\mu}^{\dagger}, a_{\nu}^{\dagger} \} =a_{\mu}^{\dagger}a_{\nu}^{\dagger}+a_{\nu}^{\dagger}a_{\mu}^{\dagger}=0
\nonumber \\
&& \{a_{\mu}^{\dagger}, a_{\nu} \} = \delta_{\mu \nu} \\
&&\{a_{\mu}, a_{\nu} \} =0 \nonumber
\end{eqnarray}
Using these relations, we obtain
\begin{eqnarray}
&&\{ \Psi^{\dagger}(x), \Psi(x') \} = \delta(x-x'), \nonumber \\
&&\{ \Psi^{\dagger}(x), \Psi^{\dagger}(x') \} =0, \\
&&\{ \Psi(x), \Psi(x') \} =0. \nonumber
\end{eqnarray}
\\
$<${\bf Problem}$>$ Derive the above commutation relations using the commutation relations of $a_{\mu}^{\dagger}(x)$
and $a_{\nu}(x)$. \\ \\
The Hamiltonian operator is given by
\begin{eqnarray}
H&=&\int \Psi^{\dagger}(x) \left\{-\frac{\hbar^2}{2m} \nabla^2 + V(x) \right\} \Psi(x) d^3 x \\
&\equiv& \sum_{\mu} E_{\mu} a_{\mu}^{\dagger} a_{\mu}. \nonumber
\end{eqnarray}
The corresponding Schr\"{o}dinger equation is
\begin{eqnarray}
H \Phi = E \Phi.
\end{eqnarray}
We obtain as eigenfunctions
\begin{eqnarray}
\Phi_{(n)} = \prod_{\mu} (a_{\mu}^{\dagger})^{n_\mu} \Phi_0 ,
\end{eqnarray}
with $N=\sum_{\mu}N_{\mu}$ and $E=\sum_{\mu} E_{\mu} n_{\mu}$. \\
First consider a one-particle state
\begin{eqnarray}
\Phi=\sum_{\mu} c_{\mu} a_{\mu}^{\dagger} \Phi_0
\end{eqnarray}
which is a superposition of one-particle states $a_{\mu}^{\dagger} \Phi_0$. \\
We replace $a_{\mu}^{\dagger}$ by $\Psi^{\dagger}$ using
\begin{eqnarray}
\int \psi_{\mu}(x) \Psi^{\dagger}(x)d^3 x = a_{\mu}^{\dagger} .
\end{eqnarray}
\begin{eqnarray}
\Phi &=& \int \sum_{\mu} c_{\mu} \psi_{\mu}(x) \Psi^{\dagger}(x) d^3 x \Phi_0 \nonumber \\
     &=& \int f(x) \Psi^{\dagger}(x) d^3 x \Phi_0
\end{eqnarray}
$\Phi$ represents a superposition of the single particle, which maybe thought as having been created at
different places with a \underline{probability amplitude $f$($x$)}. \\ \\
$<${\bf Problem}$>$ Show that $\Psi^{\dagger}(x')$ creates a particle at the place $x'$. \\
$<${\bf Sol.}$>$ $\Psi^{\dagger}(x)  \Psi(x) \Psi^{\dagger}(x') \Phi_0 = \Psi^{\dagger}(x) \delta(x-x') \Phi_0
=  \delta(x-x')\Psi^{\dagger}(x')\Phi_0 $. \\ \\
We now have
\begin{eqnarray}
H \Phi &=& \int \Psi^{\dagger}(x) \left(-\frac{\hbar^2}{2m} \nabla_{x}^2 + V(x)\right) \Psi(x)d^3 x \int f(x')
\Psi^{\dagger}(x') d^3 x' \Phi_0 \nonumber \\
&=& \int \int \Psi^{\dagger}(x) \left(-\frac{\hbar^2}{2m} \nabla_{x}^2 + V(x)\right) f(x') \delta(x-x') d^3 x d^3 x' \Phi_0
\nonumber \\
&=& \int \Psi^{\dagger}(x) \Phi_0 \left(-\frac{\hbar^2}{2m}
\nabla_{x}^2 + V(x)\right) f(x) d^3 x.
\end{eqnarray}
Here, we use the relation $H \Phi = E \Phi = E \int f(x)
\Psi^{\dagger}(x) \Phi_0 d^3 x$. This is valid only when
\begin{eqnarray}
\left(-\frac{\hbar^2}{2m} \nabla^2 + V(x)\right)f(x)=Ef(x) .
\end{eqnarray}
This one-particle example shows that the second quantization formalism is identical to the usual Schr\"{o}dinger theory. \\ \\
\hspace*{-0.2cm}$\bullet$ {\bf What about a 2-particle state?} \\ \\
Consider the most general two-particle state
\begin{eqnarray}
\Phi = \sum_{\mu_1 \mu_2} c_{\mu_1
\mu_2}a_{\mu_1}^{\dagger}a_{\mu_2}^{\dagger}\Phi_0 ,
\end{eqnarray}
which can be rearranged
\begin{eqnarray}
\Phi &=& \int \int (\sum_{\mu_1 \mu_2}
\psi_{\mu_1}(x)\psi_{\mu_2}(x') c_{\mu_1 \mu_2})\Psi^{\dagger}(x)
\Psi^{\dagger}(x') \Phi_0 d^3 x d^3 x' \nonumber \\
&=& \int \int f(x,x') \Psi^{\dagger}(x) \Psi^{\dagger}(x') \Phi_0
d^3 x d^3 x'.
\end{eqnarray}
Now we show that $f(x,x')$ is antisymmetric. Interchanging $x$
with $x'$, we have
\begin{eqnarray}
&&\int \int f(x',x) \Psi^{\dagger}(x')
\Psi^{\dagger}(x) \Phi_0 d^3 x' d^3 x \nonumber \\
&=&-\int \int f(x',x) \Psi^{\dagger}(x) \Psi^{\dagger}(x') \Phi_0
d^3 x' d^3 x.
\end{eqnarray}
Thus, we note that $f(x,x')=-f(x',x)$ by comparing to Eq.(40). \\
Now, we assume the Hamiltonian operator of interacting electrons
in second quantization is given by
\begin{eqnarray}
H&=&\int \Psi^{\dagger}(x)  \left(-\frac{\hbar^2}{2m} \nabla^2 + V(x)\right) \Psi(x) d^3 x  \nonumber \\
  &+& \frac{1}{2} \int \int \Psi^{\dagger}(x)\Psi^{\dagger}(x') \frac{e^2}{|x-x'|} \Psi(x')\Psi(x) d^3 x d^3
  x'.
\end{eqnarray}
Combining Eq.(40) with (42), we obtain
\begin{eqnarray}
\hspace*{-0.5cm}\left\{ -\frac{\hbar^2}{2m} \nabla_{x_1}^2 + V(x_1
) -\frac{\hbar^2}{2m} \nabla_{x_2}^2 + V(x_2 )+ \frac{e^2}{|x_1
-x_2 |} \right\} f(x_1 ,x_2 )= E f(x_1 ,x_2 ).
\end{eqnarray}
\\
$<${\bf Problem}$>$ Prove Eq.(43). \\ \\
\hspace*{-0.2cm}$\bullet$ {\bf n-particle state} \\ \\
If we apply the field Hamiltonian operator to an n-particle state,
\begin{eqnarray}
\Phi=\int d^3 x_{1} \cdot \cdot \cdot d^3 x_{n} f_n (x_1 , \cdot \cdot \cdot
 , x_n ) \Psi^{\dagger}(x_1 ) \cdot \cdot \cdot \Psi^{\dagger}(x_n )
 \Phi_0,
\end{eqnarray}
we will find the n-body wavefunction must satisfy
\begin{eqnarray}
\hspace*{-0.7cm}\left\{ \sum_{j=1}^{n} \left(-\frac{\hbar^2}{2m} \nabla_{x_j}^2 + V(x_j )\right) +  \frac{1}{2}
 \sum_{i \neq j} \frac{e^2}{|x_i -x_j |} \right\} f(x_1 , \cdot \cdot \cdot
 , x_n ) = E f(x_1 , \cdot \cdot \cdot
 , x_n ).
\end{eqnarray}
In this way, all of the n-body first quantized systems can be
contained in the corresponding second quantized theory (quantum
field theory). In this sense, `many-body theory' is basically a `nonrelativistic quantum field theory'. \\
For interacting electron systems, the Hamiltonian is given by
\begin{eqnarray}
H&=& \int d^3 x \Psi^{\dagger}(x) T \Psi(x) \nonumber \\
 &+& \frac{1}{2} \int \int d^3 x d^3 x' \Psi^{\dagger}(x)\Psi^{\dagger}(x') V(x, x')
 \Psi(x')\Psi(x)  \\
 &=& \sum_{r,s} a_{r}^{\dagger} \langle r | T | s \rangle a_{s} + \frac{1}{2} \sum_{r,s,t,u}
 a_{r}^{\dagger}a_{s}^{\dagger} \langle r s | V | t u \rangle
 a_{u}a_{t}.
\end{eqnarray}
\hspace*{5cm} (Note the order of operators.)
\subsection{Pictures in Quantum Mechanics}
{\bf Schr\"{o}dinger Picture} \\ \\
The usual elementary description of quantum mechanics assumes that
the state vectors are time dependent, whereas the operators are
time independent. The Schr\"{o}dinger equation in the
Schr\"{o}dinger picture takes the form
\begin{eqnarray}
i \hbar \frac{\partial}{\partial t} | \Psi_S (t) \rangle = H |
\Psi_S (t) \rangle.
\end{eqnarray}
A formal solution is given by
\begin{eqnarray}
| \Psi_S (t) \rangle = e^{-i H(t-t_0 )/\hbar} | \Psi_S (t_0 )
\rangle.
\end{eqnarray}
Generally, undergraduate quantum mechanics dwells on the
Schr\"{o}dinger picture formalism, although the Heisenberg
picture, which we will discuss shortly, is regarded as a
convenient tool for a formal development of quantum mechanics.
However, in many-body physics, the situation is completely
reversed. It will be shown that the Heisenberg and the interaction
pictures naturally lead into perturbative expansions of the
physical quantities,
which are the basis of the Green's function formalism. \\ \\
Therefore, almost all the many-body theoretical schemes are based
on the Heisenberg (interaction) picture. Only recently, several
quantum field theorists found that the Schr\"{o}dinger picture
provides a convenient scheme to calculate the physical quantities
in selected quantum field theoretical problems. Last few years,
the authors and their collaborators started to develop a many-body
theory based on the Schr\"{o}dinger picture and have succeeded to
provide a general theoretical framework. In this lecture,
the many-body theoretical scheme based on the functional Schr\"{o}dinger picture will be introduced. \\ \\
{\bf Heisenberg Picture} \\ \\
In the Heisenberg picture, we pass the burden of the
time-dependency from the wavefunction to the operator. The state
vector (wavefunction) in the Heisenberg picture is defined as
\begin{eqnarray}
| \Psi_H (t) \rangle  \equiv e^{iHt/\hbar} | \Psi_S (t) \rangle.
\end{eqnarray}
Combining Eq.(49) with $t_0 = 0$, we obtain
\begin{eqnarray}
| \Psi_H (t) \rangle  =| \Psi_S (t=0) \rangle
\end{eqnarray}
which is independent of $t$. \\
Considering the fact that the expectation value should remain same
in both pictures,
\begin{eqnarray}
&&\langle \Psi_S (t) | O_S | \Psi_S (t) \rangle  \nonumber \\
&&=\langle \Psi_H | O_H | \Psi_H \rangle,
\end{eqnarray}
we obtain
\begin{eqnarray}
O_H (t) = e^{iHt/\hbar} O_S e^{-iHt/\hbar}.
\end{eqnarray}
We observe that the state vector in the Heisenberg picture is now independent on $t$, whereas
the time-dependency is now carried by the operator. This fact is summarized as follows
\begin{center}
\begin{tabular}{ccc} \hline
          & Schr\"{o}dinger picture    & Heisenberg picture \\ \hline
State Ket &       T-dep.               &     T-ind.         \\
Operator  &       T-ind.               &     T-dep.         \\ \hline
\end{tabular}
\end{center}
The time derivative of Eq.(53) yields the Heisenberg equation of
motion,
\begin{eqnarray}
i \hbar \frac{\partial}{\partial t} O_H (t) = [O_H (t), H].
\end{eqnarray}
\\
$<${\bf Problem}$>$ Prove Eq.(54). \\ \\
This result determines the equation of motion for any operators in
the Heisenberg picture.
In particular, if $O_S $ commutes with $H$, then $O_H$ is a constant of the motion. \\ \\
{\bf Interaction Picture} \\ \\
When the Hamiltonian can be expressed as
\begin{eqnarray}
H=H_0 + H_I  ,
\end{eqnarray}
where $H_0$ yields a soluble problem, we can introduce the interaction picture, which is convenient for
perturbative treatment of many-particle systems. \\
Define the interaction state vector in the following way
\begin{eqnarray}
| \Psi_I (t) \rangle  \equiv e^{iH_0 t/\hbar} | \Psi_S (t) \rangle
\end{eqnarray}
(* Note that in the exponent, it is not $H$ but $H_0 $.). Then
\begin{eqnarray}
i \hbar \frac{\partial}{\partial t} | \Psi_I (t) \rangle &=& -H_0 e^{iH_0 t/\hbar} | \Psi_S (t) \rangle
 + e^{iH_0 t/\hbar} i \hbar \frac{\partial}{\partial t} | \Psi_S (t) \rangle \nonumber \\
 &=& e^{iH_0 t/\hbar} [-H_0 +H] e^{-iH_0 t/\hbar} | \Psi_I (t) \rangle \nonumber \\
 &=& e^{iH_0 t/\hbar} H_I e^{-iH_0 t/\hbar} | \Psi_I (t) \rangle
 \nonumber.
\end{eqnarray}
We, therefore, obtain the following equations.
\begin{eqnarray}
&& i \hbar \frac{\partial}{\partial t} | \Psi_I (t) \rangle = H_I (t) | \Psi_I (t) \rangle \\
&& H_I (t) \equiv  e^{iH_0 t/\hbar} H_I e^{-iH_0 t/\hbar}
\end{eqnarray}
For any operators,
\begin{eqnarray}
&&\langle \Psi_S (t) | O_S | \Psi_S (t) \rangle  \nonumber \\
&& =\langle \Psi_I (t) | e^{iH_0 t/\hbar} O_S e^{-iH_0 t/\hbar} |
\Psi_I (t) \rangle .
\end{eqnarray}
Thus, we have
\begin{eqnarray}
O_I (t) \equiv  e^{iH_0 t/\hbar} O_S e^{-iH_0 t/\hbar}.
\end{eqnarray}
Differentiating the above equation with respect to time, one
obtains
\begin{eqnarray}
i \hbar \frac{\partial}{\partial t} O_I (t) &=& e^{iH_0 t/\hbar} (O_S H_0 -H_0 O_S ) e^{-iH_0 t/\hbar} \nonumber \\
&=& [O_I (t), H_0 ].
\end{eqnarray}
\\
We now solve the equation of motion in the interaction picture. Define a unitary operator that
determines the state vector at time $t$ in terms of the state vector at the time $t_0$
\begin{eqnarray}
| \Psi_I (t) \rangle = U(t,t_0 ) | \Psi_I (t_0 ) \rangle.
\end{eqnarray}
$U$ satisfies
\begin{eqnarray}
U(t_0 , t_0 )=1.
\end{eqnarray}
Using the Schr\"{o}dinger picture Eq.(49),
\begin{eqnarray}
&&| \Psi_I (t) \rangle =e^{iH_0 t/\hbar} | \Psi_S (t) \rangle
= e^{iH_0 t/\hbar} e^{-iH(t-t_0 )/\hbar} | \Psi_S (t_0 ) \rangle \nonumber \\
&&= e^{iH_0 t/\hbar} e^{-iH(t-t_0 )/\hbar} e^{-iH_0 t_0 /\hbar} | \Psi_I (t_0 ) \rangle  ,
\end{eqnarray}
which identifies
\begin{eqnarray}
U(t , t_0 ) = e^{iH_0 t/\hbar} e^{-iH(t-t_0 )/\hbar} e^{-iH_0 t_0
/\hbar}.
\end{eqnarray}
We can immediately show that \\
\hspace*{0.05cm} 1. $U^{\dagger}(t , t_0 )U(t , t_0 ) = U(t , t_0 )U^{\dagger}(t , t_0 ) = 1$ \\
\hspace*{0.7cm}which implies $U$ is unitary \\
\vspace{-0.5cm}
\begin{eqnarray}
\hspace{-12.6cm} U^{\dagger}(t , t_0 )=U^{-1}(t , t_0 ).
\end{eqnarray}
\vspace{-0.5cm} 
\begin{eqnarray}
\hspace{-12cm}2. \ \ U(t_1 , t_2 )U(t_2 , t_3 )=U(t_1 , t_3 ).
\end{eqnarray}
\vspace{-0.5cm} 
\begin{eqnarray}
\hspace{-13.4cm}3. \ \ U(t , t_0 )U(t_0 , t )=1.
\end{eqnarray}
Although Eq.(65) is the formal solution of Eq.(62), it is not very useful. \\
From Eq.(57), one can write
\begin{eqnarray}
i \hbar \frac{\partial}{\partial t}U(t , t_0 )| \Psi_I (t_0 ) \rangle =
H_I (t) U(t , t_0 )| \Psi_I (t_0 ) \rangle , \nonumber
\end{eqnarray}
which is true for arbitrary $| \Psi_I (t_0 ) \rangle$. \\
Thus, we obtain
\begin{eqnarray}
i \hbar \frac{\partial}{\partial t}U(t , t_0 ) = H_I (t) U(t , t_0
).
\end{eqnarray}
Integrating this equation from $t_0$ to $t$, it is shown
\begin{eqnarray}
U(t , t_0 )-U(t_0 , t_0 ) = -\frac{i}{\hbar} \int_{t_0}^{t} dt' H_I (t') U(t' , t_0 ) , \nonumber
\end{eqnarray}
which gives the relation,
\begin{eqnarray}
U(t , t_0 ) = 1 - \frac{i}{\hbar} \int_{t_0}^{t} dt' H_I (t') U(t'
, t_0 ).
\end{eqnarray}
Thus, we arrived at an integral equation for the unitary operator $U(t , t_0 )$. \\
The above equation is solved by iteration.
\begin{eqnarray}
\hspace*{-0.2cm}U(t , t_0 ) = 1 + \left( \frac{-i}{\hbar} \right) \int_{t_0}^{t} dt' H_I (t') +
\left( \frac{-i}{\hbar} \right)^2 \int_{t_0}^{t} dt' \int_{t_0}^{t'} dt'' H_I (t') H_I (t'') + \cdot \cdot \cdot
\end{eqnarray}
Consider the third term in this expansion. It may be written as
\begin{eqnarray}
&&\int_{t_0}^{t} dt' \int_{t_0}^{t'} dt'' H_I (t') H_I (t'') \nonumber \\
&&=\frac{1}{2}\int_{t_0}^{t} dt' \int_{t_0}^{t'} dt'' H_I (t') H_I
(t'') +\frac{1}{2} \int_{t_0}^{t} dt'' \int_{t''}^{t} dt' H_I (t')
H_I (t'').
\end{eqnarray}
\\ \hspace*{4cm}
\epsfig{figure=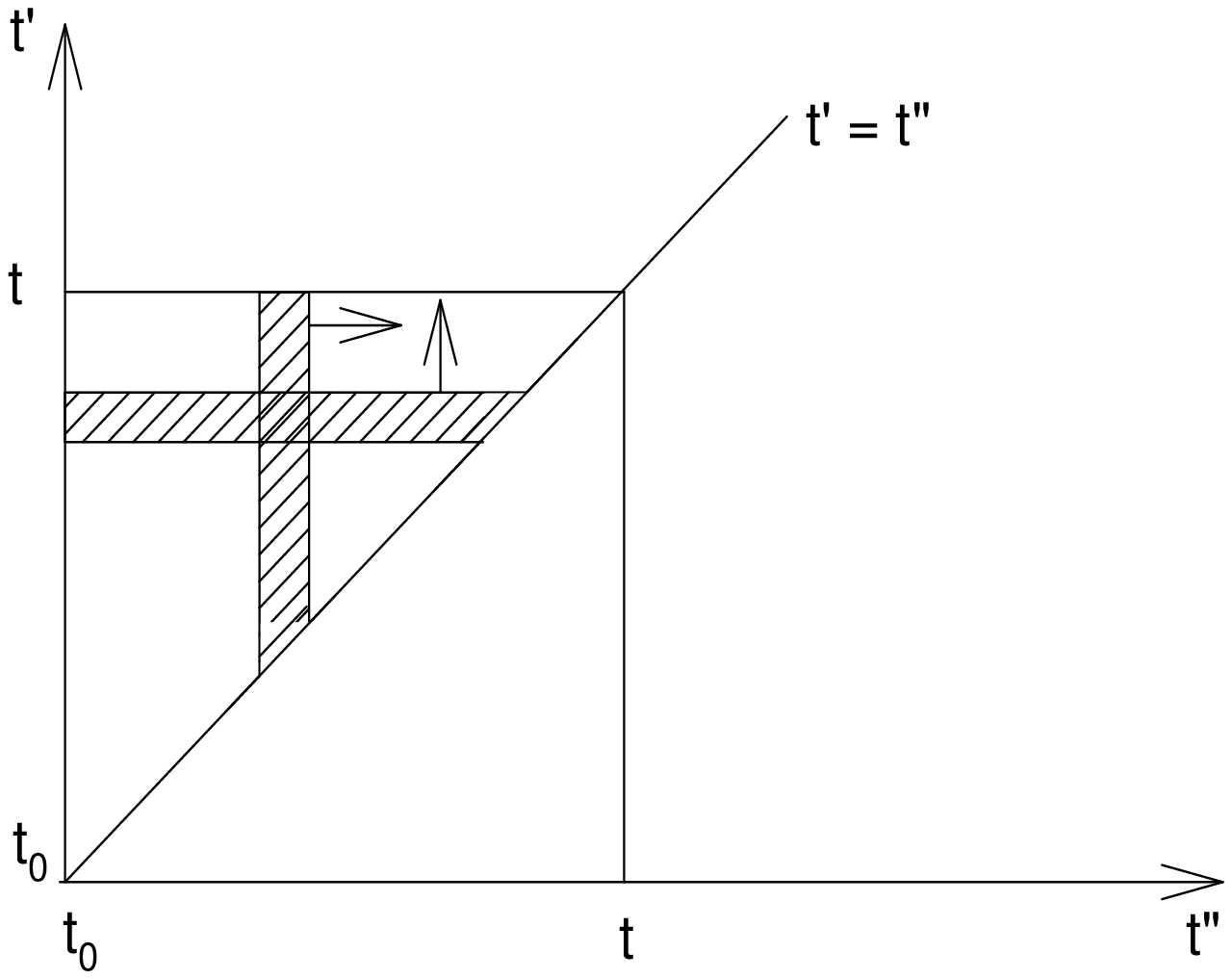,height=6cm,width=8cm} \\ \\
Now the second term becomes (with $t' \leftrightarrow t''$)
\begin{eqnarray}
\frac{1}{2} \int_{t_0}^{t} dt'' \int_{t''}^{t} dt' H_I (t') H_I
(t'') =\frac{1}{2} \int_{t_0}^{t} dt' \int_{t'}^{t} dt'' H_I (t'')
H_I (t'). \nonumber
\end{eqnarray}
Thus, we have altogether
\begin{eqnarray}
&&\int_{t_0}^{t} dt' \int_{t_0}^{t'} dt'' H_I (t') H_I (t'') \nonumber \\
&&=\frac{1}{2} \int_{t_0}^{t} dt' \int_{t_0}^{t} dt'' [H_I (t')
H_I (t'') \theta(t'-t'') +H_I (t'') H_I (t') \theta(t''-t')], \ \
\end{eqnarray}
where $\theta(t)$ is the step function. We note that the operator
containing the latest time stands farthest to the left. We call
this a time-ordered product of operators, denoted by the symbol
$T$. Thus the above expression can be rewritten
\begin{eqnarray}
\int_{t_0}^{t} dt' \int_{t_0}^{t'} dt'' H_I (t') H_I (t'')  \equiv
\frac{1}{2} \int_{t_0}^{t} dt' \int_{t_0}^{t} dt'' T [H_I (t') H_I
(t'')].
\end{eqnarray}
Finally, a general expression for $U$ is obtained,
\begin{eqnarray}
U(t , t_0 )= \sum_{n=0}^{\infty}  \left( \frac{-i}{\hbar}
\right)^{n}\frac{1}{n!}
 \int_{t_0}^{t} dt_1 \cdot \cdot \cdot \int_{t_0}^{t} dt_n  T [H_I (t_1 ) \cdot \cdot \cdot H_I (t_n
 )].
\end{eqnarray}
Now, a Heisenberg operator can be related to an interaction
operator using $U$.
\begin{eqnarray}
O_H (t) &=& e^{iHt/\hbar} O_S e^{-iHt/\hbar} \nonumber \\
&=& e^{iHt/\hbar} e^{-iH_0 t/\hbar} O_I e^{iH_0 t/\hbar}
e^{-iHt/\hbar}.
\end{eqnarray}
Eq.(65) yields
\begin{eqnarray}
O_H (t) = U(0 , t ) O_I (t) U(t , 0 ).
\end{eqnarray}
In addition
\begin{eqnarray}
| \Psi_H \rangle =| \Psi_S (0) \rangle =| \Psi_I (0) \rangle \ \  , \ \ O_S = O_H (0) = O_I (0) ,
\end{eqnarray}
so that all three pictures coincide at time $t=0$. \\
These relations provide a formal way to obtain the exact eigenstates of the interacting system
from the noninteracting ones. \\ \\
Since $| \Psi_H \rangle$ is independent of time, it satisfies the time-independent form of the
Schr\"{o}dinger equation
\begin{eqnarray}
H | \Psi_H \rangle = E | \Psi_H \rangle .
\end{eqnarray}
These state vectors are therefore the exact eigenstates of the system and are naturally very
complicated for an interacting system. This can be expressed
\begin{eqnarray}
| \Psi_H \rangle =| \Psi_I (0) \rangle =U(0 , t_0 )| \Psi_I (t_0 )
\rangle.
\end{eqnarray}
If we choose $| \Psi_I (t_0 ) \rangle = | \Psi_I (-\infty )
\rangle $ and $H= H_0 + e^{-\epsilon |t|} H_I $ and use the
concept of "adiabatic switching on", we can express an exact
eigenstate of the interacting system in terms of an eigenstate of
$H_0$.
\begin{eqnarray}
&&| \Psi_H \rangle =| \Psi_I (0) \rangle =U_{\epsilon}(0 , -\infty )| \Psi_0 \rangle , \\
&& H_0  | \Psi_0 \rangle = E_0 | \Psi_0 \rangle .
\end{eqnarray}
(For details, see Chapt. 3 of Gen. Ref. 3.) \\
The above relation allows a convenient and systematic calculation
of physical quantities using Green's function method in the
Heisenberg picture.

\subsection{Green's Function Method}
{\bf Zero-Temperature Formalism} \\ \\
The concept of a Green's function (or propagator) plays a fundamental role in the many-particle theory. \\
The single particle Green's function is defined by
\begin{eqnarray}
i G_{\alpha \beta} (xt, x't') = \frac{\langle \Psi_0 | T [\psi_{H \alpha}(xt)\psi_{H \beta}^{\dagger}(x't')]
|\Psi_0 \rangle }{\langle \Psi_0 |\Psi_0 \rangle} ,
\end{eqnarray}
where $|\Psi_0 \rangle$ is the Heisenberg ground state of the interacting system satisfying
\begin{eqnarray}
H  | \Psi_0 \rangle = E | \Psi_0 \rangle
\end{eqnarray}
and $\psi_{H \alpha}(x,t)$ is a Heisenberg operator,
\begin{eqnarray}
\psi_{H \alpha}(x,t) = e^{iH t/\hbar} \psi_{\alpha}(x) e^{-iH t/\hbar} .
\end{eqnarray}
Here, the indices $\alpha$ and $\beta$ label the components of the field operators;
$\alpha$ and $\beta$ may represent two values for spin $\frac{1}{2}$ fermions, whereas there are no
indices for spin zero bosons. \\
The $T$ product represents the time ordering,
\begin{eqnarray}
T[\psi_{H \alpha}(xt) \psi_{H \beta}^{\dagger}(x't')] =&&
\psi_{H \alpha}(xt) \psi_{H \beta}^{\dagger}(x't') \ \ \ t>t' , \nonumber \\
&\pm& \psi_{H \beta}^{\dagger}(x't')\psi_{H \alpha}(xt) \ \ \ t'>t ,
\end{eqnarray}
where the upper (lower) sign refers to bosons (fermions). \\ \\
Here, we ask why the Green's function is so useful. The answers are as follows: \\
\begin{itemize}
\item[1.] The Green's function offers a convenient basis for the perturbation expansion in combination
with the Feynman rules.
\item[2.] Although the ground-state expectation value implies the loss of much detailed information about
the ground state, still the single-particle Green's function contains the observable properties of great
interest:
\begin{itemize}
\item[(a)] The expectation value of any single-particle operator
in the ground state of the system. \item[(b)] The ground-state
energy of the system. \item[(c)] The excitation spectrum of the
system.
\end{itemize}
\end{itemize}
The single-particle operator is expressed as
\begin{eqnarray}
J= \int d^3 x  {\cal J}(x) ,
\end{eqnarray}
where
\begin{eqnarray}
{\cal J}(x) = \sum_{\alpha \beta} \psi_{\beta}^{\dagger}(x) J_{\beta \alpha}(x) \psi_{\alpha}(x) .
\end{eqnarray}
$J_{\beta \alpha}(x)$ is the first-quantized operator. \\
The ground-state expectation value of the operator density is given by
\begin{eqnarray}
\langle {\cal J}(x) \rangle &\equiv& \frac{\langle \Psi_0 | {\cal J}(x) | \Psi_0 \rangle}
{\langle \Psi_0 |\Psi_0 \rangle}  \nonumber \\
&=& \lim_{x' \rightarrow x} \sum_{\alpha \beta} J_{\beta \alpha }(x)
\frac{\langle \Psi_0 | \psi_{\beta}^{\dagger}(x') \psi_{\alpha}(x) | \Psi_0 \rangle}
{\langle \Psi_0 |\Psi_0 \rangle} \nonumber \\
&=&\pm i \lim_{t' \rightarrow t^{+}} \lim_{x' \rightarrow x} \sum_{\alpha \beta} J_{\beta \alpha }(x)
G_{\alpha \beta}(xt, x't') \nonumber \\
&=&\pm i \lim_{t' \rightarrow t^{+}} \lim_{x' \rightarrow x} {\rm
tr} [J(x) G(xt, x't')].
\end{eqnarray}
For example, the number density operator, $n(x) =
\psi^{\dagger}(x) \psi(x)$, becomes
\begin{eqnarray}
\langle n(x) \rangle = \pm i {\rm tr} G(xt, xt^{+}) .
\end{eqnarray}
The kinetic energy $\langle T \rangle$ is
\begin{eqnarray}
\langle T \rangle = \pm i\int d^3 x  \lim_{x' \rightarrow x} \left[   -\frac{\hbar^2}{2m} \nabla^2
{\rm tr} G(xt, x't^{+}) \right]
\end{eqnarray}
and the potential energy is
\begin{eqnarray}
\langle V \rangle = \pm \frac{1}{2} i\int d^3 x \lim_{t'
\rightarrow t^{+}} \lim_{x' \rightarrow x} \sum_{\alpha}\left[  i
\hbar \frac{\partial}{\partial t} + \frac{\hbar^2 \nabla^2}{2m}
\right] G_{\alpha \alpha}(xt, x't').
\end{eqnarray}
\\
$<${\bf Problem}$>$ Prove Eq.(92). \\ \\
The total ground state energy is solely given in terms of the single-particle Green's function.
\begin{eqnarray}
E&=& \langle T+V \rangle =\langle H \rangle  \nonumber \\
&=& \pm \frac{1}{2} i\int d^3 x \lim_{t' \rightarrow t^{+}} \lim_{x' \rightarrow x}
\left[  i \hbar \frac{\partial}{\partial t} - \frac{\hbar^2 \nabla^2}{2m} \right]
{\rm tr} G(xt, x't')
\end{eqnarray}
In addition to the above useful quantities, we can also obtain information on the excitation
spectrum of the system [Section 7, Gen. Ref. 3. ]. \\ \\
Now the usefulness of the above relations boils down to the point
that how accurately we can calculate the Green's function.  In
this lecture, it is not intended to explain the Green's function
method. Here, we only mention that the Green's function method leads into a perturbative expansion without proof. \\ \\
The exact Green's function may be written as
\begin{eqnarray}
i G_{\alpha \beta}(x,y) &=& \sum_{\nu=0}^{\infty} \left( \frac{-i}{\hbar} \right)^{\nu} \frac{1}{\nu !}
\int_{-\infty}^{\infty} dt_1 \cdot \cdot \cdot \int_{-\infty}^{\infty} dt_\nu  \nonumber \\
&\times& \frac{\langle \Psi_0 | T[H_{1}(t_{1}) \cdot \cdot \cdot
 H_{1}(t_{\nu}) \psi_{\alpha}(x) \psi_{\beta}^{\dagger}(y)] |\Psi_0 \rangle}{\langle \Psi_0 | S |\Psi_0
 \rangle}.
\end{eqnarray}
Here, $ S=U_{\epsilon}(\infty , -\infty )$. The above expression
clearly shows that we can calculate the Green's function order by
order of the
perturbation, $H_{1}$, up to the desired order. This is greatly facilitated by the Feynman diagram method. \\ \\
{\bf Dyson's equation} \\ \\
The concept of the Green's function method can be beautifully
represented by the Dyson's equation pictorially.
\begin{eqnarray}
G = G_0 + G_0  \Sigma G_0  ,
\end{eqnarray}
where $\Sigma$ represents the self-energy due to the interaction and $G_0$ the free particle
 Green's function. This equation is graphically represented. \\ \\
\hspace*{2.5cm} \epsfig{figure=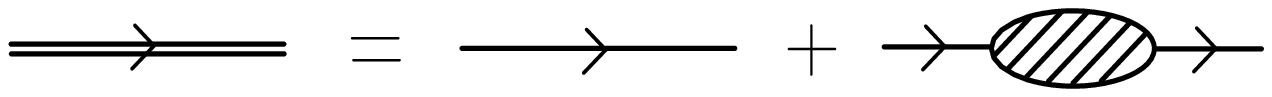,height=1.2cm,width=12.5cm}
\\ \\
 Using the concept of the proper self-energy, this expression can be modified as
\begin{eqnarray}
G = G_0 + G_0  \Sigma^{*} G  ,
\end{eqnarray}
which is graphically given \\ \\
\hspace*{2.5cm} \epsfig{figure=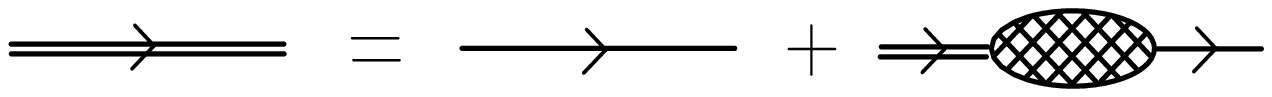,height=1.2cm,width=12.5cm}
\\ \\
A proper self-energy is the self-energy insertion that can not be
separated into two pieces by cutting a
single particle line. \\
Eq.(96) and its graphic representation clearly shows the perturbative nature of the formal solution
as follows
\begin{eqnarray}
G = G_0 + G_0  \Sigma^{*} G_0 + G_0  \Sigma^{*} G_0 \Sigma^{*} G_0
+ \cdot \cdot \cdot .
\end{eqnarray}
This perturbation expansion allows one to calculate the Green's
function and consequently associated physical observables as
accurately as possible when the perturbation (or the mutual
interaction)
is weak. \\ \\
{\bf Finite-Temperature Formalism} \\ \\
At finite temperatures, it is necessary to separate the
calculation into two parts. The first step is the introduction of
a temperature Green's function ${\cal G}$. This function has a
simple perturbation expansion similar to that for $G$ at $T=0$ and
also enable us to evaluate the equilibrium thermodynamic
properties of the system. The second step then relates ${\cal G}$
to a time-dependent Green's function that describes the  linear
response of the system to an external perturbation and provides
the excitation energies of the system containing one more
or one less particle. \\ \\
At finite temperature, it is most convenient to use the grand canonical ensemble, whose grand Hamiltonian is
given by
\begin{eqnarray}
K=H-\mu N .
\end{eqnarray}
The grand partition function is written as
\begin{eqnarray}
Z_G &\equiv& \sum_{N} \sum_{j} e^{-\beta (E_j -\mu N )}   \nonumber \\
&=& \sum_{N} \sum_{j} \langle N_j | e^{-\beta (H-\mu N)} | N_j \rangle  \nonumber \\
&=& {\rm Tr} \left(  e^{-\beta (H-\mu N)} \right)  \nonumber \\
&=& {\rm Tr} \left(  e^{-\beta K} \right) .
\end{eqnarray}
The thermodynamic potential is given by
\begin{eqnarray}
\Omega(T,V,\mu) = - k_B T \ln Z_G .
\end{eqnarray}
The statistical  operator $\rho_G$ is given by
\begin{eqnarray}
\rho_G &=& Z_G^{-1}  e^{-\beta (H-\mu N)} \nonumber \\
&=& e^{\beta (\Omega -K)}.
\end{eqnarray}
For any operator, the ensemble average $\langle {\cal O} \rangle$ is obtained with the prescription
\begin{eqnarray}
\langle {\cal O} \rangle &=& {\rm Tr} \left( \rho_G {\cal O} \right) =
{\rm Tr} \left( e^{\beta (\Omega -K)} {\cal O} \right)  \nonumber \\
&=& \frac{{\rm Tr} \left( e^{-\beta K} {\cal O} \right)}{{\rm
Tr}e^{-\beta K}}.
\end{eqnarray}
We introduce the modified Heisenberg picture operator
\begin{eqnarray}
{\cal O}_K (x,t) = e^{K \tau /\hbar} {\cal O}_S (x) e^{-K \tau
/\hbar}.
\end{eqnarray}
In particular, the field operators assume the form
\begin{eqnarray}
&&\psi_{K \alpha} (x, \tau) =  e^{K \tau /\hbar} \psi_{\alpha}(x) e^{-K \tau /\hbar} \nonumber, \\
&&\psi_{K \alpha}^{\dagger}(x, \tau) =  e^{K \tau /\hbar}
\psi_{\alpha}^{\dagger}(x) e^{-K \tau /\hbar}.
\end{eqnarray}
Note that $\psi_{K \alpha}^{\dagger}(x, \tau)$ is not the adjoint of $\psi_{K \alpha}(x, \tau)$
as long as $\tau$ is real. \\
The single-particle temperature Green's function is defined as
\begin{eqnarray}
{\cal G}_{\alpha \beta}(x \tau, x' \tau') \equiv {\rm Tr} \left\{
\rho_G T_{\tau} [\psi_{K \alpha}(x, \tau) \psi_{K
\beta}^{\dagger}(x', \tau')]  \right\}.
\end{eqnarray}
Here the symbol $T_{\tau}$ orders the operators with smaller $\tau$ at the right. \\ \\
The temperature Green's function is useful because it enables us to calculate the
thermodynamic behavior of the system. If $H$ is time independent, then ${\cal G}$ depends only on the
combination $\tau - \tau'$ and not on $\tau$ and $\tau'$ separately. \\ \\
By definition,
\begin{eqnarray}
{\rm tr}{\cal G}(x \tau, x \tau^{+}) &=& \mp \sum_{\alpha} {\rm Tr} [\rho_G \psi_{K \alpha}^{\dagger}(x, \tau)
\psi_{K \alpha}(x, \tau)] \nonumber \\
&=& \mp e^{\beta \Omega} \sum_{\alpha} {\rm Tr} \left[ e^{-\beta K} e^{K \tau /\hbar}
\psi_{\alpha}^{\dagger}(x)\psi_{\alpha}(x) e^{-K \tau /\hbar} \right] \nonumber \\
&=& \mp e^{\beta \Omega} \sum_{\alpha} {\rm Tr} \left[  e^{-\beta K}
\psi_{\alpha}^{\dagger}(x)\psi_{\alpha}(x)\right]  \nonumber \\
&=& \mp \langle n(x)  \rangle .
\end{eqnarray}
A single-particle operator can be written
\begin{eqnarray}
\langle J  \rangle &=&  {\rm Tr} (\rho_G J)  \nonumber \\
&=& \sum_{\alpha \beta} \int d^3 x \lim_{x' \rightarrow x} \lim_{\tau' \rightarrow \tau^{+}}
J_{\beta \alpha }(x) {\cal G}_{\alpha \beta}(x \tau, x' \tau')   \nonumber \\
&=& \mp \int d^3 x \lim_{x' \rightarrow x} \lim_{\tau' \rightarrow
\tau^{+}} {\rm tr}[J(x){\cal G}(x \tau, x' \tau') ].
\end{eqnarray}
Particular examples of interest are
\begin{eqnarray}
\langle \sigma  \rangle &=& \mp \int d^3 x {\rm tr} [\sigma {\cal G}(x \tau, x \tau^{+})], \\
\langle T  \rangle &=& \mp \int d^3 x \lim_{x' \rightarrow x}
\frac{-\hbar^2 \nabla^2}{2m}
  {\rm tr}{\cal G}(x \tau, x' \tau^{+}).
\end{eqnarray}
The ensemble average of two body operators usually requires a two-particle temperature Green's function. \\
However, the mean potential energy of two-body interactions can be solely expressed in terms of ${\cal G}$ as
in the zero-temperature case. The expression is given by
\begin{eqnarray}
\langle V  \rangle &=& \frac{1}{2} \int d^3 x \int d^3 x'' V(x-x'') {\rm Tr}[\rho_G \psi_{\alpha}^{\dagger}(x)
\psi_{\gamma}^{\dagger}(x'') \psi_{\gamma}(x'')\psi_{\alpha}(x)] \nonumber \\
&=& \mp \frac{1}{2} \int d^3 x \lim_{x' \rightarrow x} \lim_{\tau'
\rightarrow \tau^{+}} \left[ -\hbar \frac{\partial}{\partial \tau}
+ \frac{\hbar^2 \nabla^2}{2m} + \mu \right] {\rm tr}{\cal G}(x
\tau, x' \tau').
\end{eqnarray}
\\
$<${\bf Problem}$>$ Starting from the Heisenberg equation of motion on $\psi_{K \alpha}(x \tau)$, prove
Eq.(110). \\ \\
The internal energy is given by
\begin{eqnarray}
E&=&\langle H \rangle =\langle T+V \rangle \nonumber \\
&=& \mp \frac{1}{2} \int d^3 x \lim_{x' \rightarrow x} \lim_{\tau'
\rightarrow \tau^{+}} \left[ -\hbar \frac{\partial}{\partial \tau}
-\frac{\hbar^2 \nabla^2}{2m} + \mu  \right] {\rm tr}{\cal G}(x
\tau, x' \tau').
\end{eqnarray}
Here, we give the expression for the thermodynamic potential in
terms of the single-particle Green's function without proof (For
details see Section 23, Gen. Ref. 3 ).
\begin{eqnarray}
\Omega(T,V,\mu) &=& \Omega_0 (T,V,\mu) \mp \int_{0}^{1} \lambda^{-1} d\lambda \int d^3 x \\
&& \lim_{x' \rightarrow x} \lim_{\tau' \rightarrow \tau^{+}}
\frac{1}{2} \left[ -\hbar \frac{\partial}{\partial \tau}
+\frac{\hbar^2 \nabla^2}{2m} + \mu  \right] {\rm tr}{\cal
G}^{\lambda}(x \tau, x' \tau'),
\end{eqnarray}
where the Hamiltonian is written as
\begin{eqnarray}
H(\lambda)=H_0 + \lambda H_1 .
\end{eqnarray}
Again, here we observe that the problem of obtaining useful physical quantities boils down to the point
how one can obtain exact Green's function. \\
We end this section without going into the Green's function
formalism. Instead, we just give the expression for the Green's
function which shows the perturbative nature of the formalism.
\begin{eqnarray}
&&\hspace{-1cm}{\rm tr}{\cal G}_{\alpha \beta}(x \tau, x' \tau') = -{\rm Tr} \left\{ e^{-\beta K_0} \sum_{n=0}^{\infty}
\frac{(-\hbar)^{-n}}{ n!}  \right. \nonumber \\
&&\hspace{-1cm} \left. \int_{0}^{\beta \hbar} d\tau_1 \cdot \cdot \cdot \int_{0}^{\beta \hbar} d\tau_n
T_{\tau}[K_{1}(\tau_1 ) \cdot \cdot \cdot K_{1}(\tau_n ) \psi_{I \alpha}(x \tau)
\psi_{I \beta}(x' \tau')] \right\} \nonumber \\
&& \hspace{-1cm}\Big/ {\rm Tr} \left\{ e^{-\beta K_0} \sum_{n=0}^{\infty}
\frac{(-\hbar)^{-n} }{n!}  \int_{0}^{\beta \hbar} d\tau_1 \cdot \cdot \cdot \int_{0}^{\beta \hbar} d\tau_n
T_{\tau}[K_{1}(\tau_1 ) \cdot \cdot \cdot K_{1}(\tau_n )]  \right\}
\end{eqnarray}
The denominator is just the perturbation expansion of $e^{-\beta
\Omega}$. It serves to eliminate all disconnected diagrams in the
numerator. It should be reminded that the Dyson's equations in the
zero-temperature are equally valid in the finite temperature
formalism. \\ \\
In this section, we showed that the Green's functions in the zero
and finite temperature formalism naturally lead into perturbative
expansions. The perturbative calculations are greatly facilitated
by the Feynman diagram rules. \\ \\
As will be shown later in this lecture, in general, perturbation
theory yields an asymtotic rather than convergent series. Under
appropriate circumstances, it may yield a useful physical
approximation but it cannot be applied to arbitrary precision.
Therefore, when the perturbation, which often represents
interactions between particles is large, then the outcome of the
Green's function method is in doubt. It should be noted that many
of challenging problems in condensed matter physics originate from
strongly correlated systems.  In order to handle such strongly
correlated systems, various non-perturbative schemes have been
developed. However, in many cases, the approaches lack systematic
means of improvement beyond the first obtained result. In later
chapters, we address this problem and present a method, which
provides a systematic improvement scheme beyond the mean field
results for strongly correlated systems.

\section{Schr\"{o}dinger Picture in Many-Body Theory }

\idn The Schr\"{o}dinger picture for many-body (field) theory is a natural extension of
one- or few-body quantum mechanics.

For ordinary quantum mechanics, we start with a Hamiltonian
operator and canonically quantize by postulating commutation
relations between coordinate of position operators and their
conjugate momenta. We reach the coordinate Schr\"{o}dinger picture
representation by representing the diagonal position operator with
its eigenvalues, and use a differential representation of the
commutators by replacing the conjugate momenta with derivatives.
Coordinate representation of state vectors are called
wavefunctions. The Schr\"{o}dinger equation becomes a differential
equation.

For many-body (field) theory in the Schr\"{o}dinger picture,
wavefunction should carry infinite number of degrees of freedom.
This can be accomplished by substituting 'function' by
'functional'. Differential representations of the canonical
commutators are obtained by replacing conjugate momenta with
functional derivatives. Coordinate representations of state
vectors or elements of Fock space are wavefunctionals. The
Schr\"{o}dinger equation is a functional differential equation,
whose solutions, the eigenfunctionals of the Hamiltonian
functional differential operator, represent possible states of the
system.

In addition to the functional Schr\"{o}dinger picture method, the
path integral formulation of many-body theory also requires
functional calculus. In this connection, this approach is called
functional integral method of many-particle (field) theory. In
this lecture, we will not treat functional integral method.
Instead, we refer to Gen. Ref. 2.

First, we learn about functional and its calculus.

\subsection{Functional Calculus}
{\bf Functional} \\ \\
\idn A functional space is an infinite-dimensional space where each point in the space is
a function on space-time. That is, each point in the function space is a mapping of
space-time into the real or complex numbers. A particular point in a function space will
map points in space-time to scalars, spinors or vectors.

\underline{ A mapping of the points in the function space to numbers is called a functional}.
Thus, a functional associates a number with each function on space-time.

For example, let $a$ be a point in a scalar function space ${\cal A}$. $a$ is a scalar
function on space-time, $a=a(x) \in {\cal C }$ or ${\cal R}$. Let $F$ be a functional on
${\cal A}$. $F$ maps points in ${\cal A}$ into numbers. $F=F[a]\in {\cal C }$ or ${\cal R}$.
A simple example of a functional is $F[a]= \int a(x) dx$. 

{\bf Functional Derivatives} \\

The functional derivative is formally defined by
\begin{eqnarray}
\frac{\delta F[a]}{\delta a} = \lim_{\epsilon \rightarrow 0} \frac{F[a+\epsilon \delta]
-F[a]}{\epsilon} ,
\end{eqnarray}
where $\delta$ is the Dirac $\delta$-function distribution. Similarly, we can define
a functional directional derivative in the direction of the function $\lambda(x)$ as
\begin{eqnarray}
\frac{\delta_{\lambda} F[a]}{\delta_{\lambda} a} = \lim_{\epsilon
\rightarrow 0} \frac{F[a+\epsilon {\lambda}] -F[a]}{\epsilon}.
\end{eqnarray}
\\ 
\hspace*{4cm} \epsfig{figure=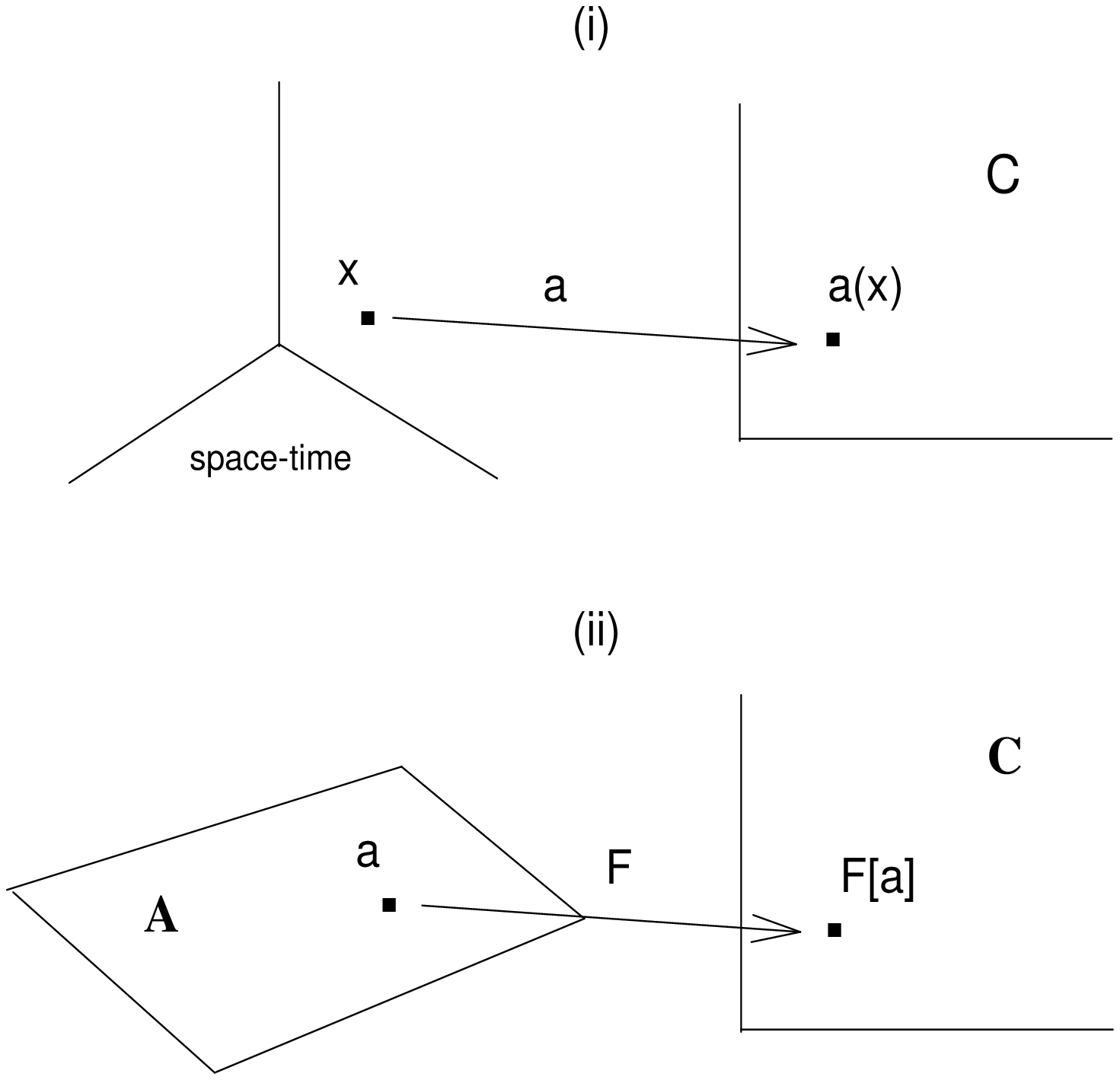,height=9cm,width=9.5cm}
{\small
\begin{itemize}
\item[] {\bf Fig. II.1}  (i) $a$ is a function on space-time. It maps points in space-time
        to real or complex numbers. \ \
        (ii) $a$ is a point in function space ${\cal A}$. The functional $F[a]$
        maps points in the function space ${\cal A}$ to real or complex numbers.
        The functional $F[a]$ turns functions into numbers.
\end{itemize}}
\vspace{0.6cm}
Also by ${\delta F[a]}/{\delta a(x)}$, we mean the change in $F$ with respect to a change
in the function $a$ at the point $x$ only. 

Formally
\begin{eqnarray}
\frac{\delta F[a]}{\delta a(x)} = \lim_{\epsilon \rightarrow 0} \frac{F[\hat{a}]
-F[a]}{\epsilon} .
\end{eqnarray}
Here, $\hat{a}$ is equal to $a$ except at the point $x$ where $\hat{a}= a+\epsilon$.
For example, if $F[a]= \int a(z) \delta(z-y) dz = a(y)$, then $F[\hat{a}]=\int
(a(z)+ \epsilon \delta(x-z)) \delta(z-y) dz = a(y) + \epsilon \delta(x-y)$.

From above, we observe that
\begin{eqnarray}
\frac{\delta a(y)}{\delta a(x)} &=& \lim_{\epsilon \rightarrow 0}
\frac{a(y)+\epsilon \delta(x-y)-a(y)}{\epsilon} \nonumber \\
&=& \delta(x-y) .
\end{eqnarray}
Similarly,
\begin{eqnarray}
\frac{\delta_{\lambda} a(y)}{\delta_{\lambda} a(x)} =
\lambda(x-y).
\end{eqnarray}
Important examples include the action functional ${\cal S} = \int
L(q_{i}, \dot{q}_{i}, t)dt$, the Lagrangian functional $L = \int
{\cal L}(\phi, \nabla \phi, \dot{\phi},t) d^3 x$ and the
Hamiltonian functional $H= \int (\pi(\vec{x},t)\cdot \partial_{t}
\phi(x,t)- {\cal L})d^3 x$. In fact, in deriving Eq.(8) from
Eq.(7) we have implicitly used the functional calculus.

As an exercise, we consider the action for the $\lambda \phi^4$
model in the scalar field theory (We use the scalar field theory
as a prototype to develop the functional Schr\"{o}dinger theory.).
The action is given by
\begin{eqnarray}
{\cal S}[\phi] &=& \int dt \int d^3 y {\cal L}(x)  \nonumber \\
&=& \int d^4 y \left[ \frac{1}{2} (\partial_{\mu} \phi
\partial^{\mu} \phi - m_{0}^{2} \phi^2 ) - \frac{\lambda_0}{4 !}
\phi^4 \right].
\end{eqnarray}
Here the notation
\begin{eqnarray}
ds^2 = dx^{\mu}dx_{\mu}=dt^2 -(dx^2 + dy^2 + dz^2 )
\end{eqnarray}
is used with $c=1$. We used
\begin{eqnarray}
&&x^{\mu} = (x^0 , x^1 , x^2 , x^3 )= (ct, x, y, z) , \nonumber \\
&&x_{\mu} = (x_0 , x_1 , x_2 , x_3 )= (ct, -x, -y, -z) .
\end{eqnarray}
Also, the differential operators are defined
\begin{eqnarray}
\partial_{\mu} = \frac{\partial}{\partial x^{\mu}} &=& (\partial_{0},\partial_{1},
\partial_{2},\partial_{3}) = \left( \frac{1}{c}\frac{\partial}{\partial t},
\frac{\partial}{\partial x},\frac{\partial}{\partial y},\frac{\partial}{\partial z}  \right)
\nonumber \\
&=& \left( \frac{1}{c}\frac{\partial}{\partial t}, \nabla  \right) , \nonumber \\
\partial^{\mu} = \frac{\partial}{\partial x_{\mu}} &=&
\left( \frac{1}{c}\frac{\partial}{\partial t}, -\nabla  \right) .
\end{eqnarray}
For future reference, we also list
\begin{eqnarray}
 \Box= \partial^{\mu}\partial_{\mu} &=& \frac{1}{c^2}\frac{\partial^2}{\partial t^2}
 -\left( \frac{\partial^2}{\partial x^2} + \frac{\partial^2}{\partial y^2}+
 \frac{\partial^2}{\partial z^2}  \right) \nonumber \\
 &=& \frac{1}{c^2}\frac{\partial^2}{\partial t^2} -\nabla^2 ,
\end{eqnarray}
called the d'Alembertian operator. \\
The energy-momentum 4-vector of a particle is
\begin{eqnarray}
p^{\mu}= \left( \frac{E}{c}, \vec{p}  \right) , \ \
p_{\mu}= \left( \frac{E}{c}, -\vec{p}  \right)
\end{eqnarray}
given the invariant
\begin{eqnarray}
p^2 = p^{\mu}p_{\mu} = \frac{E^2}{c^2} - \vec{p}\cdot \vec{p} = m^2 c^2
\end{eqnarray}
or, when $c=1$
\begin{eqnarray}
p^2 = E^2 - \vec{p}^2 = m^2 .
\end{eqnarray}
Often $p \cdot x$ is used for $p_{\mu} x^{\mu}$
\begin{eqnarray}
p \cdot x = p_{\mu} x^{\mu} = Et- \vec{p} \cdot \vec{r}.
\end{eqnarray}
From Eq.(121), we have
\begin{eqnarray}
&&\frac{\delta}{\delta \phi(x)} \int d^4 y \frac{1}{2}
(\partial^{\mu} \phi \partial_{\mu} \phi)
=-\frac{\delta}{\delta \phi(x)} \int d^4 y \frac{1}{2} ( \phi \partial^{\mu}\partial_{\mu} \phi) \nonumber \\
&&= -\int d^4 y  \delta^4 (x-y)\partial^{\mu}\partial_{\mu}
\phi(y) = -\partial^{\mu}\partial_{\mu} \phi(x) ,
\end{eqnarray}
and
\begin{eqnarray}
\frac{\delta}{\delta \phi(x)} \int d^4 y  \phi^n (y) &=& n \int d^4 y  \phi^{n-1} (y)\delta^4 (x-y) \nonumber \\
&=& n \phi^{n-1} (x) .
\end{eqnarray}
Thus
\begin{eqnarray}
\frac{\delta {\cal S}[\phi]}{\delta \phi(x)} = -\left(
\partial_{\mu}\partial^{\mu} \phi(x) + m_{0}^{2}\phi(x)  +
\frac{\lambda_0}{3!} \phi^3 (x) \right) = 0 ,
\end{eqnarray}
which is the Klein-Gordon equation with the $\phi^4$ interaction. \\ \\
{\bf Functional Differential Equations} \\

In the Schr\"{o}dinger picture of many-body theory, the Schr\"{o}dinger equation is a functional differential
equation. For an action ${\cal S}[\psi]$, the field equations are
\begin{eqnarray}
\frac{\delta {\cal S}[\psi]}{\delta \psi(x)} =0 .
\end{eqnarray}
A functional differential equation can be thought of as an infinite set of coupled partial differential
equations or a partial differential equation on an infinite-dimensional vector space.

For example, consider
\begin{eqnarray}
\int dy \left( \frac{\delta}{\delta a(y)} \frac{\delta}{\delta
a(y)} -b(y) \right) F[a] =0 .
\end{eqnarray}
This equation is satisfied by
\begin{eqnarray}
F[a]= \eta {\rm exp} ( \pm \int dx \sqrt{b(x)} a(x)).
\end{eqnarray}
\\
$<${\bf Problem}$>$ Show that the above $F[a]$ satisfies Eq.(134). \\

A more realistic form which may appear in the real problem is given by
\begin{eqnarray}
\int dy \frac{\delta}{\delta a(y)} \frac{\delta}{\delta a(y)} F[a]
+ \int dx dy f(x,y) a(x)a(y) F[a] = \mu F[a] .
\end{eqnarray}
Assume $f(x,y)$ is symmetric in $x$ and $y$.

First we substitute
\begin{eqnarray}
F[a]= \eta {\rm exp} (G[a]),
\end{eqnarray}
and solve the resulting equation for $G[a]$.
\begin{eqnarray}
&& \int dy \left( \frac{\delta}{\delta a(y)} \frac{\delta}{\delta
a(y)} G[a] +
 \left( \frac{\delta}{\delta a(y)} G[a] \right)^2 \right) \nonumber \\
 &&= \mu - \int dx dy f(x,y) a(x)a(y).
\end{eqnarray}
To proceed, we must match the power of $a(y)$ for both sides.
Thus, we write
\begin{eqnarray}
G[a]=\frac{1}{2} \int dx dy g(x,y) a(x)a(y).
\end{eqnarray}
$G[a]$ will be a solution provided the symmetric $g(x,y)$ satisfies
\begin{eqnarray}
\mu= \int dy g(y,y) \ \ {\rm and} \ \ f(x,y)= \int dz g(x,z) g(y,z) .
\end{eqnarray}
This technique of comparing the powers of both sides is called the 'power counting' method. \\ \\
{\bf Functional Integrals - Bosonic Variables} \\

Functional integrals are integrals of functionals over function
spaces and, in general, are not well defined. There are only two
types that can be done exactly. The first type is the
$\delta$-functional, $\delta[a-\xi]$.
\begin{eqnarray}
\int Da \delta[a-\xi] =1 .
\end{eqnarray}
$D$ in the measure indicates that the integral is over the
function space containing $a(\vec{x})$, which can be expressed
\begin{eqnarray}
Da = \prod_{\vec{x}} da(\vec{x}) .
\end{eqnarray}
Correspondingly,
\begin{eqnarray}
\delta[a-\xi] = \prod_{\vec{x}} \delta(a(\vec{x})-\xi(\vec{x})) .
\end{eqnarray}
Thus the integral in Eq.(141) is an infinite product of independent integrals, one for each $\vec{x}$.
Thus, we have
\begin{eqnarray}
&&\int Da F[a] \delta[a-\xi] \approx \int  \prod_{\vec{x}}  da(\vec{x}) \delta(a(\vec{x})-\xi(\vec{x})) F[a] \nonumber \\
&&= \prod_{\vec{x}} \int da(\vec{x}) \delta(a(\vec{x})-\xi(\vec{x})) F[a] \nonumber \\
&&= F[\xi].
\end{eqnarray}
Often it helps to think of the functions, such as $a(x)$, as
infinite dimensional (column) vectors where $x$ plays the role of
an index. With this view, a functional integral looks like the
natural infinite-dimensional limit of ordinary finite-dimensional
integrals. The second type of functional that we can integrate
exactly is the Gaussian functional.
\begin{eqnarray}
\int Da e^{-F[a]} , \ \ {\rm where} \ \ F[a]= \int dx a^2 (x)
\end{eqnarray}
We have
\begin{eqnarray}
&&\int Da e^{-\int dx a^2 (x)}  = \int \prod_{x} da(x) e^{-\sum_{x} a^2 (x)} \nonumber \\
&&=\prod_{x} \int da(x) e^{- a^2 (x)} = \prod_{x} \sqrt{\pi} ,
\end{eqnarray}
where we used $\int_{-\infty}^{\infty} dx e^{-\alpha x^2} = \sqrt{\frac{\pi}{\alpha}}$.

Following the same steps, we have
\begin{eqnarray}
\int Da e^{-\int dx f(x)a^2 (x)}  =\prod_{x} \sqrt{\frac{\pi}{f(x)}} .
\end{eqnarray}
We regard $f(x)$ as an infinite-dimensional diagonal matrix
\begin{eqnarray}
f(x,y)=f(x) \delta(x-y) .  \nonumber
\end{eqnarray}
$f(x)$ also represents the diagonal elements of the diagonal matrix. They are also the eigenvalues of the matrix. \\
The product of the eigenvalues is the determinant of the matrix, hence
\begin{eqnarray}
\prod_{x} \sqrt{\frac{\pi}{f(x)}} &\longrightarrow& (\sqrt{\pi})^{\infty} \prod_{x} \frac{1}{\sqrt{f(x)}}
= \frac{(\sqrt{\pi})^{\infty}}{\sqrt{{\rm det} f}}  \nonumber \\
&=& {\rm det}^{-1/2} \left(  \frac{f}{\pi} \right) ,
\end{eqnarray}
so
\begin{eqnarray}
 \int Da e^{-\int dx f(x)a^2 (x)} =  {\rm det}^{-1/2} \left(  \frac{f}{\pi}
 \right).
\end{eqnarray}
If the operator is not diagonal, the result is still the same.
\begin{eqnarray}
\int Da e^{-\int dx dy a(x) g(x,y) a(y)} =  \frac{(\sqrt{\pi})^{\infty}}{\sqrt{{\rm det} g}}
\end{eqnarray}
Finally, consider
\begin{eqnarray}
\int Da e^{F[a,J]} ,
\end{eqnarray}
where
\begin{eqnarray}
F[a,J] =  \int dx dy a(x) g(x,y) a(y) + \int dx  J(x) a(x)
\end{eqnarray}
is the Gaussian functional with the addition of a source term. \\
For the 1-dimensional case
\begin{eqnarray}
&& \int_{-\infty}^{\infty} dx e^{-\alpha x^2 +b x} = \int_{-\infty}^{\infty} dx e^{-\alpha (x-b/2)^{2}}
e^{1/4 (b \frac{1}{\alpha}b)} \nonumber \\
&&= e^{1/4 (b \frac{1}{\alpha}b)} \int dx e^{-\alpha x^{2}} =  e^{1/4 (b \frac{1}{\alpha}b)}
\sqrt{\frac{\pi}{\alpha}}  .
\end{eqnarray}
Thus,
\begin{eqnarray}
\int Da e^{-F[a,J]} = {\rm exp} \left( \frac{1}{4} \int dx dy J(x) g^{-1}(x-y) J(y)  \right)
 \frac{(\sqrt{\pi})^{\infty}}{\sqrt{{\rm det} g}},
\end{eqnarray}
where $g^{-1}(x-y)$ is the inverse operator (matrix) of $g(x-y)$,
\begin{eqnarray}
\int dz g^{-1}(x-z) g(z-y) = \delta(x-y).
\end{eqnarray}

The source term in the Gaussian allows us to easily compute the functional integral of any moment of
the Gaussian. Namely,
\begin{eqnarray}
&&\int Da a(x_1 ) \cdot \cdot \cdot a(x_n ) {\rm exp} ( -\int dx dy a(x) g(x,y) a(y)) \nonumber \\
&& = \frac{\delta}{\delta J(x_1 )} \cdot \cdot \cdot
\frac{\delta}{\delta J(x_n )} \int Da e^{-F[a,J]} \Big|_{J=0}.
\end{eqnarray}
Using the above relation, we obtain
\begin{eqnarray}
\int Da a(x_1 ) {\rm exp} (-\int dx dy a(x) g(x,y) a(y)) =0 .
\end{eqnarray}
and
\begin{eqnarray}
&&\int Da a(x_1 ) a(x_2 ){\rm exp} (-\int dx dy a(x) g(x,y) a(y)) \nonumber \\
&& = \frac{1}{2} g^{-1}(x_1 -x_2 )
\frac{(\sqrt{\pi})^{\infty}}{\sqrt{{\rm det} g}}.
\end{eqnarray}
In fact, any odd moment vanishes. Here, we note that the potentially bothersome infinite number
in the numerator is often absorbed in or canceled by a normalization. \\ \\
{\bf Fermionic Variables - Grassmann Algebra} \\

Since fermionic operators anticommute, we must introduce
anticommuting numbers and functions, which are called 'Grassmann
variables'. Consider an eigenvalue problem for the fermion field
operator,
\begin{eqnarray}
\psi(\vec{x}) | \phi \rangle = \phi(\vec{x}) | \phi \rangle .
\end{eqnarray}
For two such operators, a naive calculation shows
\begin{eqnarray}
&&\psi_1 \psi_2 | \phi \rangle  = \psi_1 \phi_2 | \phi \rangle =\phi_1 \phi_2 | \phi \rangle \nonumber \\
&=& -\psi_2 \psi_1 | \phi \rangle = -\psi_2 \phi_1 | \phi \rangle =-\phi_2 \phi_1 | \phi \rangle , \nonumber
\end{eqnarray}
which gives $\phi_1 \phi_2 =-\phi_2 \phi_1$. Also we note that $\phi^2 =0$ following the Pauli exclusion principle.

Let $\eta$ be such an anticommuting variable. This means that
\begin{eqnarray}
\{ \eta, \eta \} =0 ,
\end{eqnarray}
thus $\eta^2 =0$.  Let us define the derivative operator for an anticommuting variable in an analogous way
to the bosonic case,
\begin{eqnarray}
\left\{ \frac{d}{d\eta} , \eta  \right\} =1 .
\end{eqnarray}
For any $f(\eta)$, the power series expansion of $f$ must be of the form
\begin{eqnarray}
f(\eta) = a + b \eta ,
\end{eqnarray}
since $\eta^2 =0$.  Therefore, $d^2 f/ d\eta^2 =0$, so
\begin{eqnarray}
\left\{ \frac{d}{d\eta} , \frac{d}{d\eta}  \right\} = 0 .
\end{eqnarray}
When $a$ and $b$ are anticommuting numbers, then
\begin{eqnarray}
\frac{df}{d\eta} = -b .
\end{eqnarray}
We introduce integration by defining two following integrals :
\begin{eqnarray}
&& \int d\eta \equiv 0 ,  \nonumber \\
&& \int d\eta \eta \equiv 1 .
\end{eqnarray}
This definition is chosen so that $\int d\eta$ is translationally invariant and linear. It is not the inverse
operation of $d / d\eta$ ; in fact, integration acts like differentiation. For example, if $\eta_1$ and $\eta_2$
are two anticommuting variables, then
\begin{eqnarray}
\int d\eta_1 d\eta_2 \eta_1 \eta_2 &=& -\int  d\eta_1 d\eta_2 \eta_2 \eta_1 \nonumber \\
&=&-\int  d\eta_2 d\eta_1 \eta_1 \eta_2  \nonumber \\
&=& -1
\end{eqnarray}
Let's consider a space of anticommuting functions, $\eta(x)$, namely
\begin{eqnarray}
\left\{  \frac{\delta}{\delta \eta(x)}, \eta(y) \right\} = \delta(x-y) .
\end{eqnarray}
We carry out the functional integrals as in the bosonic case. From
the definition of integration, Eq.(165), we see that the
$\delta$-functional $\delta[\eta]$ is
\begin{eqnarray}
\delta[\eta] = \prod_{x} \eta(x) ,
\end{eqnarray}
so that
\begin{eqnarray}
\int D\eta \delta[\eta] =1 .
\end{eqnarray}
Here, we choose the ordering of the $d\eta(x)$'s in $D\eta = \prod_{x} d\eta(x)$ to be the opposite
of the ordering in the functional so that no extra minus signs are introduced in doing the integrals. Similarly
\begin{eqnarray}
 \delta[\eta-\sigma] = \prod_{x} (\eta(x)-\sigma(x)) ,
\end{eqnarray}
then,
\begin{eqnarray}
\int D\eta  \delta[\eta-\sigma]F[\eta] &=& \int \prod_{x} d\eta(x)   \prod_{x} (\eta(x)-\sigma(x))F[\eta] \nonumber \\
&=& F[\sigma].
\end{eqnarray}
Note that $F[\eta]$ is placed on the right side of the delta
function ( See p.28, Gen. Ref. 2. ) not to spoil the ordering
convention discussed above.
\\

$<${\bf Problem}$>$ Prove the above equation using one component
delta function  $\delta[\eta-\sigma] = \eta-\sigma$. \\

Now consider the Gaussian integral,
\begin{eqnarray}
\int D\eta {\rm exp} (-\int dx dy \eta(x) g(x,y) \eta(y)) .
\end{eqnarray}
$g(x,y)=-g(y,x)$ in order to get a non-vanishing result. For the 2-dimensional integral, we have
$\eta_1$ and $\eta_2$  so that
\begin{eqnarray}
&&\int d\eta_1 d\eta_2 e^{-\eta_1 g_{12} \eta_2 -\eta_2 g_{21} \eta_1} \nonumber \\
&&=\int d\eta_1 d\eta_2 e^{-2\eta_1 g_{12} \eta_2 }
\end{eqnarray}
where we used $\eta_2 g_{21} \eta_1 =-\eta_1 g_{21} \eta_2 =\eta_1 g_{12} \eta_2$. \\
Expand the exponential in a power series,
\begin{eqnarray}
e^{-2\eta_1 g_{12} \eta_2 } &=& 1-2\eta_1 g_{12} \eta_2 + 2(\eta_1 g_{12} \eta_2 )^2 + \cdot \cdot \cdot \nonumber \\
&=& 1-2\eta_1 g_{12} \eta_2 .
\end{eqnarray}
Finally, we obtain
\begin{eqnarray}
&&\int d\eta_1 d\eta_2 e^{-2\eta_1 g_{12} \eta_2 } \nonumber \\
&& = \int d\eta_1 d\eta_2 (1-2\eta_1 g_{12} \eta_2) = 2 g_{12} .
\end{eqnarray}
Since for an antisymmetric $2 \times 2$ matrix, ${\rm det} g = -g_{12} g_{21} = (g_{12})^2 $, we rewrite
\begin{eqnarray}
\int d\eta_1 d\eta_2 e^{-2\eta_1 g_{12} \eta_2 } = 2 \sqrt{{\rm
det} g}.
\end{eqnarray}
\\
$<${\bf Problem}$>$ Show that for the 4-dimensional case,
\begin{eqnarray}
&& \int d\eta_1 \cdot \cdot \cdot d\eta_4 e^{-2( \eta_1 g_{12} \eta_2 + \cdot \cdot \cdot
 +\eta_3 g_{34} \eta_4 )}  \nonumber \\
&& = \frac{2^3}{2!} (g_{12} g_{34} -g_{13} g_{24}+g_{14} g_{23}) =
4 \sqrt{{\rm det} g} .
\end{eqnarray}
We can show by induction
\begin{eqnarray}
&&\int d\eta_1 \cdot \cdot \cdot d\eta_N e^{-2( \eta_1 g_{12} \eta_2 + \cdot \cdot \cdot
 +\eta_{N-1} g_{N-1,N} \eta_N )}  \nonumber \\
&&=  (2)^{N/2} \sqrt{{\rm det} g}.
\end{eqnarray}
We assert that in the $N \rightarrow \infty$ limit
\begin{eqnarray}
\int D \eta {\rm exp} (-\int dx dy \eta(x) g(x,y) \eta(y)) = (
\sqrt{2} )^{\infty} \sqrt{{\rm det} g}.
\end{eqnarray}
Often, it is necessary to consider the fields to be complex. For bosonic functional integrals,
the generalization is straightforward. Let $a(x)$ be an ordinary complex function with
$a(x)=(a_1 (x) + i a_2 (x))/\sqrt{2}$ where  $a_1 (x)$ and $a_2 (x)$ are real valued. \\
If we switch variables from $(a^*, a)$ to $(a_1 , ia_2 )$, the Gaussian integral can be readily performed to be
\begin{eqnarray}
&&\int Da^* Da {\rm exp} (-\int dx dy a^*(x) f(x,y) a(y) ) \nonumber \\
&&=\int Da_1 D(ia_2 ) {\rm exp} (-\int dx dy (a_1 (x)-ia_2 (x)) f(x,y)(a_1 (y)+i a_2 (y) ) )  \nonumber \\
&&=\frac{(i \pi)^{\infty}}{{\rm det} f} .
\end{eqnarray}
In the second step, $f(x,y)$ is assumed symmetric and can be
diagonalized so that the integral gives two copies of Eq. (150).

The fermionic Gaussian functional integral over complex anticommuting functions requires a little more attention. \\
Let $\eta$ be a complex anticommuting variable and write $\eta = (\xi_1 + i \xi_2 )/ \sqrt{2}$
where $\xi_1$ and $\xi_2$ are real-valued. Then $\eta^* = (\xi_1 -i \xi_2 )/\sqrt{2}$ . \\
Let
\begin{eqnarray}
&&\int d\eta \eta =1 \ \ , \ \ \int d\eta^* \eta^* =1 ,  \nonumber \\
&& \int d\eta = \int d\eta^* = \int d\eta \eta^* = \int d\eta^* \eta = 0 .
\end{eqnarray}
This means that $d\eta = (d\xi_1 - i d\xi_2 )/ \sqrt{2}$ and $d\eta^* = (d\xi_1 + i d\xi_2 )/ \sqrt{2}$ . \\ \\
$<${\bf Problem}$>$ Show that using Eq.(181), $d\eta = (d\xi_1 - i d\xi_2 )/ \sqrt{2}$ and
$d\eta^* = (d\xi_1 + i d\xi_2 )/ \sqrt{2}$ . \\

Thus, we have $d\eta^* d\eta = -i d\xi_1 d\xi_2$. The 1-dimensional Gaussian integral is simply
( $m$ : an ordinary commuting number)
\begin{eqnarray}
\int d\eta^* d\eta {\rm exp} (-\eta^* m \eta) = \int d\eta^* d\eta
(1 -\eta^* m \eta) = m.
\end{eqnarray}
The 2-dimensional Gaussian integral is
\begin{eqnarray}
&&\int d\eta_{1}^* d\eta_{2}^*  d\eta_{1} d\eta_{2}  {\rm exp}
\left(-\sum_{ij=1}^{2} \eta_{i}^* m_{ij} \eta_j \right) \nonumber \\
&&= \int d\eta_{1}^* d\eta_{2}^*  d\eta_{1} d\eta_{2} \Big[ 1-
(\eta_{1}^* m_{11} \eta_{1}
+\eta_{1}^* m_{12} \eta_{2} +\eta_{2}^* m_{21} \eta_{1} +\eta_{2}^* m_{22} \eta_{2})   \nonumber \\
&&\left. + \frac{1}{2!} (
 \ \ \ \ \ \ \ _{''} \ \ \ \ \ \ \ )^2 + \cdot \cdot \cdot \right]  \nonumber \\
&&= - \int d\eta_{1}^* d\eta_{2}^*  d\eta_{1} d\eta_{2}
\Big(\eta_{2}^{*}\eta_{1}^{*}\eta_{2}\eta_{1}
(m_{11}m_{22} -m_{12}m_{21})\Big) \nonumber \\
&&= -{\rm det} m_{ij}.
\end{eqnarray}
In order to avoid the annoying sign problem, we define
\begin{eqnarray}
D\eta^* D\eta = \prod_x d\eta^{*}(x) d\eta(x) .
\end{eqnarray}
Therefore, we should order
$ d\eta_{1}^* d\eta_{1} d\eta_{2}^*   d\eta_{2} \cdot \cdot \cdot d\eta_{N}^*   d\eta_{N} $.
Then, we have
\begin{eqnarray}
\int D\eta^* D\eta  {\rm exp} \left(-\int dx dy \eta^{*}(x) g(x,y) \eta(y) \right) =  {\rm det} g.
\end{eqnarray}
The hyperbolic fermionic Gaussian integral $\int D\eta^* D\eta
{\rm exp} \left(+\int dx dy \eta^{*}(x) g(x,y) \eta(y) \right)$
has no well-defined bosonic analogue. In order to get a
well-defined limit, we should reverse the order, so that $D\eta^*
D\eta = \prod_x (d\eta(x) d\eta^{*}(x))$. Thus,
\begin{eqnarray}
\int D\eta^* D\eta  {\rm exp} \left(+\int dx dy \eta^{*}(x) g(x,y)
\eta(y) \right) = {\rm det} g .
\end{eqnarray}
Functional integral techniques using the above results have
emerged in last two decades as a powerful tool for study of
many-particle systems. Unfortunately, this topic is out of the
scope of this lecture. Instead, we refer to Gen. Ref. 2 for a
detailed treatment on this topic.

\subsection{Functional Schr\"{o}dinger Picture -Schr\"{o}dinger Picture in The Second Quantized Formalism}

We have learned that the Heisenberg picture method provides a
convenient tool to study the many-particle system through Green's
function. However, when the mutual interaction is strong,
perturbative approaches are known less than reliable. Also when
one studies systems such as nonequilibrium many-particle systems
and evolution of a system in which time variation is important,
Green's function method may not be a good starting point.

In this  connection, field theorists have been studying the
functional Schr\"{o}dinger picture field theory to provide an
alternative theory to the Green's functions formalism.

In quantum mechanics (first quantized system), coordinate and
conjugate momentum are represented by $q_{i}(t)$ and $p_{i}(t)$.
In second quantized system, the field $\psi(\vec{x}, t)$ and its
conjugate momentum are used to
represent the infinite degrees of freedom. \\ \\
{\bf Bosonic Theory} \\

We now work in a coordinate Schr\"{o}dinger picture but with an
infinite degree of freedom. In analogy to $\hat{x} | x \rangle = x
| x \rangle$ in quantum mechanics, we consider
\begin{eqnarray}
\psi(\vec{x}) | \phi \rangle = \phi(\vec{x}) | \phi \rangle  ,
\end{eqnarray}
where $\phi(\vec{x})$ is the eigenvalue function corresponding to
the field operator $\psi(\vec{x})$. $| \phi \rangle$ represents an
eigenstate. The coordinate representation of the state $| \Psi
\rangle$, now time dependent, is the wave functional,
$\Psi[\phi]$.
\begin{eqnarray}
\Psi[\phi] = \langle \phi |\Psi\rangle
\end{eqnarray}
In analogy to the quantum mechanical commutation relation $[q_i , p_j ]= i \hbar \delta_{i j}$, we require
\begin{eqnarray}
[\phi(\vec{x}) , \pi(\vec{x'})] = i \hbar \delta (\vec{x} -\vec{x'}) .
\end{eqnarray}
This requirement is easily satisfied by
\begin{eqnarray}
\pi(\vec{x}) = \frac{\hbar}{i} \frac{\delta}{\delta \phi(\vec{x})} .
\end{eqnarray}
Compare this expression with the corresponding quantum mechanical one $p= \frac{\hbar}{i}
\frac{\partial}{\partial x}$ .

In terms of the coordinate basis,
\begin{eqnarray}
\langle \phi' | \pi(\vec{x}) | \phi \rangle = \frac{\hbar}{i}
\frac{\delta}{\delta \phi(\vec{x})} \delta[\phi -\phi'].
\end{eqnarray}
With this prescription, the Schr\"{o}dinger equation
\begin{eqnarray}
i\hbar \frac{\partial}{\partial t} | \Psi \rangle = H | \Psi \rangle
\end{eqnarray}
turns into a functional differential equation
\begin{eqnarray}
i\hbar \frac{\partial}{\partial t} \Psi[\phi,t] = H[\phi, \frac{\hbar}{i} \frac{\delta}{\delta \phi}]\Psi[\phi,t] .
\end{eqnarray}
In a time-independent case,
\begin{eqnarray}
H[\phi, \frac{\hbar}{i} \frac{\delta}{\delta \phi}]\Psi[\phi] =
E \Psi[\phi] ,
\end{eqnarray}
with $\Psi[\phi,t]= e^{-iEt/\hbar} \Psi[\phi]$.

As an example of the bosonic formalism developed above, the free
scalar theory will be studied. As it is a convention in the field
theory, we set $\hbar = c=1$. The action for the free scalar
theory is given (See Eq.(121).)
\begin{eqnarray}
{\cal S}[\psi] &=& \frac{1}{2} \int d^4 x {\cal L}(x) \nonumber \\
&=& \frac{1}{2} \int d^4 x \left(  \partial^{\mu} \phi(x)
\partial_{\mu} \phi(x) - m^{2} \phi^{2}(x)  \right) .
\end{eqnarray}
The corresponding Hamiltonian is defined by the Legendre transform of the Lagrangian
\begin{eqnarray}
H= \int d^3 x \left( \pi(\vec{x},t) \cdot \partial_{t}
\phi(\vec{x},t) - {\cal L}  \right).
\end{eqnarray}
Compare this with the quantum mechanical relation $H(\vec{q},
\vec{p}, t)= \sum_i \dot{q}_i p_i - L(\vec{q}, \dot{\vec{q}}, t)$.
Here, we note that the field momentum conjugate to $\phi$ is
\begin{eqnarray}
\pi(x)= \frac{{\partial \cal L}}{\partial (\partial_t \phi(x))} =
\dot{\phi}(x).
\end{eqnarray}
Using these relations, the Hamiltonian is written as
\begin{eqnarray}
H=\frac{1}{2}  \int d^3 x (\pi^2 + |\nabla \phi|^2 +m^2 \phi^2 ).
\end{eqnarray}
Using Eq.(190) for $\pi(\vec{x})$, the time-independent Schr\"{o}dinger functional equation is
expressed as
\begin{eqnarray}
\frac{1}{2}  \int d^3 x  \left( - \frac{\delta^2
\Psi[\phi]}{\delta \phi^2 (\vec{x})} + (|\nabla \phi |^2 + m^2
\phi^2 ) \Psi[\phi] \right) = E \Psi[\phi].
\end{eqnarray}
This equation can be solved exactly since there is no coupling term. \\ \\
$<${\bf Problem}$>$ Solve the functional differential equation (199) using a trial functional for the
ground state, $\Psi_0 [\phi]= \eta {\rm exp}(-G[\phi])$. Show that the ground state energy is given by
$E_0 = \frac{1}{2}  \int d^3 k \omega_k \delta^3 (0)$. (See Chapt. 10, Gen. Ref. 1. ). \\ \\
{\bf Fermionic Theory} \\

The functional Schr\"{o}dinger picture for fermionic
many-particles is more complicated than the bosonic case because
we should handle anticommuting Grassmann variables.

In order to deal with the situation, we examine the nature of the
fermionic field theory which provides the anticommutivity of the
field operators. In the fermionic field theory, one can use either
hermitian or non-hermitian field operators (For a somewhat deeper
discussion, see Appendix A.).

It should be noted that for neutral (chargeless) Majorana
fermions, hermitian operators should always be used, whereas for
charged spin 1/2 particles, both hermitian and non-hermitian
operators are acceptable. This comes from the fact that
non-hermitian operators can always be written as a sum of two
hermitian operators. For details of discussion, we refer to
Appendix A and references listed there. Here, we just present the
results.

When one uses hermitian field operators, the following equal-time
anticommutation relation is obtained
\begin{eqnarray}
\{ \psi_a (x), \psi_b (x') \}_{ x^{0} = {x^{0}}'}
= \delta_{ab} \delta( \vec{x}-\vec{x}' ) .
\end{eqnarray}
Here, $\psi^{\dagger}(x) = \psi(x)$. When non-hermitian field
operators are used, we obtain the equal-time anticommutation
relation,
\begin{eqnarray}
\{ \psi_a (x), \psi_b^{\dagger} (x') \}_{x^0 = {x^0}'} = \delta_{ab} \delta(\vec{x}-\vec{x}') ,   \\
\{ \psi_a (x), \psi_b (x') \}_{x^0 = {x^0}'} = 0 =
\{ \psi_a^{\dagger} (x), \psi_b^{\dagger} (x') \}_{x^0 = {x^0}'}
\end{eqnarray}
where $\psi^{\dagger}(x) \neq \psi(x)$.

As mentioned above, since non-hermitian field operator can be
obtained through a linear combination of hermitian operators, it
is convenient to start with hermitian operators. Since, in this
case, the anticommutation relation, Eq.(200), indicates that the
field operator itself corresponds to the conjugate momentum (with
some constant factor), it is not possible to obtain a simple
prescription as in the bosonic case. However, they can be
represented by a linear combination of their eigenvalues and the
functional differential operators ;
\begin{eqnarray}
\psi_a (\vec{x}) = \alpha \theta_a (\vec{x}) + \beta \frac{\delta}{\delta \theta_a (\vec{x})}
\end{eqnarray}
where $\theta_a (\vec{x})^* =\theta_a (\vec{x})$ and $ \{ \theta_a
(\vec{x}) , \theta_b (\vec{x}') \} =0$.  It must satisfy the
equal-time anticommutation relations, Eq.(200) and this imposes a
condition for the constants $\alpha$ and $\beta$,
$\alpha\beta= \frac{1}{2}$. \\ \\
$<${\bf Problem}$>$ Show that $\alpha\beta= \frac{1}{2}$. \\

Thus, a simple choice for the hermitian fermion field operators is
to take $|\alpha|=|\beta|= 1/\sqrt{2}$, which gives two equivalent
representations
\begin{eqnarray}
\psi_a (\vec{x}) = \frac{1}{\sqrt{2}}  \left[ \theta_a (\vec{x}) +  \frac{\delta}{\delta \theta_a (\vec{x})}
\right]
\end{eqnarray}
and
\begin{eqnarray}
\psi_a (\vec{x}) = \frac{1}{\sqrt{2} i}  \left[ \theta_a (\vec{x}) -  \frac{\delta}{\delta \theta_a (\vec{x})}
\right] .
\end{eqnarray}

The representation for the non-hermitian spinor field operators,
which are charge eigenvalues, can be constructed combining two
hermitian field $\psi_1$ and $\psi_2$,
\begin{eqnarray}
\psi = \psi_1 + i \psi_2 .
\end{eqnarray}
We also define a complex Grassmann variable $\theta(\vec{x})$ ;
\begin{eqnarray}
\theta(\vec{x})=\theta_1 (\vec{x})+ i \theta_2 (\vec{x}) .
\end{eqnarray}
Using Eq.(204) for both $\psi_1$ and $\psi_2$,
\begin{eqnarray}
&&\psi_1 = \frac{1}{\sqrt{2}} \left( \theta_1 + \frac{\delta}{\delta \theta_1}  \right) , \nonumber \\
&&\psi_2 = \frac{1}{\sqrt{2}} \left( \theta_2 + \frac{\delta}{\delta \theta_2}  \right) ,
\end{eqnarray}
we have
\begin{eqnarray}
&&\psi = \frac{1}{\sqrt{2}}  \left[ \theta(\vec{x}) +  \frac{\delta}{\delta \theta^* (\vec{x})} \right]  , \nonumber \\
&&\psi^{\dagger} = \frac{1}{\sqrt{2}}  \left[ \theta^* (\vec{x}) +  \frac{\delta}{\delta \theta(\vec{x})}
\right]  .
\end{eqnarray}
This prescription was first proposed by Floreanini and Jackiw [1].

Another way to obtain a representation for the non-hermitian field
is to use Eq.(204) and (205) together with a real Grassmann
variable $\theta$ ($\theta^* (\vec{x})=\theta(\vec{x}) $).
\begin{eqnarray}
&&\psi_1 = \frac{1}{\sqrt{2}} \left( \theta + \frac{\delta}{\delta \theta}  \right)  \nonumber \\
&&\psi_2 = \frac{1}{\sqrt{2} i} \left( \theta -
\frac{\delta}{\delta \theta}  \right)
\end{eqnarray}
From these expressions, we obtain
\begin{eqnarray}
&&\psi = \theta(\vec{x}) , \nonumber \\
&&\psi^{\dagger} = \frac{\delta}{\delta \theta(\vec{x})} ,
\end{eqnarray}
which was first proposed by Duncan, Meyer-Ortmans and Roskies (DMR) [2]. It should be noted that, in this
prescription, the hermitian conjugate of $\theta(\vec{x})$ is not given by the transpose of $\theta(\vec{x})$
but
\begin{eqnarray}
[\theta(\vec{x})]^{\dagger} = \frac{\delta}{\delta \theta(\vec{x})} .
\end{eqnarray}
It can be shown easily that the above two representations
Eq.(209) and (211) both satisfy the anticommutation relations
Eq.(201) and (202). For charged particles including electrons,
any of the above two representations can be used. We show, in the
next chapter, examples of applications for both representations to
physical problems. However, for neutral spin-$\frac{1}{2}$
particles, Eq.(204) or (203) with $\alpha\beta= \frac{1}{2}$
should be used.

Using the above prescription, we can readily write down the functional Schr\"{o}dinger equation for fermionic
particles. Especially when the Hamiltonian is time-independent,
\begin{eqnarray}
H[\psi^\dagger , \psi] | \Psi \rangle = H \left[
\frac{1}{\sqrt{2}} \left( \theta^* + \frac{\delta}{\delta \theta}
\right), \frac{1}{\sqrt{2}} \left( \theta + \frac{\delta}{\delta
\theta^* } \right) \right] | \Psi \rangle
\end{eqnarray}
with the Floreanini-Jackiw prescription. When the DMR prescription is used, we have
\begin{eqnarray}
H[\psi^\dagger , \psi] | \Psi \rangle = H \left[
\frac{\delta}{\delta \theta} , \theta \right] | \Psi \rangle .
\end{eqnarray}

In the functional  Schr\"{o}dinger picture representation of fermionic many-particles, the physical
information is given by the state functional $\langle \theta, t | \Psi \rangle $ and also by the scalar
product between states,
\begin{eqnarray}
\langle \Psi_1 | \Psi_2 \rangle = \langle \Psi_2 | \Psi_1 \rangle^*  .
\end{eqnarray}
Unlike the bosonic case, the dual to the state ket, here, is not a
simple hermitian conjugate. In order to investigate this problem,
we first consider a system of one degree of freedom described by
annihilation  and creation operators, $a^\dagger$ and $a$, which
satisfy
\begin{eqnarray}
&& \{ a, a^\dagger  \} = 1 , \nonumber \\
&& \{ a, a \} = 0 = \{ a^\dagger , a^\dagger  \}.
\end{eqnarray}
The Hilbert space of the system consists of vacuum and 1-particle states, $| 0 \rangle $ and $| 1 \rangle$ ;
\begin{eqnarray}
&& a | 0 \rangle  =0 , a | 1 \rangle  = | 0 \rangle  \nonumber \\
&& a^\dagger | 0 \rangle  = | 1 \rangle  , \ \ \ a^\dagger | 1 \rangle  = 0
\end{eqnarray}
We now want to represent the Hilbert space by functional $\Psi[\theta]$ of real Grassmann number $\theta$,
which satisfies $\theta \theta =0$.  The most general functional of $\theta$ may be written as
\begin{eqnarray}
\Psi[\theta] = \alpha + \beta \theta \ \ \ ( \alpha, \beta \ = \ {\rm constants}),
\end{eqnarray}
which implies that the Hilbert space is described by the two component basis $( 1, \theta )$. Thus,
a natural choice for the $\theta$-representation is to take
\begin{eqnarray}
| 0 \rangle  \longrightarrow 1 \ ,  \ \ \ | 1 \rangle  \longrightarrow \theta  ,
\end{eqnarray}
which imply
\begin{eqnarray}
\langle \theta | 0 \rangle = 1 \ \ {\rm and} \ \ \langle \theta | 1 \rangle = \theta .
\end{eqnarray}
From the above relation, we obtain the transformation function
\begin{eqnarray}
\langle \theta' | \theta \rangle &=& \langle \theta' | 0 \rangle  \langle 0 | \theta \rangle
+ \langle \theta' | 1 \rangle  \langle 1 | \theta \rangle  \nonumber \\
&=& 1+ \theta' \theta = e^{\theta' \theta} .
\end{eqnarray}
This relation shows that the $\theta$-basis is not orthonormal.
With the above relation, we now define the $\theta$-representation
of the dual state,
\begin{eqnarray}
\langle \Psi_1 | \Psi_2 \rangle &=& \int D \theta D \theta' \langle \Psi_1 | \theta' \rangle
\langle \theta' | \theta \rangle  \langle \theta  | \Psi_2 \rangle \nonumber \\
&\equiv& \int D \theta \bar{\Psi}_1 [\theta] \Psi_2 [\theta] ,
\end{eqnarray}
where $\Psi_2 [\theta] =\langle \theta | \Psi_2 \rangle $. Thus, the dual state $\bar{\Psi} [\theta]$
of ${\Psi} [\theta]$ is defined by
\begin{eqnarray}
\bar{\Psi} [\theta] &=& \int D \theta' \langle \Psi | \theta' \rangle \langle \theta' | \theta \rangle \nonumber \\
&=& \int D \theta' \Psi^{\dagger}[\theta'] e^{\theta'\theta} ,
\end{eqnarray}
where $\Psi^{\dagger}$ is the Hermitian conjugate of $\Psi$. This can be easily generalized to the case of
infinite degree of freedom as follows
\begin{eqnarray}
\bar{\Psi}[\theta(\vec{x})]= \int D \theta' \langle \Psi | \theta'(\vec{x}) \rangle e^{\theta'(\vec{x})
\theta(\vec{x})} ,
\end{eqnarray}
where the variable $\theta$ can have components in the internal space and the integral convention is used for
the repeated variables $\vec{x}$.

As an example, we consider the dual state of a Gaussian functional
\begin{eqnarray}
\Psi[\theta] = e^{\frac{1}{2} \theta_{a}(\vec{x}) G_{ab}(\vec{x},\vec{y}) \theta_{b}(\vec{y})} .
\end{eqnarray}
It can be shown easily that the dual is given by another Gaussian
functional (See Appendix B.)
\begin{eqnarray}
\bar{\Psi}[\theta] = C e^{\frac{1}{2} \theta_{a}(\vec{x}) \bar{G}_{ab}(\vec{x},\vec{y}) \theta_{b}(\vec{y})}  ,
\end{eqnarray}
where $C$ is a constant independent of $\theta$ and $\bar{G}$ is defined by
\begin{eqnarray}
\bar{G} = [G^{\dagger}]^{-1} .
\end{eqnarray}
In the case of the Floreanini-Jakiw representation where complex
$\theta$ is used, the above prescription for the dual becomes
slightly modified to become (See Appendix B.)
\begin{eqnarray}
\bar{\Psi}[\theta,\theta^{\dagger}] =  \int D \theta' D \theta'^{\dagger}
\langle \Psi | \theta', \theta'^{\dagger} \rangle e^{\theta'^{\dagger}(\vec{x})\theta(\vec{x}) -
\theta^{\dagger}(\vec{x})\theta'(\vec{x})} ,
\end{eqnarray}
where $\theta^{\dagger}$ denotes the transpose of the complex conjugate of $\theta$ .
However, the dual of a Gaussian functional
\begin{eqnarray}
{\Psi}[\theta,\theta^{\dagger}] = e^{\theta^\dagger G \theta}
\end{eqnarray}
has the same form as in the hermitian case of Eq.(226),
\begin{eqnarray}
\bar{\Psi}[\theta,\theta^{\dagger}] = C e^{\theta^\dagger \bar{G} \theta} ,
\end{eqnarray}
where $\bar{G}= {G^{\dagger}}^{-1}$. This simple property of the
Gaussian functional renders the functional Schr\"{o}dinger picture
useful for computation of physical quantities.

In the next chapter, we show some simple applications of the formalism developed so far.

\section{Applications to Many-Body Problems - Variational Approach }

\idn Many-body theoretical schemes, which are essentially
non-relativistic field theories have been all based on the
Heisenberg (interaction) picture, as we can see easily from the
conventional Green's function method and the functional integral
method. Only, very recently the functional Schr\"{o}dinger picture
method has been applied to many-particle systems with some
successes.

In this chapter, the functional Schr\"{o}dinger picture formalism
developed in Chapter II will be applied to many-particle systems
within variational approximation.

\subsection{Electron Gas and The Hubbard Model}
{\bf Electron Gas} \\ \\
\idn The Schr\"{o}dinger picture formalism provides a convenient basis for variational
approximation, since, with a trial functional $\langle u^\dagger , u | \Psi \rangle$,
the ground state energy of a many-particle system can be calculated easily.

As an example, we first study an interacting electron gas system [1].
The Hamiltonian is given by
\begin{eqnarray}
H &=& \sum_{\alpha} \int d^3 x \psi_{\alpha}^{\dagger}(\vec{x})
\left( -\frac{\hbar^2}{2m} \nabla^2 \right) \psi_{\alpha}(\vec{x})  \nonumber \\
&+& \frac{1}{2} \sum_{\alpha, \beta} \int d^3 x d^3 y V(\vec{x},\vec{y})
\psi_{\alpha}^{\dagger}(\vec{x}) \psi_{\beta}^{\dagger}(\vec{y})
\psi_{\beta}(\vec{y}) \psi_{\alpha}(\vec{x}) ,
\end{eqnarray}
where $\alpha$ and $\beta$ are spin indices. \\
For field operators $\psi^\dagger$ and $\psi$, we use the Floreanini-Jackiw formalism,
\begin{eqnarray}
&& \psi(\vec{x}) = \frac{1}{\sqrt{2}} \left( u(\vec{x}) +
\frac{\delta}{\delta u^{\dagger}(\vec{x})}  \right) , \nonumber \\
&& \psi^{\dagger}(\vec{x}) = \frac{1}{\sqrt{2}} \left( u^{\dagger}(\vec{x}) +
\frac{\delta}{\delta u(\vec{x})}  \right) .
\end{eqnarray}
We note that $\{ \psi_{i}(\vec{x},t), \psi_{i}^{\dagger}(\vec{x}',t) \} =
\delta_{ij} \delta(\vec{x}-\vec{x}')$ and $u$ and $u^\dagger$ are anticommuting
Grassmann variables. Substitution of field operators into the Hamiltonian turns
the many-particle Schr\"{o}dinger equation into a functional differential
equation,
\begin{eqnarray}
H[(1/\sqrt{2})(u+\delta/\delta u^\dagger ), (1/\sqrt{2})(u^\dagger
+\delta/\delta u )] | \Psi_0 \rangle  = E| \Psi_0 \rangle .
\end{eqnarray}

We now calculate the ground state energy of the above
many-particle system using variational approximation. A natural
choice for the trial functional is the Gaussian, which becomes an
exact solution to non-interacting many-particle systems.
\begin{eqnarray}
| \Psi_0 \rangle  &=& {\rm exp} \left( \sum_{\alpha,\beta} \int \int d^3 x d^3 y u_{\alpha}^{\dagger}(\vec{x})
G_{\alpha \beta}(\vec{x},\vec{y}) u_{\beta}(\vec{y}) \right) \nonumber \\
&=& {\rm exp}(u^{\dagger}_{A} G_{AB} u_B ) , \\
\langle \Psi_0 | &=& {\rm exp} \left( \sum_{\alpha,\beta} \int \int d^3 x d^3 y u_{\alpha}^{\dagger}(\vec{x})
\bar{G}_{\alpha \beta}(\vec{x},\vec{y}) u_{\beta}(\vec{y}) \right) \nonumber \\
&=& {\rm exp}(u^{\dagger}_{A} \bar{G}_{AB} u_B )  ,
\end{eqnarray}
where $\bar{G}=(G^{\dagger})^{-1}$ . Here, we used Eq.(230).

The expectation value of the Hamiltonian is given by
\begin{eqnarray}
\langle H \rangle &=& \frac{\langle \Psi_0 | H |\Psi_0  \rangle}{\langle \Psi_0 |\Psi_0  \rangle} \nonumber \\
&=& h_{AB} \frac{\langle \Psi_0 | \psi_{A}^{\dagger} \psi_B |\Psi_0  \rangle}{\langle \Psi_0 |\Psi_0  \rangle} \nonumber \\
&&+\frac{1}{2} V_{AB} \frac{\langle \Psi_0 | \psi_{A}^{\dagger}\psi_{B}^{\dagger}  \psi_B  \psi_A
|\Psi_0  \rangle}{\langle \Psi_0 |\Psi_0  \rangle} ,
\end{eqnarray}
where
\begin{eqnarray}
h_{AB} &=& -\frac{\hbar^2}{2m} \delta(\vec{x}-\vec{y})  \delta_{\alpha \beta} \nabla^2, \\
V_{AB} &=& V(\vec{x}, \vec{y}) = \frac{e^2}{|\vec{x}-\vec{y}|} .
\end{eqnarray}
The normalization constant $\langle \Psi_0 |\Psi_0  \rangle$ is
calculated easily
\begin{eqnarray}
\langle \Psi_0 |\Psi_0  \rangle &=& \int Du Du^\dagger e^{u^\dagger (G+\bar{G}) u} \nonumber \\
&=& \int Du Du^\dagger e^{u^\dagger S u} \nonumber \\
&=& {\rm Det} S ,
\end{eqnarray}
where $S=G+\bar{G}$ and Eq.(186) was used.

In order to calculate $\langle \Psi_0 | \psi_{A}^{\dagger} \psi_B |\Psi_0  \rangle$,
we first consider
\begin{eqnarray}
\psi_{A}^{\dagger} \psi_B |\Psi_0  \rangle &=& \frac{1}{2} \left( u_{A}^{\dagger}+ \frac{\delta}{\delta u_A }  \right)
\left( u_{B}+ \frac{\delta}{\delta u_{B}^{\dagger} }  \right) e^{u^{\dagger} G u} \nonumber \\
&=& \frac{1}{2} (u_{A}^{\dagger} u_B + \delta_{AB} + u_B u_{C}^{\dagger} G_{CA} +u_{A}^{\dagger} G_{BC} u_C \nonumber \\
&& +G_{BA}+G_{BC} u_C  u_{D}^{\dagger} G_{DA})|\Psi_0  \rangle .   \nonumber
\end{eqnarray}
Thus,
\begin{eqnarray}
&&\langle \Psi_0 | \psi_{A}^{\dagger} \psi_B |\Psi_0  \rangle  \nonumber \\
&&=\frac{1}{2}  \left[ (1+G)_{BA} \langle \Psi_0 |\Psi_0  \rangle
+(1+G)_{BC} (1-G)_{DA} \langle \Psi_0 |u_{D}^{\dagger} u_C |\Psi_0  \rangle \right] .
\end{eqnarray}
Also
\begin{eqnarray}
&&\langle \Psi_0 |u_{D}^{\dagger} u_C |\Psi_0  \rangle  \nonumber \\
&&= \int Du Du^\dagger  u_{D}^{\dagger} u_C  e^{u^{\dagger} S u} \nonumber \\
&&= \frac{\partial}{\partial S_{DC}} \langle \Psi_0 |\Psi_0  \rangle =
\frac{\partial}{\partial S_{DC}} {\rm Det} S \nonumber \\
&&= S_{CD}^{-1} {\rm Det} S = S_{CD}^{-1}\langle \Psi_0 |\Psi_0  \rangle .
\end{eqnarray}
Finally, we have
\begin{eqnarray}
\frac{\langle \Psi_0 | \psi_{A}^{\dagger} \psi_B |\Psi_0  \rangle}{\langle \Psi_0 |\Psi_0  \rangle}
= \frac{1}{2} \Omega_{BA} ,
\end{eqnarray}
where
\begin{eqnarray}
\Omega_{BA} &=& [(I + G)S^{-1}(I+\bar{G})]_{BA} \nonumber \\
&=& [(I + G)(G+\bar{G})^{-1}(I+\bar{G})]_{BA} .
\end{eqnarray}
Here $I$ is the $2 \times 2$ identity matrix. \\
Similarly, we obtain
\begin{eqnarray}
\frac{\langle \Psi_0 | \psi_{A}^{\dagger}\psi_{B}^{\dagger} \psi_B
\psi_A |\Psi_0  \rangle}{\langle \Psi_0 |\Psi_0  \rangle} =
\frac{1}{4} [\Omega_{AA}\Omega_{BB}-\Omega_{BA}\Omega_{AB}] .
\end{eqnarray}
\\
$<${\bf Problem}$>$ Derive Eq.(244). \\

Collecting the above results, we have
\begin{eqnarray}
\langle H \rangle &=& \frac{1}{2} \sum_{\alpha} \int d^3 x d^3 y
\left[ -\frac{\hbar^2}{2m}  \delta(\vec{x}-\vec{y}) \nabla^2 \Omega_{\alpha \alpha}(\vec{x},\vec{y}) \right] \nonumber \\
&+&\frac{1}{8} \sum_{\alpha,\beta} \int d^3 x d^3 y
V(\vec{x},\vec{y}) \Omega_{\alpha \alpha}(\vec{x},\vec{x})
\Omega_{\beta \beta}(\vec{y},\vec{y}) \nonumber \\
&-&\frac{1}{8} \sum_{\alpha,\beta} \int d^3 x d^3 y V(\vec{x},\vec{y}) \Omega_{\beta \alpha}(\vec{y},\vec{x})
\Omega_{\alpha \beta}(\vec{x},\vec{y}) .
\end{eqnarray}
Here, we note that the second term corresponds to the direct term
and the third term to the exchange term, respectively.

When the system is homogeneous, it is convenient to use Fourier transformations of $\Omega_{\alpha \beta}(\vec{x},\vec{y})$
and $V(\vec{x},\vec{y})$ .
\begin{eqnarray}
\Omega_{\alpha \beta}(\vec{x},\vec{y}) &=& \int \frac{d^3 k}{(2\pi)^3} e^{i\vec{k} \cdot (\vec{x}-\vec{y}) }
\omega_{\alpha \beta}(\vec{k}) , \\
V(\vec{x},\vec{y}) &=& \int \frac{d^3 k}{(2\pi)^3} e^{i\vec{k} \cdot (\vec{x}-\vec{y}) } V(\vec{k}) .
\end{eqnarray}
$V(\vec{k}) = 4\pi e^2 / |\vec{k}|^2$ for the Coulomb gas.

These prescriptions allow one to express the expectation value of the Hamiltonian in $\vec{k}$-space.
\begin{eqnarray}
\langle H \rangle &=& \frac{1}{2} V_g \sum_{\alpha} \int dK \epsilon_0 (\vec{k}) \omega_{\alpha \alpha}(\vec{k}) \nonumber \\
&+&\frac{1}{8} V_g V(0) \sum_{\alpha,\beta} \int dK dK'
\omega_{\alpha \alpha}(\vec{k})
 \omega_{\beta \beta}(\vec{k}') \nonumber \\
&-&\frac{1}{8} V_g \sum_{\alpha,\beta} \int dK dK' V(\vec{k}- \vec{k}') \omega_{\beta \alpha}(\vec{k})
\omega_{\alpha \beta}(\vec{k}') ,
\end{eqnarray}
where $\epsilon_0 (k) = \frac{\hbar^2 k^2}{2m}$, $dK = \frac{d^3 k}{(2\pi)^3}$ and $V_g$ is the volume of the system.

In order to simplify the above expression, we introduce the particle number $N$,
\begin{eqnarray}
N=\frac{1}{2} V_g \sum_{\alpha} \int dK  \omega_{\alpha \alpha}(\vec{k}) ,
\end{eqnarray}
which follows directly from Eq.(242). \\
Thus,
\begin{eqnarray}
\langle H \rangle =\frac{1}{2} V_g \sum_{\alpha} \int dK {\rm Tr}
[h \omega] + \frac{1}{2} V(0) N^2 /V_g ,
\end{eqnarray}
with
\begin{eqnarray}
h_{\alpha \beta}(\vec{k}) = \epsilon_0 (\vec{k}) \delta_{\alpha \beta}
-\frac{1}{4} \int dK' V(\vec{k}- \vec{k}') \omega_{\alpha \beta}(\vec{k}') .
\end{eqnarray}

In order to obtain the ground state energy, it is necessary to
take a variation on $\langle H \rangle $ with respect to $g$ or
$\bar{g}$, which is the Fourier transformations of $G$ or
$\bar{G}$ respectively.
\begin{eqnarray}
\frac{\delta \langle H \rangle }{\delta \bar{g}(\vec{k})} =
\frac{1}{2} V_g h_{\alpha \beta} \frac{\delta \omega_{\beta
\alpha}}{\delta \bar{g}} = 0 .
\end{eqnarray}
Using $\omega = (I+g)(g+\bar{g})^{-1}(I+\bar{g})$, we obtain
\begin{eqnarray}
\frac{1}{(g+\bar{g})^2} (I-g) h (I+g) = 0 . \nonumber
\end{eqnarray}
Similarly, we can obtain the condition for $g$. Altogether,
\begin{eqnarray}
(I-g) h (I+g) = 0 , \ \ (I+\bar{g}) h (I-\bar{g}) = 0.
\end{eqnarray}
From Eq.(253), we obtain 3 types of solutions, namely, $g=-I$,
$g=I$ and the last type $g \neq \pm I$.

$g=-I$ is a trivial solution which gives $\omega=0$. Next we consider $g=I$. It is known that
it gives a trivial solution in the free Dirac theory and the Gross-Neveu model [2, Gen. Ref. 5].
However, it will be shown that $g=I$ gives a Hartree-Fock solution in the electron gas model.
We believe that this interesting phenomenon arises due to the difference in the nature of the ground states
in the electron gas from that in the Dirac theory. In the Dirac theory, the vacuum is the Dirac sea with an
infinite energy which should be renormalized away, whereas in the electron gas, the vacuum ground state
is the fermi sphere which is a very much relevant quantity in condensed matter physics.

When $g=I$, we obtain $\omega =2I$. Therefore, at T=0K, the particle density $\omega_{\alpha \beta}(\vec{k})$
is given by
\begin{eqnarray}
\omega_{\alpha \beta}(\vec{k}) = 2 \delta_{\alpha \beta} \theta(k_F -k) ,
\end{eqnarray}
where $k_F$ is the fermi wavevector and $\theta(k_F -k)=1$ for $k \leq k_F$ and $0$ for $k > k_F$.

When this result is substituted in Eq.(250), we obtain the variational ground state energy in the
Gaussian functional as follows,
\begin{eqnarray}
\langle H \rangle = (2S+1) V_g \int dK [\epsilon_0 (\vec{k})+ \Sigma (\vec{k})] \theta(k_F -k) ,
\end{eqnarray}
where $S$ represents the spin. The self-energy $\Sigma (\vec{k})$ is defined as
\begin{eqnarray}
\Sigma (\vec{k}) = \frac{1}{2} n V(0) - \frac{1}{2} \int dK' V(\vec{k}-\vec{k}') \theta(k_F -k) .
\end{eqnarray}
Here $n=N/V_g$ is the electron density. We note the above result is the standard Hartree-Fock result
(Eq.(10.22) in Gen. Ref. 3). \\ \\
{\bf Hubbard Model} \\

Most of the theories of strongly correlated electron systems begin with the Hubbard model [3]. Therefore,
it is a natural choice to be studied by the new many-body theory via the functional  Schr\"{o}dinger
picture.

The Hubbard model Hamiltonian is given by
\begin{eqnarray}
H &=& -\sum_{ij \alpha} \epsilon_{ij} c_{i \alpha}^{\dagger} c_{j\alpha} + h.c.
+ U \sum_{i} n_{i \uparrow} n_{i \downarrow}  \nonumber \\
&=& - \sum_{A,B} t_{AB} c_{A}^{\dagger}c_{B} + h.c. + \frac{1}{2}
U \sum_{AB} T_{AB} n_{A} n_{B} .
\end{eqnarray}
Here, the first term represents the hopping term, the second the hermitian conjugate of the first term,
and the third the on-site Coulomb repulsion. We used the notation
$A=(i, \alpha), \ B=(j, \beta), \ t_{AB} = \epsilon_{ij} \delta_{\alpha \beta}$, and
$T_{AB} = \delta_{ij} (1-\delta_{\alpha \beta})$. $i$ and $j$ represent the lattice site and $\alpha$ and
$\beta$ the spin indices.

Since, now, we are dealing with discrete lattice sites, we should generalize the Floreanini-Jackiw formalism
to the discrete case by writing
\begin{eqnarray}
c_A = \frac{1}{\sqrt{2}} \left( u_A + \frac{\delta}{\delta u_{A}^{\dagger}}  \right) , \ \
c_A^{\dagger} = \frac{1}{\sqrt{2}} \left( u_A^{\dagger} + \frac{\delta}{\delta u_{A}}  \right) .
\end{eqnarray}
Here, $u_A$ and $u_{A}^{\dagger}$ are Grassmann variables which satisfy $\{ u_A , u_{B}^{\dagger} \}=
\delta_{AB}$ . The Gaussian trial wavefunction assumes the same form as Eqs.(234) and (235),
\begin{eqnarray}
| \Psi_0 \rangle &=& e^{u_{A}^{\dagger} G_{AB} u_B} , \nonumber \\
\langle \Psi_0 | &=& e^{u_{A}^{\dagger} \bar{G}_{AB} u_B} ,
\end{eqnarray}
but with the summations on $i$ and $j$ instead of integrals on $\vec{x}$ and $\vec{y}$. Following the
same steps, Eqs.(236) $\sim$ (244), we obtain
\begin{eqnarray}
\langle H \rangle &=& \frac{\langle \Psi_0 | H |\Psi_0  \rangle}{\langle \Psi_0 |\Psi_0  \rangle}
= -\frac{1}{2} \sum_{A,B} t_{AB} \Omega_{BA} + h.c.  \nonumber \\
&+& \frac{1}{8} U \sum_{A,B} T_{AB} (\Omega_{AA}\Omega_{BB}-\Omega_{BA}\Omega_{AB}) ,
\end{eqnarray}
where $\Omega = (I+G)(G+\bar{G})^{-1}(I+\bar{G})$ . \\
In terms of $i,j$ and $\alpha, \beta$, the expectation value becomes
\begin{eqnarray}
\langle H \rangle &=& -\frac{1}{2} \sum_{ij\alpha\beta} \epsilon_{ij} \delta_{\alpha\beta}
(\Omega_{ij})_{\alpha\beta} + h.c. \nonumber \\
&&+ \frac{1}{8} U \sum_{i, \alpha \neq \beta} [(\Omega_{ii})_{\alpha\alpha}(\Omega_{ii})_{\beta\beta}-
(\Omega_{ii})_{\beta\alpha}(\Omega_{ii})_{\alpha\beta}] \nonumber \\
&=&-\frac{1}{2} \sum_{ij\alpha} \epsilon_{ij}
(\Omega_{ij})_{\alpha\alpha} + h.c. \nonumber \\
&&+ \frac{1}{8} U \sum_{i, \alpha \neq \beta} [(\Omega_{ii})_{\alpha\alpha}(\Omega_{ii})_{\beta\beta}-
(\Omega_{ii})_{\beta\alpha}(\Omega_{ii})_{\alpha\beta}]  .
\end{eqnarray}
Here, we assume the translational symmetry and consider only the nearest neighbor hopping $t_0$.
Then, the $k$-space expression is given by
\begin{eqnarray}
\langle H \rangle &=& - t_0 \sum_{k,\alpha} \gamma_{k}[\omega_k ]_{\alpha \alpha} \nonumber \\
&&+\frac{1}{8} \frac{U}{N} \sum_{k k' \alpha \beta} [(\omega_k )_{\alpha \alpha}(\omega_{k'} )_{\beta \beta}-
(\omega_k )_{\beta \alpha}(\omega_{k'} )_{\alpha \beta}] ,
\end{eqnarray}
where, $N$ is the number of the lattice sites, $\omega = (I+g)(g+\bar{g})^{-1}(I+g)$ with $g$ the Fourier
transformation of $G$. $\gamma_k$ is defined
\begin{eqnarray}
\gamma_k = \sum_{r} e^{i \vec{k} \cdot \vec{r}}
\end{eqnarray}
where $\vec{r}$ is the vector to the nearest neighbor.

In order to study various magnetic ground state, we rewrite $\langle H \rangle$ using
\begin{eqnarray}
n&=& \frac{1}{2}\frac{1}{N} \sum_{k, \alpha} [\omega_k ]_{\alpha \alpha} \nonumber \\
&=& \frac{1}{2N} \sum_{k} {\rm Tr}[\omega] ,
\end{eqnarray}
which turns out to be the particle number per site. $\langle H \rangle$ is now rewritten as,
\begin{eqnarray}
\langle H \rangle = \frac{1}{2} \sum_{k} {\rm Tr}[h \omega] - \frac{1}{2} UNn^2 ,
\end{eqnarray}
where
\begin{eqnarray}
h_k = [-2 t_0 \gamma_k + Un]I-\frac{1}{4} \frac{U}{N} \sum_{k'} \omega_{k'} .
\end{eqnarray}
Taking a variation on $\langle H \rangle$ with respect to $\bar{g}$ as done in Eq.(252), we obtain
\begin{eqnarray}
(I-g)h(I+g)=0 ,
\end{eqnarray}
which has the identical form for the electron gas model.

As we did for the electron gas model, we first consider $g=I$. In
such a case, as in Eq.(254), we readily find that
\begin{eqnarray}
\omega = 2 \theta(k_F -k)I .
\end{eqnarray}
Using this expression and the relation $\sum_{k \sigma}\theta(k_F -k) = Nn$, we obtain
\begin{eqnarray}
h_k = [-2 t_0 \gamma_k + \frac{3}{4} Un]I , \nonumber
\end{eqnarray}
and
\begin{eqnarray}
h\omega = 2[-2 t_0 \gamma_k + \frac{3}{4} Un]\theta(k_F -k)I .
\end{eqnarray}
Finally, $\langle H \rangle$ is given by
\begin{eqnarray}
\langle H \rangle = -2 t_0 \sum_{k \sigma} \gamma_k \theta(k_F -k) + \frac{1}{4} n^2 NU,
\end{eqnarray}
which is the well known paramagnetic mean field result of the Hubbard model [3,4]

In order to study magnetic ground states, we introduce the following order parameter
\begin{eqnarray}
m = \frac{1}{2N} \sum_k {\rm Tr}[\sigma_3 \omega] ,
\end{eqnarray}
which is the difference between the up-spin and the down-spin electron divided by the number of the sites.

The expectation value of the Hamiltonian can now be rewritten in terms of $n$ and $m$,
\begin{eqnarray}
\langle H \rangle = \frac{1}{2} \sum_k {\rm Tr}[h\omega] -\frac{1}{4} UN[n^2 -m^2 ],
\end{eqnarray}
where
\begin{eqnarray}
h=[-2t_0 \gamma_k + \frac{1}{2} Un]\sigma_0 -\frac{1}{2} Um\sigma_3 .
\end{eqnarray}
\\
$<${\bf Problem}$>$ Show that Eq.(272) is identical to Eq.(265). \\

Again taking a variation with respect to $\bar{g}$, one obtains
\begin{eqnarray}
(I-g)h(I+g)=0 .
\end{eqnarray}
The above equation can be solved by expanding $g$ and $h$ in terms of the Pauli spin matrices.
However, since we seek a solution relevant to ferromagnetic ground state, we assume that
$g$ and $h$ have $\sigma_0$ and $\sigma_3$ components only. Thus, $[g, h]=0$ and
$(I-g^2 )h=0$. Therefore we have
\begin{eqnarray}
g=\pm \sigma_3 \ \ {\rm or} \ \ \pm I .
\end{eqnarray}
Since $g=I$ has been studied already, $g=\pm \sigma_3$ will be studied here.
\begin{eqnarray}
g &=& \sigma \sigma_3 , \\
\sigma &=& \Big\{ \begin{array}{cl}
1 & {\rm up-spin}\\
-1& {\rm down-spin}
\end{array}
\end{eqnarray}
Now, in general $\omega = \sigma_0 + \sigma \sigma_3 $ and can be written
\begin{eqnarray}
\omega=\theta(k_{F\sigma}-k)(\sigma_0 + \sigma \sigma_3 )
\end{eqnarray}
at zero temperature. $h\omega$ in Eq.(272) is now expressed as a diagonal $2 \times 2$ matrix
\begin{eqnarray}
h\omega = \left( \begin{array}{cc}
(1+\sigma)(h_0 + h_3 )\theta(k_{F \uparrow}-k) & 0 \\
0 &  (1-\sigma)(h_0 - h_3 )\theta(k_{F \downarrow}-k)
\end{array} \right) ,
\end{eqnarray}
where $h_0 = -t_0 \gamma_k + \frac{1}{2} Un$ and $h_3 = -\frac{1}{2} Um$ .
This expression allows to write the
\begin{eqnarray}
\langle H \rangle &=& \sum_k E_{k \uparrow} \theta(k_{F \uparrow}-k) +
\sum_k E_{k \downarrow} \theta(k_{F \downarrow}-k)  \nonumber \\
&& -\frac{1}{4} UN (n^2 -m^2 ) ,
\end{eqnarray}
where
\begin{eqnarray}
E_{k \uparrow} &=&  -2 t_0 \gamma_k + \frac{1}{2} U(n-m) , \nonumber \\
E_{k \downarrow} &=&  -2 t_0 \gamma_k + \frac{1}{2} U(n+m) .
\end{eqnarray}
The above result clearly shows that the splitting due to the mean-field magnetization.

It can also be written in a more familiar form using
$\sum_{k\sigma} \frac{1}{2} Un =\frac{1}{2} UNn^2$.
\begin{eqnarray}
\langle H \rangle = \sum_{k \sigma} E_{k\sigma} + \frac{1}{4} UN (n^2 +m^2 ) ,
\end{eqnarray}
where
\begin{eqnarray}
E_{k\sigma} = -2t_0 \gamma_k -\frac{1}{2} \sigma Um ,
\end{eqnarray}
which is the ferromagnetic Hartree-Fock result [4].

A more interesting case is provided by the antiferromagnetic ordering. For simplicity, only one
dimensional case will be considered. The order parameter in this case is the staggered magnetization,
\begin{eqnarray}
m_l = (-1)^{l} m = m e^{i Q r},
\end{eqnarray}
where $Q= \pi/a$ and $r=la$.
This is readily reduced to the paramagnetic state by setting $m=0$, and to the ferromagnetic state
by setting $Q=0$. The Hamiltonian can be expressed in terms of this order parameter as follows,
\begin{eqnarray}
\langle H \rangle = -\frac{1}{2}N \int dK {\rm Tr}(h_k \omega_k ) + \frac{1}{4} Um^2 +\frac{1}{4} Un^2 ,
\end{eqnarray}
where
\begin{eqnarray}
h_k = \epsilon_k \sigma_0 + M_Q \sigma_3 .
\end{eqnarray}
Here, $dK = dk/2\pi$ and $M_Q = \frac{1}{2} Um e^{iQr}$. \\
For the antiferromagnetic ordering, the Brillouine zone should be
folded in half. This zone folding can be carried out by separating
the $k$-contributions into two parts
\begin{eqnarray}
\langle H \rangle =-\frac{1}{2}N \int ^ {'} dK \left(
\begin{array}{cc}
\epsilon_k \sigma_0 & M_Q \sigma_3 \\
M_Q \sigma_3 & \epsilon_{k+Q} \sigma_0
\end{array} \right) \left(
\begin{array}{c}
\omega_k \\ \omega_{k+Q}
\end{array} \right)
+\frac{1}{4} Um^2 +\frac{1}{4} Un^2 ,
\end{eqnarray}
where the prime on the integral signifies the zone-folding. This
Hamiltonian is diagonalized under the nesting condition
$\epsilon_{k+Q}=-\epsilon_{k}$. Now, we take a variation in
$\bar{g}$ and obtain $(\sigma_0 -g)h(\sigma_0 +g)=0$. Among the
solutions, we choose $g=\sigma_0$, because the antiferromagnetic
gap occurs due to the nesting and the zone-folding and, thus,
requires no further symmetry breaking in terms of spin variables.
Finally, we obtain
\begin{eqnarray}
\langle H \rangle = \pm N \sum_{\sigma} \int dK' \sqrt{\epsilon_{k}^{2}+ \left( \frac{1}{2} Um \right)^2 }
+  \frac{1}{4} UN(m^2 +n^2 ) ,
\end{eqnarray}
which gives two antiferromagnetic bands degenerate with spin
indices [4]. The paramagnetic ground state is readily obtained by
requiring $m=0$, and the ferromagnetic state is obtained by
setting $Q=0$ in Eq.(286).

When the on-site Coulomb interaction $U$ is large ($U \gg t$), doubly occupied sites are energetically
very expensive. Thus, the restricted Hilbert space now consists of configurations made of empty sites (holes),
and up and down spins. The effective Hamiltonian which describes this situation is called t-J model and has
been studied intensively in possible connection with high $T_C$ superconductivity and 2-dimensional
antiferromagnetism [5]. Functional Schr\"{o}dinger picture approaches coupled with the slave-boson
and the slave-fermion techniques have been shown to produce correct mean-field results on the
2-dimensional t-J systems [6].

The above results shows versatility of the functional
Schr\"{o}dinger picture in obtaining various ground state energies
within the Gaussian approximation. However, at the same time, one
may notice that the variational method coupled with a Gaussian
trial functional invariably leads into mean-field or Hartree-Fock
results. This has been early noted that in the field theory and
the mean-field results are often called Gaussian. In the later
sections and chapters, efforts going beyond the Gaussian
approximation will be described.

\subsection{BCS Superconductivity }

 The phenomenon of superconductivity is a rare
 example of a simple field theoretic many-body
 theory being able to explain most of the essential features.
 Since Bardeen, Cooper, and Schrieffer(BCS) introduced their variational theory in 1957,
 several alternative methods including the canonical transformation technique and
 the Green's function method have been successfully applied to superconductivity [Gen. Ref.
 3]. Therefore, it appears imperative that any new many-body theoretical technique should stand
the test of the BCS superconductivity.

 In this section, the functional Schr\"{o}dinger picture formalism is applied to the BCS
superconductivity [1]. Here, instead of the Floreanini-Jackiw
formalism, we employ the DMR formalism, Eq.(211) to solve the BCS
equation. It was shown that the DMR formalism require a
non-conventional definition of the hermitian conjugate. However,
it will be shown that in order to obtain the gap equation, we do
not require the dual of the ground state. Therefore, no
complications due to non-standard hermitian conjugate arise. The
same problem can also be solved using the Floreanini-Jackiw
formalism [2]. But it will be left as a problem.

The model grand canonical Hamiltonian for the BCS superconductivity is given by [Gen. Ref. 3]
\begin{eqnarray}
H  & = & \sum_{\alpha} \int d^3x \psi_{\alpha}^{\dagger}(\vec{x}) \left(- \frac{\hbar^2}{2m} \nabla^2- \mu \right)
\psi_{\alpha}(\vec{x}) \nonumber \\
& + & \frac{1}{2}g \sum_{\alpha \beta} \int d^3x \psi_{\alpha}^\dagger(\vec{x})
i \sigma_{\alpha\beta}^{(2)} \psi_{\beta}^\dagger(\vec{x}) \psi_{\beta}(\vec{x})
i \sigma_{\beta\alpha}^{(2)}  \psi_{\alpha}(\vec x) , \label{3.2.1}
\end {eqnarray}
where $\mu$ is the chemical potential and
\begin{eqnarray}
\sigma^{(2)}=\left(\begin{array}{rr}
0& -i\\
i & 0  \end{array} \right) \nonumber
\end{eqnarray}
is introduced to insure the s-type pairing. In the process of
factorizing the quadratic interaction, the usual Hartee-Fock terms
are neglected. Eq.(\ref{3.2.1}) is, then, reduced to
\begin{eqnarray}
H & = & \sum_{\alpha} \int d^3x \ \xi (\vec{x})\psi^{\dagger}_{\alpha}(\vec{x})\psi_{\alpha}(\vec{x}) \nonumber \\
 & + & \frac{1}{2} \Delta^{\ast}\sum_{\alpha\beta}\int d^3x \ \psi_{\beta}(\vec{x})i\sigma^{(2)}_{\beta\alpha}\psi_{\alpha}(\vec{x}) \nonumber \\
& - & \frac{1}{2}\Delta \sum_{\alpha\beta} \int d^3x \ \psi^{\dagger}_{\alpha}(\vec{x})i\sigma^{(2)}_{\alpha\beta}\psi^{\dagger}_{\beta}(\vec{x}) \nonumber \\
& + & V\frac{|\Delta|^2}{g} , \label{3.2.2}
\end{eqnarray}
when
$\xi(\vec{x})=-(\hbar^2/2m)\nabla^2\delta(\vec{x'}-\vec{x})-\mu$.\
$\vec{x'}$ denotes a coordinate infinitesimally larger then
$\vec{x}$, which ensures the proper operation of $\xi(\vec{x})$ on
$\psi_\alpha(\vec{x})$. The gap function is defined by
\begin{eqnarray}
\Delta & = & -
g\langle\psi_{\alpha}(\vec{x})i\sigma^{(2)}_{\alpha\beta}\psi_{\beta}(\vec{x})\rangle
\nonumber \\ \nonumber
& = & -g\langle\psi_{\uparrow}(\vec{x})\psi_{\downarrow}(\vec{x})\rangle  \\
& = & g\langle\psi_{\downarrow}(\vec{x})\psi_\uparrow(\vec{x})\rangle. \label{3.2.3}
\end{eqnarray}
In the DMR formalism, the fermion field operators are expressed as
\begin{eqnarray}
&&\psi_{\alpha}(\vec{x})=\varphi_{\alpha}(\vec{x}), \nonumber \\
&&\psi^{\dagger}_{\alpha}(\vec{x})=\frac{\delta}{\delta\varphi_{\alpha}(\vec{x})},  \label{3.2.4}
\end{eqnarray}
where $\varphi(\vec{x})$ is a real Grassmann function. With the above prescription, the time-independent Sch\"{o}dinger
equation is given by
\begin{eqnarray}
&&\left(\sum_{\alpha}\int d^3x \ \xi(\vec{x})\frac{\delta}{\delta\varphi_{\alpha}(\vec{x})}\varphi_{\alpha}(\vec{x})+\frac{1}{2}\Delta^{\ast}
\sum_{\alpha\beta}\int d^3x \ i\sigma^{(2)}_{\beta\alpha}\varphi_{\beta}(\vec{x})\varphi_{\alpha}(\vec{x}) \right. \nonumber \\
&&\left. -\frac{1}{2}\Delta\sum_{\alpha\beta}\int d^3x \ i\sigma^{(2)}_{\alpha\beta}\frac{\delta}{\delta\varphi_{\alpha}(\vec{x})}
\frac{\delta}{\delta\varphi_{\beta}(\vec{x})}+V\frac{|\Delta|^2}{g} \right)\Phi[\varphi]=E\Phi[\varphi], \label{3.2.5}
\end{eqnarray}
where $\Phi[\varphi]$ is the wavefunctional which satisfies the functional differential equation. \\
Here, it should be noted that we do not need the troublesome dual
of $\Phi[\varphi]$, since it is not necessary to evaluate $\langle
H\rangle$. Power counting indicates that $G[\varphi]$ should
minimally be quadratic in $\varphi$ so that the second functional
derivative of $G$ yields a number while the square of the first
derivative is still
quadratic in $\varphi$. \\
Therefore, we write $G[\varphi]$ as a quadratic form in $\varphi$
( For the power counting technique,
 see Chapter. 10, Gen. Ref. 1.).

In the mean field theory of the electron gas and the Hubbard model
in the previous section, $G$ has a form, $\psi^{\dagger}g\psi$.
However, the DMR formalism employed in this section requires $G$
to be written in the form $\psi F\psi$, which reflects the
pairing. With this prescription, $G[\varphi]$ is written
\begin{eqnarray}
G[\varphi]=\sum_{\alpha\beta}\int d^3xd^3y \
\varphi_{\alpha}(\vec{x})
F_{\alpha\beta}(\vec{x},\vec{y})\varphi_{\beta}(\vec{y}).
\label{3.2.6}
\end{eqnarray}
We note that $F_{\alpha\beta}(\vec{x},\vec{y})$ must be symmetric
in coordinate space and antisymmetric in spin space so that
$\Phi[\varphi]$ represents the s-wave pairing. Substituting
Eq.(\ref{3.2.6}) into Eq.(\ref{3.2.5}) and equating terms with
same order, we obtain two equations,
\begin{eqnarray}
&&E=\sum_{\alpha\beta}\int d^3x \
[\xi(\vec{x})\delta_{\alpha\beta}+i\Delta\sigma_{\alpha\beta}^{(2)}F_{\beta\alpha}(\vec{x},\vec{x})]+\frac{|\Delta|^2}{g}V,
  \label{3.2.7} \\
&&\sum_{\alpha\beta}\int d^3xd^3y \ \Big(2\xi(\vec{x})F_{\alpha\beta}(\vec{x},\vec{y})+\frac{i}{2}\Delta^{\ast}\sigma_{\alpha\beta}^{(2)}\delta(\vec{x}-\vec{y})
\nonumber  \\
&&+ 2\sum_{\gamma\delta}\int d^3z \ F_{\alpha\gamma}(\vec{x},\vec{z})i\Delta\sigma_{\gamma\delta}^{(2)}
F_{\delta\beta}(\vec{z},\vec{y}) \Big)=0. \label{3.2.8}
\end{eqnarray}
Eq.(\ref{3.2.8}) can be written in a simple matrix form
\begin{eqnarray}
2\xi K+\frac{1}{2} i\Delta^{\ast}\sigma^{(0)} + 2i\Delta K^2=0, \label{3.2.9}
\end{eqnarray}
where $K=F\sigma^{(2)}$. It is convenient to solve Eq.(\ref{3.2.9}) in momentum space. For this purpose, we define $K(\vec{k})$, the Fourier transformation of
$K(\vec{x},\vec{y})$,

\begin{eqnarray}
K(\vec{x},\vec{y})=\int dK \ \exp[-i\vec{k} \cdot (\vec{x}-\vec{y})] \sum^3_{i=0}K_{i}(\vec{k})\sigma^{(i)},  \label{3.2.10}
\end{eqnarray}
where $dK=d^3k/(2\pi)^3$. This Fourier transformation reduces Eq.(\ref{3.2.9}) to a set of coupled equations,
\begin{eqnarray}
&&2\xi_{k} K_{0} + \frac{i}{2} \Delta^{\ast} + 2i\Delta K^2_{0} + 2i\Delta\sum_{i=1}^3 K_{i}^2=0,  \nonumber \\ \label{3.2.11}
&&2\xi_{k} K_{l} + 4i\Delta K_{0} K_{l}=0,
\end{eqnarray}
where $\xi_{k}=E_{k}-\mu = \hbar^2k^2/2m - \mu$. \\
The above equations are readily solved to yield
\begin{eqnarray}
K_{1}=K_2=K_3=0, \ K_0=i(\xi_k+\sqrt{\xi^2_k +|\Delta|^2}) \Big/2\Delta. \label{3.2.12}
\end{eqnarray}
Now $F(\vec x,\vec y)$ is given as
\begin{eqnarray}
F(\vec x,\vec y)=i\sigma^{(2)} \int dK \ \exp[-i\vec{k} \cdot (\vec{x}-\vec{y})] \frac{ \xi_k + E_{k}}{2\Delta}, \label{3.2.13}
\end{eqnarray}
where $E_k = \sqrt{ \xi^2_k + |\Delta|^2}$ represents the excitation energy. The wavefunctional which satisfies Eq.(\ref{3.2.5}) is given by
\begin{eqnarray}
\Phi[\varphi]= \exp \left( \frac{1}{2} \int dK \ \frac{\xi_k +E_k}{\Delta} \Big[\varphi_{\uparrow}(\vec k)
\varphi_{\downarrow}(-\vec k)-\varphi_{\downarrow}(\vec k)\varphi_{\uparrow}(-\vec k) \Big] \right) . \label{3.2.14}
\end{eqnarray}
It is obvious that the above wavefunctional has the correct BCS state features. Also the ground state energy is obtained from Eq.(\ref{3.2.7})
and given by
\begin{eqnarray}
E=V\int dK \ (\xi_k - E_k) + V \frac{\Delta^2}{g}  \label{3.2.15}
\end{eqnarray}
Here, $\Delta$ is assumed real. Since the characteristic energy range in forming the Cooper pairs is $\hbar w_c$,
the ground state energy per volume is
\begin{eqnarray}
\epsilon_s=E_{s}/V=N(0)\int_{kw_c}^{\hbar w_c} d\epsilon \ \Big(\xi-\sqrt{\xi^2 +\Delta^2} \Big)+\frac{\Delta^2}{g}, \label{3.2.16}
\end{eqnarray}
where $\hbar w_c $ is the phonon energy and $N(0)$ the density of states at the Fermi energy .\\
The gap equation can be obtained by minimizing Eq.(\ref{3.2.15}) with respect to $\Delta$,
\begin{eqnarray}
\Delta=\frac{g}{2} \int \frac{d^3k}{(2\pi)^3} \frac{\Delta}{\sqrt{\xi^2+\Delta^2}}, \label{3.2.17}
\end{eqnarray}
which can be reduced to the standard BCS result [3]
\begin{eqnarray}
\Delta= \frac{\hbar w_c}{\sinh [1/N(0)g]}. \label{3.2.18}
\end{eqnarray}
It is shown that the functional Schr\"{o}dinger picture method via the DMR formalism can be used to
construct a consistent theory of the BCS superconductivity. It should be noted that we could avoid
the problem of the nonconventional hermitian conjugate in the DMR representation since one
does not have to calculate the expectation value of the Hamiltonian.
It is because that the factorized form of the BCS Hamiltonian is only quadratic and, thus, can be
solved exactly without resorting to the variational process.

Since, the Floreanini-Jackiw formalism is also a valid representation for fermions,
it should be possible to solve the BCS superconductivity problem using this formalism. This
is done in the literature and will be left as a problem.\\ \\
$<${\bf Problem}$>$  Obtain the gap equation using the Floreanini-Jackiw formalism (Reference.2). \\

\subsection{Dilute Bose Gas and Bose-Einstein Condensation}

\idn It has been shown that, in Section II.B, the functional
Schr\"{o}dinger picture formalism assumes an easy and
straightforward form in the bosonic scalar field theory. However,
when this formalism is applied to boson particles, occurrence of
the Bose-Einstein condensation makes the calculation less
straightforward.

In this section, the bosonic functional Sch\"{o}dinger picture is generalized to include the Bose-Einstein
condensation. Also, we try a shifted Gaussian trial functional in order to reach beyond the Gaussian approximation [1].

The standard model Hamiltonian for an interacting Bose gas is given by [Chap. 10, Gen. Ref. 3]
\begin{eqnarray}
H=\sum_{k} \frac{\hbar^2k^2}{2m} a^{\dagger}_k a_k + \frac{U}{2V}
\sum_{k_1k_2k_3k_4} a^{\dagger}_{k_{1}} a^{\dagger}_{k_{2}}a_{k_3}a_{k_4}
\delta_{k_1 +k_2 , k_3 +k_4 } . \label{3.3.1}
\end{eqnarray}
The constant matrix element $U$ can be determined by requiring
that $H$ correctly reproduce the two-body scattering properties in
vacuum. We assume that the scattering length is much less than the
inter-particle spacing $n^{1/3}$, where $n=N/V$. Therefore, the
validity of the present calculation is restricted by the condition
$na^3 \ll 1$. Under this assumption, the zero momentum operators
$a_0$ and $a_0^\dagger$ become nearly classical due to occurrence
of condensate. The interacting part of the Hamiltonian is
rewritten explicitly in terms of $a_0$ and $a_0^\dagger$ as
\begin{eqnarray}
H_{int}&=&\frac{U}{2V}  a^\dagger_0  a^\dagger_0 a_0 a_0 +\frac{U}{2V} \sum_{k\neq 0}
 \Big[2(a^{\dagger}_k a^\dagger_0 a_k a_0 + a^{\dagger}_{-k} a^\dagger_0 a_{-k} a_0) \nonumber \\ \nonumber
&+& a^{\dagger}_k a^{\dagger}_{-k} a_0 a_0 + a^\dagger_0 a^\dagger_0 a_{k} a_{-k} \Big] \\
&+& \frac{U}{V} \sum_{k\neq q\neq 0} \Big[ a^\dagger_{k+ q}  a^\dagger_0 a_k a_q + a^\dagger_{k+q}
 a^\dagger_{-q} a_k a_0 \Big], \label{3.3.2}
\end{eqnarray}
where only terms of order of $N_0^2$, $N_0$ and $\sqrt{N_0}$ have been retained.
The three-zero-momentum operator terms do not exist, because momentum conservation makes the fourth term also
have zero momentum. It should be noted that the last term, which is generally neglected in text book
calculations originates from the interaction of particles with condensate, thus representing the condensation
fluctuation.

In order to handle the contribution from the last term, we introduce a new variable $\gamma_k$,
\begin{eqnarray}
\gamma_k=\sum_{q\neq 0} a_{k+q}^{\dagger} a_q  .     \label{3.3.3}
\end{eqnarray}
We approximate $\gamma_k$ as a c-number. Replacing the operators $a_0$ and $a_0^\dagger$ by $\sqrt{N_0}$,
$H_{int}$ becomes
\begin{eqnarray}
H_{int} &=&\frac{U}{2V}N_0^2 +\frac{U}{2V}N_0 \sum_{k \neq 0} \Big[2(a^{\dagger}_k a_k+a^{\dagger}_{-k} a_{-k}) \nonumber \\
&+&a^{\dagger}_k a^{\dagger}_{-k} +a_k a_{-k} \Big] +
\frac{U}{V} \sqrt{N_0} \sum_{k\neq 0} \gamma_k(a_k+a^{\dagger}_{-k}) . \label{3.3.4}
\end{eqnarray}
The number operator satisfies the relation
\begin{eqnarray}
N_0= N- \frac{1}{2} \sum_{k \neq 0} (a^{\dagger}_k a_k + a^{\dagger}_{-k} a_{-k}) . \label{3.3.5}
\end{eqnarray}
Using this relation and keeping up to the $\sqrt N$ terms, the
model Hamiltonain becomes
\begin{eqnarray}
H&=&\frac{1}{2}n^2 VU +\frac{1}{2} \sum_{k \neq 0} \Big[ (\epsilon^0_k +nU)(a^{\dagger}_k a_{k}+a^{\dagger}_{-k} a_{-k}) \nonumber \\
&+&nU(a^{\dagger}_k a^{\dagger}_{-k}+a_k a_{-k}) \Big]
+ \frac{nU}{\sqrt{N}} \sum_{k\neq 0} \gamma_{k}(a_k +a^{\dagger}_{-k}),  \label{3.3.6}
\end{eqnarray}
where $\epsilon_k^0= \hbar^2 k^2/2m$.

When there is no Bose-Einstein condensation, the bosonic operators can  be simply written $a_k= \phi(\vec k)$ and
$a^{\dagger}_k = -\delta / \delta \phi(\vec k)$. However, due to Bose-Einstein condensation, such simple prescription does
not work. With reasons which will become transparent below, we introduce Bose-Einstein weighting factor, $\alpha_k(k \neq 0)$
and express the operators,
\begin{eqnarray}
a_k= \alpha_k \phi(k), \  a^{\dagger}_k = - \alpha_k \frac{\delta}{\delta \phi(k)}. \label{3.3.7}
\end{eqnarray}
When these expressions are introduced to the commutation relation
$[a_k,a^{\dagger}_k]= \delta_{k k^{\prime}}$, we obtain a constraint relation for $\alpha_k$, \\
namely
\begin{eqnarray}
\alpha_k \alpha_{k^\prime}\delta(k-k^\prime)=\delta_{k k^\prime}.
\label{3.3.8}
\end{eqnarray}
Note that $\alpha_k^2 \delta(0)=1$ except $k=0$.
The Hamiltonian is now given by
\begin{eqnarray}
H \left(\phi(k),-\frac{\delta}{\delta \phi(k)} \right)& =& \frac{1}{2} n^2 V U \nonumber \\ \nonumber
&+& \frac{1}{2} \sum_{k \neq 0} \alpha_k^2 \left\{ -(\epsilon^0_k + nU) \left[\frac{\delta}{\delta \phi(k)} \phi(k)
+\frac{\delta}{\delta \phi(-k)} \phi(-k) \right]  \right. \\ \nonumber
&+& \left. nU \left[\frac{\delta}{\delta \phi(k)}\frac{\delta}{\delta \phi(-k)}+ \phi(k) \phi(-k) \right] \right\} \\
&+& \frac{nU}{\sqrt{N}}\sum_{k \neq 0} \alpha_k\gamma_k \left[\phi(\vec k)- \frac{\delta}{\delta \phi(-k)} \right].    \label{3.3.9}
\end{eqnarray}
Here, we assume that the ground state functional is real and positive and has no nodes. Thus we write
\begin{eqnarray}
\Phi_0[\phi]=C e^{G[\phi]},  \label{3.3.10}
\end{eqnarray}
where $C$ is the normalization constant. \\
The Sch\"{o}dinger equation becomes
\begin{eqnarray}
&&\frac{1}{2} n^2VU-\frac{1}{2}\sum_{k \neq 0} \alpha_{k}^2 (\epsilon_{k}^{0} + nU) \nonumber \\
&&\times \left[ 2 \delta(0) + \phi(k) \frac{\delta G}{\delta \phi(k)} +\phi(-k) \frac{\delta G}{\delta \phi(-k)} \right] \nonumber \\
&& +\frac{nU}{2} \sum_{k \neq 0} \alpha_{k}^2 \left[
\frac{\delta^2 G}{\delta \phi(k)\delta \phi(-k)} +
\frac{\delta G}{\delta \phi(k)}\frac{\delta G}{\delta \phi(-k)}+\phi(k)\phi(-k)\right] \nonumber \\
&& +\frac{nU}{\sqrt{N}}\sum_{k \neq 0} \alpha_{k} \gamma_k \left[ \phi(k)-\frac{\delta G}{\delta \phi(-k)} \right]
= E_g . \label{3.3.11}
\end{eqnarray}
The power counting technique indicates that $G[\phi]$ should be at least quadratic in $\phi(k)$. Therefore,
a simple form of $G[\phi]$ can be expressed as a shifted Gaussian in $\phi$,
\begin{eqnarray}
G[\phi] = \int d^3 k [g(k) \phi(k)\phi(-k) + f(k)\phi(k)] , \label{3.3.12}
\end{eqnarray}
where $\phi(k)=-\phi(-k)$ to ensure convergence and $g(k)$ and
$f(k)$ are function parameters that will be determined below. It
should be noted that the linear term in the above expression has
no contribution if the fluctuation contribution is not included in
the calculation. Substituting Eq.(\ref{3.3.12}) into
Eq.(\ref{3.3.11}), we obtain the following relation,
\begin{eqnarray}
E_g &=& \frac{1}{2}n^2 VU + \sum_{k \neq 0} \alpha_{k}^2  \delta(0) [nU g(k) -(\epsilon_{k}^{0} + nU)] \nonumber \\
&& + \frac{nU}{2} \sum_{k \neq 0} \left[ \alpha_{k}^2 f(k) f(-k) - \frac{2 \alpha_k \gamma_{-k}}{\sqrt{N}} f(k)
\right] \nonumber \\
&&+ \sum_{k \neq 0} \alpha_{k}^2  \left[ -(\epsilon_{k}^{0} + nU)f(k) + 2nUg(k)f(k)
+ \frac{nU \gamma_k}{\alpha_k \sqrt{N}}(1-2g(k)) \right] \phi(k)   \nonumber \\
&& \frac{1}{2} \sum_{k \neq 0} \alpha_{k}^2   \Big[ -4(\epsilon_{k}^{0} + nU) g(k) + 4nUg(k)^2 +nU \Big]
\phi(k)\phi(-k)
\end{eqnarray}
Power counting of the above equation readily yields,
\begin{eqnarray}
&& 4nUg(k)^2 -4(\epsilon_{k}^{0} + nU) g(k) +nU =0 ,  \\
&& -(\epsilon_{k}^{0} + nU)f(k) + 2nUg(k)f(k)
+ \frac{nU \gamma_k}{\alpha_k \sqrt{N}}(1-2g(k)) =0 , \\
&&E_g = \frac{1}{2}n^2 VU + \sum_{k \neq 0} [nU g(k) -(\epsilon_{k}^{0} + nU)] \nonumber \\
&& + \frac{nU}{2} \sum_{k \neq 0} \left[ \alpha_{k}^2 f(k) f(-k) - \frac{2 \alpha_k \gamma_{-k}}{\sqrt{N}} f(k)
\right] .
\end{eqnarray}
In deriving the above relations, we used Eq.(\ref{3.3.8}). Function parameters are now determined to be
\begin{eqnarray}
&&g(k)= \frac{\epsilon_{k}^{0} +nU+E(k)}{2nU} , \label{3.3.17} \\
&&f(k)= \frac{\gamma_k}{\alpha_k \sqrt{N}}\frac{\epsilon_{k}^{0} +E(k)}{E(k)} ,
\end{eqnarray}
where
\begin{eqnarray}
E(k) = \sqrt{(\epsilon_{k}^{0} +nU)^2 -(nU)^2 } .
\end{eqnarray}
Finally, the ground state energy is given by
\begin{eqnarray}
E_g &=& E_0 + E_1 + E_2  \nonumber \\
&=& \frac{1}{2}n^2 VU -\frac{1}{2}\sum_{k \neq 0} [(\epsilon_{k}^{0} + nU)-E(k)]  \nonumber \\
&& -\frac{nU^2}{V} \sum_{k \neq 0} \frac{\gamma_{k}\gamma_{-k}}{\epsilon_{k}^{0} + 2nU} . \label{3.3.20}
\end{eqnarray}
The first two terms, $E_0$ and $E_1$, are the standard results which can be obtained by a canonical transformation
[p.317, Gen. Ref. 3]. The last term, $E_2$, is the fluctuation contribution from the condensation. Following
the standard step relating the scattering length to the interaction [p.314, Gen. Ref. 3]
\begin{eqnarray}
\frac{4\pi \hbar^2 a}{m}  = U - \frac{U^2}{ 2V}\sum_{k \neq 0} \frac{1}{\epsilon_{k}^{0}} + \cdot \cdot \cdot , \label{3.3.21}
\end{eqnarray}
we obtain the ground state energy density
\begin{eqnarray}
E/V = \frac{2\pi \hbar^2 a n^2}{m} \left[ 1+ \frac{128}{15 \sqrt{\pi}}(na^3 )^{1/2} \right] ,
\end{eqnarray}
which is the standard textbook result.  Note that we have not yet
evaluated the fluctuation contribution from the condensation.

Before we evaluate the new correction term to the ground state energy, it is interesting to check whether
the present formalism predicts the particle depletion from the zero momentum correctly. \\
The depletion is defined by
\begin{eqnarray}
\frac{N-N_0 }{N} = \frac{1}{N} \sum_{k \neq 0} \langle
a_{k}^{\dagger} a_k \rangle
\end{eqnarray}
This value is equal to $\gamma_0 /N$ and can be calculated using
Eq.(\ref{3.3.7}), (\ref{3.3.8}) and (\ref{3.3.10}). We obtain for
$n_k = \langle  a_{k}^{\dagger} a_k \rangle$
\begin{eqnarray}
n_k &=& \Big{\langle} -\alpha_{k}^{2} \left[ \delta(0)+ \phi(k) \frac{\delta G}{\delta \phi(k)} \right] \Big{\rangle} \nonumber \\
&=& \Big{\langle} -1+2\alpha_{k}^{2} g(k) \phi(k)^2
-\alpha_{k}^{2} f(k) \phi(k) \Big{\rangle} ,
\end{eqnarray}
where we used $\phi(k)=-\phi(-k)$. \\
The expectation value of $\phi(k)^2$ and $\phi(k)$ can be evaluated by carrying out the Gaussian integrals
\begin{eqnarray}
\frac{\langle \Phi_0 |\phi(k)^2 |\Phi_0 \rangle}{\langle \Phi_0 | \Phi_0 \rangle} = \frac{1}{4 g(k)}
+ \frac{f(k)^2}{4g(k)^2} , \label{3.3.25}
\end{eqnarray}
and
\begin{eqnarray}
\frac{\langle \Phi_0 |\phi(k) |\Phi_0 \rangle}{\langle \Phi_0 | \Phi_0 \rangle} =  \frac{f(k)}{2g(k)} . \label{3.3.26}
\end{eqnarray}
\\
$<${\bf Problem}$>$  Prove Eqs.(\ref{3.3.25}) and (\ref{3.3.26}). \\

Using these expressions, we obtain
\begin{eqnarray}
n_k = -1 + \frac{\alpha_{k}^{2}}{2} , \label{3.3.27}
\end{eqnarray}
where $k \neq 0$ . \\
In order to obtain $\alpha_{k}$, which is introduced as a measure in the k-space to conserve the particle
numbers, we use a simple physical argument. We note that $n_k$ should satisfy the following conditions
\begin{eqnarray}
\lim_{k \rightarrow 0} n_k = \infty \ \ {\rm and} \ \  \lim_{k \rightarrow \infty} n_k = 0 .
\end{eqnarray}
Also, we obtain from Eq.(\ref{3.3.17}) the following properties
\begin{eqnarray}
g(0) \simeq \frac{1}{2} \ \ {\rm and} \ \  E(0) \simeq 0 \ \ {\rm for} \ \ k \rightarrow 0
\end{eqnarray}
and
\begin{eqnarray}
nUg(k) \sim E(k)  \ \ {\rm for} \ \ k \rightarrow \infty .
\end{eqnarray}
The simplest function to satisfy the above properties for $n_k$ is
given by
\begin{eqnarray}
\frac{\alpha_{k}^{2}}{2} = \frac{nUg(k)}{E(k)}. \label{3.3.29}
\end{eqnarray}
Indeed, we note that $n_k$ has the desired properties of a very sharp peak at $k=0$ and almost zero elsewhere.

The particle depletion is now calculated using Eqs.(\ref{3.3.27}) and (\ref{3.3.29}) to yield
\begin{eqnarray}
\frac{N-N_0 }{N} &=& \frac{1}{N} \sum_{k \neq 0}  \left[ \frac{nUg(k)}{E(k)}-1 \right]  \nonumber \\
&=& \frac{1}{2N} \sum_{k \neq 0} \left[
\frac{\epsilon_{k}^{0}+nU}{E(k)}-1 \right]
\end{eqnarray}
which gives the standard textbook result, $\frac{8}{3}\left(
\frac{na^3}{\pi} \right)^{1/2}$ after carrying out the $k$
summation [p.317, Gen. Ref. 3]. So far, we have shown that the
functional Schr\"{o}dinger picture can handle dilute bose gas
successfully even when the Bose-Einstein condensation occurs.
Next, it will be shown that the functional Schr\"{o}dinger picture
scheme coupled with a shifted Gaussian function enables us to
calculate the fluctuation contribution from the condensation.

The ground state energy of dilute bose gas is generally given (See
for example Eq.(22.21), Gen. Ref. 3.)
\begin{eqnarray}
\frac{E}{V} &=& \frac{2\pi n^2 a \hbar^2}{m} \left[  1+ \frac{128}{15}
\left( \frac{na^3}{\pi} \right)^{1/2} \right.  \nonumber \\
&& \left. + 8 \left( \frac{4}{3} \pi - \sqrt{3}  \right)(na)^3 \ln (na^3 ) + {\cal O}(na^3 ) + \cdot \cdot \cdot \right] .
\end{eqnarray}
We note that the first two terms are the Gaussian results discussed above. The next order correction term
originates from the second term of $H_{int}$, Eq.(\ref{3.3.2}). However, the coefficient of the last term has
never been determined.

Below, we show that the fluctuation contribution, the last term in Eq.(\ref{3.3.20}) produces a term in $na^3$.
Indeed, a rough estimate of this contribution showed that this contribution would be of the order $na^3$
[p.318, Gen. Ref. 3]. The fluctuation contribution, $E_2$, carries the same ultraviolet divergence as in $E_1$.
The divergence in $E_1$ was handled through the expression of the scattering length $a$ to order of $U^2$ as
given in Eq.(\ref{3.3.21}). However, for the calculation of $E_2$, such a simple cancellation scheme does not
exist, because it is now necessary to expand $a$ up to $U^3$. Therefore, we resort to an alternative cutoff
procedure called minimal subtraction [2]. In minimal subtraction, ultraviolet divergences are removed as
part of the regularization power of $k$ from the momentum space integrated. Following this prescription, we obtain
$E_2$ in a dimensionless form
\begin{eqnarray}
\frac{E_2}{E_c} = -\frac{2U}{NV} \sum_{k \neq 0} \gamma_k \gamma_{-k}
\left[ \frac{1}{\epsilon_{k}^{0} +2nU} - \frac{1}{\epsilon_{k}^{0}} \right] ,
\end{eqnarray}
where $E_c = 2\pi \hbar^2 a n^2 V/m$ . Now, it is necessary to evaluate $\gamma_k \gamma_{-k}$. It is known that
a dominant contribution arises when a particle interacts with itself, which happens when $\vec{q}' = \vec{k} +
\vec{q}$. Therefore,
\begin{eqnarray}
\gamma_k \gamma_{-k} &=& {\sum_{qq'}}' a_{k+q}^{\dagger}a_{q}a_{-k+q'}^{\dagger}a_{q'} \nonumber \\
&\simeq& {\sum_{q}}' a_{k+q}^{\dagger}a_{k+q}a_{q}^{\dagger}a_{q} \nonumber \\
&\simeq& \sum_{q \neq 0} n_{q}^{2} \nonumber \\
&=&  \sum_{q \neq 0} \left[ \frac{\epsilon_{q}^{0}+nU -E(q)}{2E(q)} \right]^2  . \label{3.3.33}
\end{eqnarray}
This approximation gives a simple result for the condensation fluctuation correction $E$,
\begin{eqnarray}
E_2 /E_c = 16 \left( \pi - \frac{8}{3}  \right) na^3 .
\end{eqnarray}

Here, we note that the above result is obtained through a rather
drastic approximation of Eq.(\ref{3.3.33}). Also, there are
indications that there might be $na^3$ order terms from other
sources [p.319, Gen. Ref. 3]. A calculation which produces a
$na^3$ order term from the particle-particle interaction has been
reported [3].

In this section, we generalized the bosonic functional Schr\"{o}dinger picture to the case when
Bose-Einstein condensation exists. Also, a shifted Gaussian functional is employed to calculate the
fluctuation contribution from the condensation.

\subsection{Finite Temperature Formalism of The Electron Gas}
\idn So far, we have presented only zero-temperature formalism. In
order that the functional Schr\"{o}dinger picture many-body theory
be truly useful, it requires extension to finite temperature. In
this section, a finite temperature formulation of many-body theory
based on the functional Schr\"{o}dinger picture is presented [1].
It is shown that a Gaussian approximation of the theory produces
the finite temperature Hartree-Fock results both for the para- and
ferromagnetic phases in a simple and convenient fashion. It is
also shown that the self-consistent equation for the ferromagnetic
splitting yields the Stoner condition for the ferromagnetism.

A finite temperature quantum field theory using the functional Schr\"{o}dinger picture has been recently
formulated by generalizing the Gaussian approximation of the variational calculation in quantum mechanics [2].
The method is based on density matrix formalism in statistical mechanics [3].

Here, we briefly introduce the concept of the density matrix. When
$| \phi_i \rangle$ is an eigenket, $E_i$ the corresponding
eigenvalue of the Hamiltonian $H$ of the system and $ Q $ the
partionfunction, the probability that the system is in the state
$| \phi_i \rangle$ is $(1/Q) e^{-\beta E_i}$. Thus, the density
matrix is defined to be
\begin{eqnarray}
\rho = \sum_n \omega_n | \phi_n \rangle \langle \phi_n | ,
\label{4.1}
\end{eqnarray}
where $\omega_n =(1/Q) e^{-\beta E_n}$ and $\beta = 1/ k_B T$ . \\
We can readily show that $\langle A \rangle = {\rm Tr} \rho A$ and
${\rm Tr} \rho =1$. Because $H | \phi_n \rangle = E_n | \phi_n
\rangle$, we can write
\begin{eqnarray}
\rho = \frac{1}{Q} \sum_n e^{-\beta E_n} | \phi_n \rangle \langle
\phi_n | = \frac{e^{-\beta H}}{Q} ,
\end{eqnarray}
where
\begin{eqnarray}
e^{-\beta F}  = Q =  \sum_n e^{-\beta E_n} = {\rm Tr}e^{-\beta H} ,
\end{eqnarray}
and, thus,
\begin{eqnarray}
\rho = \frac{e^{-\beta H}}{{\rm Tr}e^{-\beta H}} .
\end{eqnarray}
Now we regard the density matrix as a function of $\beta$. The unnormalized $\rho$ is defined by
\begin{eqnarray}
\rho_U (\beta) = e^{-\beta H} .
\end{eqnarray}
In place of $\rho_U $, we will hereafter write $\rho$.
$\rho(\beta)$ obeys the differential equation
\begin{eqnarray}
-\frac{\partial \rho}{\partial \beta} = H \rho .  \label{4.6}
\end{eqnarray}
\\
$<${\bf Problem}$>$ Using the energy representation, $H|i\rangle = E_i |i\rangle$, prove Eq.(\ref{4.6}). \\

Solving this differential equation, one can obtain $\rho(\beta)$, $F$ and other physical properties of the
model system represented by $H$ [3]. In order to develop a finite temperature many-body theory, we first
study non-interacting fermion system which has a grand Hamiltonian given by
\begin{eqnarray}
K&=& H -\mu N \nonumber \\
&=& \sum_\alpha \int d^3 x \psi_{\alpha}^{\dagger}(\vec{x}) \left[ -\frac{\hbar^2}{2m} \nabla^2  \right]
\psi_{\alpha}(\vec{x}) -\mu N , \label{4.7}
\end{eqnarray}
where $N$ represents the number operator and $\mu$ the chemical potential. Using the
Floreanini-Jackiw prescription, the grand Hamiltonian is expressed as
\begin{eqnarray}
K = \frac{1}{2} \sum_{AB} \xi_{AB} \left( u_{A}^{\dagger} + \frac{\delta}{\delta u_A}  \right)
\left( u_{B} + \frac{\delta}{\delta u_{B}^{\dagger}}  \right) ,
\end{eqnarray}
where $\xi_{AB} = -(\hbar^2 /2m) [\delta(\vec{x}-\vec{y}) \nabla_{y}^{2} ] \delta_{\alpha \beta}
-\mu \delta_{AB}$ and $A=(\vec{x}, \alpha), \ B=(\vec{y}, \beta) $.

We choose a Gaussian trial density matrix
\begin{eqnarray}
\rho = {\rm exp}(u_{A}^{\dagger}F_{AB} u_{B} + J_A u_A + u_{A}^{\dagger} L_A + C ) . \label{4.9}
\end{eqnarray}
to satisfy Eq.(\ref{4.6}). \\
Substituting Eq.(\ref{4.9}) into Eq.(\ref{4.6}), we obtain the following relation.
\begin{eqnarray}
&&-u^\dagger \frac{\partial F}{\partial \beta} u - \frac{\partial J}{\partial \beta} u
- u^\dagger \frac{\partial L}{\partial \beta} -\frac{\partial C}{\partial \beta} \nonumber \\
&&=\frac{1}{2} [u^\dagger (\xi -F\xi F)u + (-J\xi -J\xi F) u + u^\dagger (\xi L -F\xi L) \nonumber \\
&&\ \ \ +(\xi + F\xi-J\xi L)] .
\end{eqnarray}
Since the above equation is valid for any Grassmann variables $u$ and $u^\dagger$, we obtain
four differential equations,
\begin{eqnarray}
-\frac{\partial F}{\partial \beta} &=& \frac{1}{2} (I-F^2 )\xi , \\
-\frac{\partial J}{\partial \beta} &=& - \frac{1}{2} J(I+F )\xi , \\
-\frac{\partial L}{\partial \beta} &=& \frac{1}{2} (I-F )\xi L , \\
-\frac{\partial C}{\partial \beta} &=& {\rm Tr}[(I+F-JL)\xi] .
\end{eqnarray}
These equations are readily solved to yield
\begin{eqnarray}
F &=& - \coth\left(\frac{1}{2}\beta \xi\right) \label{4.15} \\
J &=& - v^\dagger e^{\frac{1}{2}\beta \xi} \left[\sinh\left(\frac{1}{2}\beta \xi\right) \right]^{-1} \\
L &=& - v e^{-\frac{1}{2}\beta \xi} \left[\sinh\left(\frac{1}{2}\beta \xi\right) \right]^{-1} \\
C &=& {\rm Tr} \left[-\frac{1}{2}\beta \xi+ \ln\left[\sinh\left(\frac{1}{2}\beta \xi\right)\right]
 - v^\dagger v \coth\left(\frac{1}{2}\beta \xi\right)\right] .
\end{eqnarray}
In the above Grassmann variables $v$ and $v^\dagger$ are introduced as constants of integrations in order to
make $\rho$ as a Gaussian type density matrix. \\
Collecting these expressions, we obtain
\begin{eqnarray}
(\rho_{nor})_{uv} &=& \frac{1}{N} e^{{\rm Tr} \left[-\frac{1}{2}\beta \xi \right]}
{\rm Det}\left[\sinh\left(\frac{1}{2}\beta \xi\right) \right] e^{u^\dagger v - v^\dagger u} \nonumber \\
&& e^{-(u^\dagger + v^\dagger ) \left[ \coth\left(\frac{1}{2}\beta \xi\right) \right](u+v)} \label{4.19},
\end{eqnarray}
where the normalization constant $N$ is to be obtained by the relation, ${\rm Tr}[\rho_{nor}]=1$. \\ \\
$<${\bf Problem}$>$ Prove Eqs.(\ref{4.15}) $\sim$ (\ref{4.19}). \\

We note that $N={\rm Tr}[\rho]$. Carrying out the integral, we obtain
\begin{eqnarray}
N &=& \int Du^\dagger Du \rho_{uu}  \nonumber \\
&=& e^{{\rm Tr} \left[-\frac{1}{2}\beta \xi \right]}  {\rm Det}\left[\sinh\left(\frac{1}{2}\beta \xi\right) \right]
\int Du^\dagger Du e^{-u^\dagger \left[ 4 \coth\left(\frac{1}{2}\beta \xi\right) \right]u} \nonumber \\
&=& e^{{\rm Tr} \left[-\frac{1}{2}\beta \xi \right]}  {\rm Det}
\left[4 \cosh\left(\frac{1}{2}\beta \xi\right) \right],
\end{eqnarray}
where we used Eq.(185). The normalized density corresponding to
the above density matrix is given by
\begin{eqnarray}
\hat{\rho}_{nor} = \frac{e^{{\rm Tr} \left[\frac{1}{2}\beta \xi \right]} }{{\rm Det}
\left[4 \cosh\left(\frac{1}{2}\beta \xi\right) \right]} e^{-\beta \hat{K}} . \label{4.21}
\end{eqnarray}
The Helmholtz free energy is defined by
\begin{eqnarray}
\beta F \equiv \langle \ln \hat{\rho}_{nor} \rangle + \beta  \langle H \rangle . \label{4.22}
\end{eqnarray}
Substituting Eqs.(\ref{4.7}) and (\ref{4.21}) into Eq.(\ref{4.22}), we obtain
\begin{eqnarray}
F &=& N\mu - k_B T {\rm Tr} \left[ \ln\left(  1+ e^{-\beta \xi} \right) \right] -{\rm Tr} \left[ \ln 2 \right] \nonumber \\
&=& N\mu - k_B T \sum_{k, \sigma}  \left[ \ln\left(  1+ e^{-\beta
\xi_k } \right) \right] - N \ln 2 , \label{4.23}
\end{eqnarray}
where $\xi_k = \epsilon_{k}^{0} -\mu$ and $\sigma$ is the spin index. For the second expression, we carried out
the Fourier transform. \\
The total energy is calculated through
\begin{eqnarray}
E &=& \langle H \rangle \nonumber \\
&=& \sum_{AB} \int Du^\dagger Du   h_{AB} \psi_{A}^{\dagger} \psi_{B} \rho_{uv} \Big|_{u=v} \nonumber \\
&=& \sum_{AB} \int Du^\dagger Du   h_{AB} \left( u_{A}^{\dagger} +
\frac{\delta}{\delta u_A}  \right)
\left( u_{B} + \frac{\delta}{\delta u_{B}^{\dagger}}  \right) \rho_{uv} \Big|_{u=v} \nonumber \\
&=& \frac{1}{2} {\rm Tr} \left\{ h \left[ 1 - \tanh \left(\frac{1}{2}\beta \xi\right) \right]   \right\} \nonumber \\
&=& \sum_{k \sigma}  \epsilon_{k}^{0} n_k ,  \label{4.24}
\end{eqnarray}
where $h_{AB} = \xi_{AB} + \mu$ and $n_k = 1/\left( 1+ e^{\beta (\epsilon_{k}^{0} -\mu)}  \right)$. \\ \\
$<${\bf Problem}$>$ Prove Eqs.(\ref{4.23}) and (\ref{4.24}). \\

We observe the well-known results of non-interacting fermionic
particles, Eqs.(\ref{4.23}) and (\ref{4.24}) clearly confirm that
a finite temperature many-particle theory can be successfully
formulated using the density matrix method based on the functional
Schr\"{o}dinger picture approach.

Next, we investigate the interacting electron gas system of which the grand Hamiltonian is given by
\begin{eqnarray}
K = \sum_{AB} \xi_{AB} \psi_{A}^{\dagger} \psi_{B} + \frac{1}{2} \sum_{AB} V_{AB}\psi_{A}^{\dagger}
\psi_{B}^{\dagger} \psi_{B}\psi_{A} ,
\end{eqnarray}
where $V_{AB}$ is any appropriate particle-particle interaction potential. The density matrix for
an interacting system is not a Gaussian and generally quite complicated. As a first approximation, we
choose a trial Gaussian density with a variable function, which will be chosen from minimization of the
thermodynamic potential. Thus, the normalized trial density matrix is given by
\begin{eqnarray}
[\rho_{nor}]_{uv} &=& \frac{1}{{\rm Det}[4 \coth(\beta Q /2)]} e^{u^\dagger v - v^\dagger u} \nonumber \\
&& \times e^{-(u^\dagger + v^\dagger ) \left[ \coth\left( \beta Q /2 \right) \right](u+v)} , \label{4.25}
\end{eqnarray}
where $Q$ is the adjustable parameter matrix. The corresponding normalized density operator is written
in the form
\begin{eqnarray}
\hat{\rho}_{nor} = \frac{e^{{\rm Tr} \left[ \beta Q/2 \right]} }{{\rm Det}
\left[4 \cosh\left( \beta Q/2 \right) \right]} e^{-\beta \psi^{\dagger} Q \psi} , \label{4.26}
\end{eqnarray}
in analogy to the free particle case.

The thermodynamic potential is given by
\begin{eqnarray}
\beta \Omega &=& \beta (F - \mu N) \nonumber \\
&=& \langle \ln \hat{\rho}_{nor} \rangle + \beta \langle H \rangle - \beta \mu  \beta \langle N \rangle  \nonumber \\
&=& \langle \ln \hat{\rho}_{nor} \rangle + \beta \langle K
\rangle.\label{4.27}
\end{eqnarray}
Substituting Eq.(\ref{4.26}) into Eq.(\ref{4.27}), one can obtain
the following expression
\begin{eqnarray}
\beta \Omega &=& {\rm Tr}[\beta Q/2] - \langle  \ln {\rm Det} [4 \cosh(\beta Q/2)] \rangle \nonumber \\
&& - \beta \langle \psi^{\dagger} Q \psi \rangle + \beta \langle K \rangle .
\end{eqnarray}
The third term is calculated using the same steps used in
Eq.(\ref{4.24}).
\begin{eqnarray}
\langle \psi^{\dagger} Q \psi \rangle &=& \sum_{AB} \langle \psi_{A}^{\dagger} Q_{AB} \psi_{B} \rangle \nonumber \\
&=& \sum_{AB} Q_{AB} \int Du^\dagger Du \frac{1}{2} \left( u_{A}^{\dagger} + \frac{\delta}{\delta u_A}  \right)
\left( u_{B} + \frac{\delta}{\delta u_{B}^{\dagger}}  \right) \rho_{uv} \Big|_{u=v} \nonumber \\
&=& \frac{1}{2}  \sum_{AB} Q_{AB} G_{BA} ,
\end{eqnarray}
where $G = \sigma_0 - \tanh(\beta Q/2)$.

Similarly, we obtain
\begin{eqnarray}
&& \langle \psi^{\dagger} \xi \psi \rangle = \frac{1}{2} \sum_{AB} \xi_{AB} G_{BA} , \label{4.30}\\
&& \langle \psi_{A}^{\dagger}
\psi_{B}^{\dagger} V_{AB} \psi_{B}\psi_{A} \rangle = \frac{1}{4} \sum_{AB} V_{AB} [G_{AA}G_{BB} -G_{BA}G_{AB}] . \label{4.31}
\end{eqnarray}
\\
Collecting terms, the thermodynamic potential is expressed as
\begin{eqnarray}
\beta \Omega &=& \frac{1}{2} \beta \sum_A Q_{AA} - \sum_A \left[ \ln \left[ 4 \cosh \left(
\frac{1}{2} \beta Q \right) \right] \right]_{AA}  \nonumber \\
&&- \frac{1}{2} \beta \sum_{AB} Q_{AB} G_{BA} + \frac{1}{2} \beta \sum_{AB} \xi_{AB} G_{BA} \nonumber \\
&&+ \frac{1}{8} \beta \sum_{AB} V_{AB} [G_{AA}G_{BB} -G_{BA}G_{AB}]  .
\end{eqnarray}
For a homogenous system, it is more convenient to carry out the
calculation in $\vec{k}$-space. Thus,
\begin{eqnarray}
\beta \Omega &=& -V \int dK {\rm tr} [\ln [4\cosh A(\vec{k})]] \nonumber \\
&&+ V \int dK {\rm tr} [A(\vec{k}) \tanh A(\vec{k})] \nonumber \\
&& + \frac{1}{2} V \beta \int dK {\rm tr} [\xi(\vec{k}) G(\vec{k})]  \nonumber \\
&& + \frac{1}{8} V \beta V(0) \int dK {\rm tr} [G(\vec{k})] \int dK' {\rm tr} [G(\vec{k}')] \nonumber \\
&& - \frac{1}{8} V \beta  \int dK dK' V(\vec{k}-\vec{k}') {\rm tr} [G(\vec{k})G(\vec{k}')] , \label{4.33}
\end{eqnarray}
where $A(\vec{k}) = \beta Q(\vec{k})/2$, $G(\vec{k}) = \sigma_0 -
\tanh(\beta Q(\vec{k})/2)$ and tr is the trace for the spin
indices only.

Introducing a constraint parameter $N$,
\begin{eqnarray}
N \equiv \frac{1}{2} V \int dK {\rm tr} [G(\vec{k})] ,
\end{eqnarray}
we rewrite the thermodynamic potential as follows
\begin{eqnarray}
\beta \Omega &=& \alpha \left[ N - \frac{1}{2} V \int dK {\rm tr} [G(\vec{k})]  \right] \nonumber \\
&&-V \int dK {\rm tr}[\ln [4\cosh A(\vec{k})]] \nonumber \\
&&+V \int dK {\rm tr}[A(\vec{k}) \tanh A(\vec{k})] \nonumber \\
&&+\frac{1}{2} V \beta \int dK {\rm tr} [\xi(\vec{k}) G(\vec{k})]  \nonumber \\
&&+ \frac{1}{2} \frac{V(0)}{V} \beta N^2  \nonumber \\
&&-\frac{1}{8} V \beta  \int dK dK' V(\vec{k}-\vec{k}') {\rm tr} [G(\vec{k})G(\vec{k}')] ,
\end{eqnarray}
where $\alpha$ is the Lagrange's undetermined multiplier. Taking variations on $\beta \Omega$ with respect to
$N$ and $A(\vec{k})$, we obtain
\begin{eqnarray}
\alpha &=& - V(0) \beta N/V , \\
\beta Q(\vec{k}) &=& (\beta \xi(\vec{k})-\alpha)\sigma_0 - \frac{1}{2} \beta  \int dK' V(\vec{k}-\vec{k}') G(\vec{k}') . \label{4.37}
\end{eqnarray}
\\
$<${\bf Problem}$>$ Prove Eq.(\ref{4.37}).
\\

Eq.(\ref{4.37}) is a self-consistent equation which determines
$Q(\vec{k})$. Since all functions in the above equation are $2
\times 2$ matrices, $Q(\vec{k})$ and $G(\vec{k})$ are also
expressed as linear combinations of Pauli matrices. First, we
consider the symmetry conserving paramagnetic case. In this case,
the eigenvalues of $G(\vec{k})$ are same for both spins, and,
thus, $Q(\vec{k})$ and $G(\vec{k})$ must be multiples of the $2
\times 2$ identity matrix same as in the zero-temperature case,
\begin{eqnarray}
&&Q(\vec{k}) = q(\vec{k}) \sigma_0 , \nonumber \\
&&G(\vec{k}) = g(\vec{k}) \sigma_0 . \label{4.38}
\end{eqnarray}
By defining $q(\vec{k}) \equiv \epsilon_k -\mu$, $g(\vec{k})$ can be written as
\begin{eqnarray}
g(\vec{k}) &=& 1- \tanh \left[ \frac{1}{2} \beta (\epsilon_k -\mu ) \right]  \nonumber \\
&=& \frac{2}{1+ e^{\beta(\epsilon_k -\mu )}} = 2 n_k . \label{4.39}
\end{eqnarray}
Here, $n_k$ is the Fermi distribution function of the interacting system. The energy spectrum of the interacting system
is obtained from Eq.(\ref{4.37})
\begin{eqnarray}
\epsilon_k = \epsilon_{k}^{0} + (2S+1) V(0) n - \int dK' V(\vec{k}-\vec{k}') n_{k'} ,
\end{eqnarray}
where $n=N/V$ is the number density. The above equation is the
familiar self-consistent Hartree-Fock equation, which determines
the energy spectrum of the interacting electron gas system. The
total energy is obtained using Eqs.(\ref{4.30}), (\ref{4.31}),
(\ref{4.38}) and (\ref{4.39}).
\begin{eqnarray}
E&=&\langle H \rangle \nonumber \\
&=& \frac{1}{2} V  \int dK {\rm tr} [h(\vec{k}) G(\vec{k})]  \nonumber \\
&&+\frac{1}{8} V V(0) \int dK {\rm tr} [G(\vec{k})] \int dK' {\rm tr} [G(\vec{k}')] \nonumber \\
&&-\frac{1}{8} V  \int dK dK' V(\vec{k}-\vec{k}') {\rm tr} [G(\vec{k})G(\vec{k}')] \nonumber \\
&=& (2S+1) V\int dK \epsilon_{k}^{0} n_k \nonumber \\
&&+\frac{1}{8} V V(0)(2S+1)^2 \int dK  2n_k \int dK'  2n_{k'} \nonumber \\
&&-\frac{1}{8} V (2S+1) \int dK dK' V(\vec{k}-\vec{k}') 4 n_{k}n_{k'} \nonumber \\
&=& (2S+1) V \int dK (\epsilon_{k}^{0} + \frac{1}{2} \Sigma_{k}) n_k , \label{4.41}
\end{eqnarray}
where the self-energy $\Sigma_{k}$ is given by
\begin{eqnarray}
\Sigma_{k} = (2S+1) V(0)n - \int dK' V(\vec{k}-\vec{k}') n_{k'} .
\end{eqnarray}
The thermodynamic potential is obtained using Eqs.(\ref{4.33}),
and (\ref{4.38}) $\sim$ (\ref{4.41}),
\begin{eqnarray}
\Omega = &-& N k_B T \ln 2 \nonumber \\
        &-& k_B T (2S+1) V \int dK \ln \left[ 1+ e^{-(\epsilon_k -\mu)/k_B T} \right] . \label{4.43}
\end{eqnarray}
\\
$<${\bf Problem}$>$ Prove Eq.(\ref{4.43}). \\

As we discussed in Section III.A, there exist solutions other than the simple paramagnetic solution. In order to
consider a magnetic symmetry breaking solution, we express $Q(\vec{k})$,
\begin{eqnarray}
Q(\vec{k}) = (\epsilon_k -\mu) \sigma_0 + \gamma_k \sigma_3 .
\end{eqnarray}
Assuming that $\gamma_k$ is much smaller than $\epsilon_k $ and expanding to the first order of $\gamma_k$,
we obtain
\begin{eqnarray}
&&\tanh\left[ \frac{1}{2} \beta\left[ (\epsilon_k -\mu) \sigma_0 + \gamma_k \sigma_3 \right] \right] \nonumber \\
&\simeq& \sigma_0 \tanh\left[ \frac{1}{2} \beta (\epsilon_k -\mu) \right]
+ \sigma_3 \frac{1}{2}
\beta \gamma_k {\rm sech}^2
\left[ \frac{1}{2} \beta ( \epsilon_k -\mu ) \right] .
\end{eqnarray}
Substituting this result into Eqs.(\ref{4.37}) and (\ref{4.39}),
we obtain for the ferromagnetic state
\begin{eqnarray}
&&\beta[(\epsilon_k -\mu) \sigma_0 + \gamma_k \sigma_3] = (\beta \xi_k -\alpha ) \sigma_0 \nonumber \\
&&-\frac{1}{2} \beta \int dK' V(\vec{k}-\vec{k}') \left[ \sigma_0
\left( 1-\tanh\left( \frac{1}{2}
\beta(\epsilon_{k'} -\mu)  \right)  \right) \right. \nonumber \\
&&\left. -\sigma_3 \frac{1}{2} \beta \gamma_{k'} {\rm sech}^2 \left( \frac{1}{2}
\beta(\epsilon_{k'} -\mu)  \right) \right] .
\end{eqnarray}
From this identity equation, we obtain two self-consistent equations as follows
\begin{eqnarray}
&&\epsilon_{k \sigma} = \epsilon_{k}^{0} + (2S+1) V(0) n - \int dK' V(\vec{k}-\vec{k}') n_{k' \sigma} , \\
&&\gamma_k = \frac{1}{4} \beta \int dK' V(\vec{k}-\vec{k}')
\gamma_{k'} {\rm sech}^2 \left( \frac{1}{2} \beta(\epsilon_{k'}
-\mu)  \right) . \label{4.48}
\end{eqnarray}
Solving Eq.(\ref{4.48}) self-consistently, one can obtain the
magnitude of the splitting between the up- and down- spin energy
spectrum. The zeroth order approximation of Eq.(\ref{4.48}) yields
the celebrated Stoner condition for ferromagnetism,
\begin{eqnarray}
1&=& \bar{V}(0) \int dK \left[ - \frac{\partial n(\epsilon_k )}{\partial \epsilon_k } \right] \nonumber \\
 &=& \bar{V} F(0) \approx \bar{V} N(0) .
\end{eqnarray}
Here, $F(0)$, $N(0)$, and $\bar{V}(0)$ represent the static Lindhard function, the density of states at the Fermi
surface and the average exchange interaction respectively. The total energy and the thermodynamic potential are
obtained in the same fashion as for the paramagnetic case,
\begin{eqnarray}
E&=& V \sum_{\sigma} \int dK \left[ \epsilon_{k}^{0} +  \frac{1}{2} \Sigma_{k \sigma} \right] n_{k \sigma} , \\
\Omega&=& - \frac{1}{\beta} \ln 2 - \frac{1}{\beta} V \sum_{\sigma} \int dK \ln \left[ 1+
e^{-\beta(\epsilon_{k \sigma} -\mu)} \right] ,
\end{eqnarray}
where
\begin{eqnarray}
&&\Sigma_{k \sigma} =\frac{1}{2}  V(0) n - \frac{1}{2} \int dK'  V(\vec{k}-\vec{k}') n_{k' \sigma} , \\
&&n_{k \sigma} = \frac{1}{1+ e^{\beta(\epsilon_{k \sigma} -\mu)}}
.
\end{eqnarray}
$\epsilon_{k \sigma}$ and $n_{k \sigma}$ can be determined self-consistently once $\gamma_k$ is obtained.

In this chapter, we have shown that the functional Schr\"{o}dinger
picture formalism can produce the well-known Hartree-Fock or mean
field results through variational process with Gaussian trial
functionals.

We also have shown that, in certain special cases, a shifted Gaussian can be employed to obtain higher terms
which do not appear in the Gaussian approximation. Although the present method is quite versatile and can be
applied to any models, including the electron gas, the Hubbard, and the nonlinear sigma model [4], it is
quite clear that the variational method lacks means for systematic improvement. Therefore, it is not possible
to go beyond the Hartree-Fock result within the theoretical schemes presented in this chapter.

In order to overcome this problem, a time-dependent functional Schr\"{o}dinger picture theory is
constructed [5]. It is known that infinitesimal time-dependent Hartree-Fock fluctuations around the Hartree-Fock
ground state yield the random-phase-approximation (RPA) results (p.292, Gen. Ref. 2). Indeed, it was shown that,
in the small oscillation regime around the Hartree-Fock solutions, the RPA results can be obtained. The theory
was applied to the electron gas to yield the electron-hole excitation spectrum, the collective plasma oscillation,
and the spin wave energy. However, again it should be noted that this line of approach is limited to the RPA results.
Often, for systems with strong correlations, the RPA results are not satisfactory. Therefore, it is necessary to
provide a theoretical scheme which can go beyond the Gaussian and the RPA results.

In the next chapter, we discuss such efforts using the functional Schr\"{o}dinger picture.

\section{Further Development of The Functional Schr\"{o}dinger Picture \\
         Method}
\idn A basic problem inherent to any variational calculation is that it lacks means for systematic
improvement. Therefore, in order to devise a many-body theoretical scheme applicable to wide range
of problems, it is necessary to overcome this problem.

In undergraduate quantum mechanics courses, students are taught generally two types of approximation schemes,
perturbation and variational methods. Each method is known to have its own advantages and limitations.
Perturbation theory is applicable only when the perturbing term is small, whereas the variational method
lacks systematic means for improvement as mentioned above. Efforts to combine these two approximations and
provide a new scheme which retains the advantages of both methods have been existed in various field of physics.
In this chapter, we review these efforts and introduce the variational perturbation theory and the optimized
perturbation theory in plain quantum mechanics languages. Then, we construct the variational perturbation
scheme based on the functional Schr\"{o}dinger picture. The theory is applied to the $\lambda \phi^4$ field
theory in order to compare the scheme with other existing methods. In the remaining section, the theory will be
extended to the optimized perturbation with the  $\lambda \phi^4$ as a model system.

\subsection{Variational and Optimized Perturbation Theories in Quantum Mechanics}

{\bf Perturbation Theory} \\

In order to explain the salient features of perturbation theory, we reproduce the discussion on perturbation
theory from Negele and Orland (Chapt. 2, Gen. Ref. 2). In the process, we will show that how the new
concepts of variational perturbation and optimized perturbation are related to plain perturbation theory.

Perturbation theory is based on the belief that the behavior of a
physical system is continuous in some 'small' parameter describing
the difference  between a solvable problem and the actual system.
It is crucial to note, however, that in general perturbation
theory yields an asymptotic rather than a convergent series. Only
under appropriate circumstances, it may yield a useful physical
approximation, but it cannot be applied systematically to
arbitrary precision.

The salient feature of asymptotic expansions may be illustrated by the following simple integral
\begin{eqnarray}
Z(g)= \int \frac{dx}{\sqrt{2\pi}} e^{-x^2 /2 - gx^4 /4}
\end{eqnarray}
corresponding to the classical partition function $Z=\int dx e^{-V(x)}$ for a particle in a potential,
$V(x)= x^2 /2 + gx^4 /4$.
Physically, the classical or quantum behavior in this potential changes completely when $g$ changes sign,
since a particle is localized when $g>0$ and not when $g<0$. Thus, we expect that the theory is nonanalytic
at $g=0$ and that an expansion in powers of $g$ has zero radius of convergence.

The perturbation series for $Z(g)$ is easily obtained by expanding $e^{-gx^4 /4}$, with the result
\begin{eqnarray}
Z(g)=\sum_n g^n Z_n
\end{eqnarray}
with
\begin{eqnarray}
g^n Z_n &=& \frac{(-g)^n}{n!} \frac{1}{4^n} \int \frac{dx}{\sqrt{2\pi}} e^{-x^2 /2} x^{4n} \nonumber \\
&=& \frac{(-g)^n (4n-1)!!}{n! 4^n} \nonumber \\
&=& \frac{(-g)^n (4n)!}{n! 16^n (2n)!} \nonumber \\
&\sim& \frac{1}{\sqrt{n\pi}} \left( \frac{4gn}{e}  \right)^n , \ \ \ {n \rightarrow \infty}  \label{4.1.3}
\end{eqnarray}
where the asymptotic behavior in the last line is obtained using Stirling's formula
$n! = \sqrt{2\pi} n^{n+ \frac{1}{2}} e^{-n}$. Since $Z_n$  grows like $n!$, the series diverges as expected
from the non-analyticity at $g=0$.

The crucial point for the present discussion is the fact that, under appropriate circumstances,
a finite number of terms of an asymptotic series may give an excellent approximation. The residual error after
$n$ terms is bounded by the $(n+1)$st term
\begin{eqnarray}
R_n &\equiv& \Big| Z(g)- \sum_{m=0}^{n} g^m Z_m \Big| \nonumber \\
&=& \int \frac{dx}{\sqrt{2\pi}} e^{-x^2 /2} \Big| e^{-gx^4 /4}-\sum_{m=0}^{n} \frac{1}{m!}
\left( -\frac{gx^4 }{4}  \right)^m \Big| \nonumber \\
&\leq& \int \frac{dx}{\sqrt{2\pi}} e^{-x^2 /2} \frac{1}{(n+1)!} \left( \frac{gx^4 }{4}  \right)^{n+1}  \nonumber \\
&=& g^{n+1} |Z_{n+1}| ,
\end{eqnarray}
so the approximation continues improve as long as $g^n Z_n$
decreases. Fig. IV.1 shows $R_n$ and $g^n | Z_n |$ as a function of
$n$ for several values of the coupling constant $g$. From the
asymptotic expression for $g^n Z_n$, Eq.(\ref{4.1.3}), we note
that the minimum occurs for $n \sim \frac{1}{4g}$, so that the
minimum error is $g^n Z_n |_{min} \sim \sqrt{4g/\pi} e^{-1/4g}$.
This exponential dependence on the inverse coupling constant is
characteristics of perturbation theory for weak coupling.

The real problem comes at large coupling constant where the series begins to diverge in very low order.
At $g=0.1$, for example, one must stop at the third term after only attaining an accuracy of
$R_3 /R_0 =25 \% $.
\\ \\ \\ \\ \\
\hspace*{2cm} \epsfig{figure=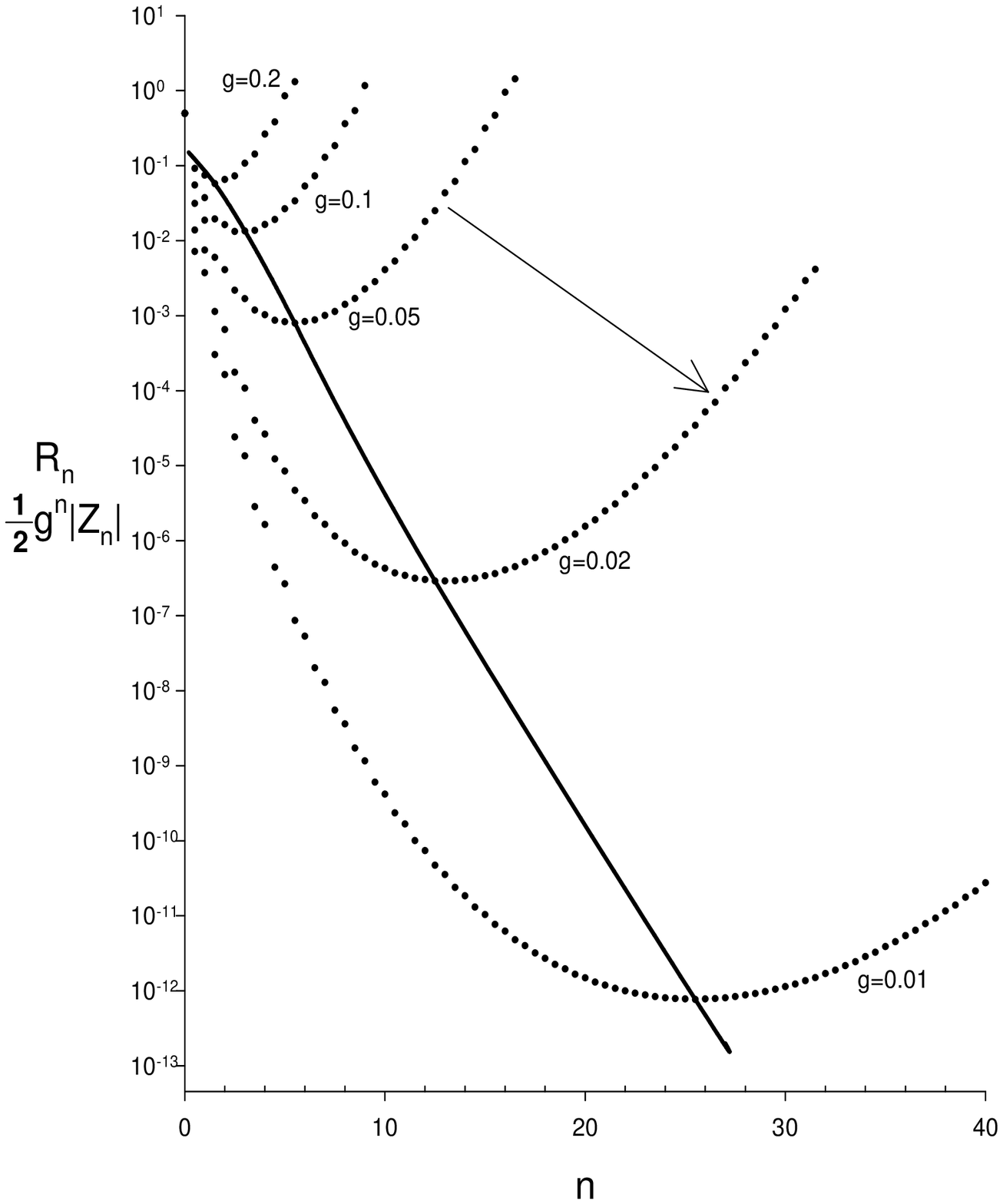,height=12cm,width=11cm}
{\small
\begin{itemize}
\item[] {\bf Fig. IV.1}  Asymptotic expansion of $Z(g)$. For five values of the coupling
constant, $\frac{1}{2} g^n | Z_n |$ is plotted at each integer $n$ and the residual error
$R_n$ is plotted at each half integer $n+ \frac{1}{2}$. The arrow symbolizes the variation
process carried out for the variational perturbation theory, whereas the solid line connecting
the minima corresponds to the optimized perturbation approximation. (from Gen. Ref. 2).
\end{itemize}}
\vspace{0.6cm}

In order to overcome this problem, many prescriptions including
the Borel summation (Section 7.5, Gen. Ref. 2) have been proposed.
In this lecture, we discuss two systematic approaches to overcome
this problem, namely the variational perturbation theory and the
optimized perturbation theory. Fig. IV.1, which is Fig. II.2 in
General Reference 2,
is reproduced here to explain the relationship between the plain perturbation and the two methods mentioned above.  
\\
\noindent
{\bf Variational Perturbation Theory} \\

Here, we explain the concept of variational perturbation theory using a simple quantum mechanical language [1].
Although this concept has been studied intensively in connection with path integrals [2], field theories and
other physics subfields [3], we believe that it had never been discussed in a plain quantum mechanical language
before Ref. 1.

In a variational calculation, one first chooses a trial function
$\Psi(\lambda)$ as a function of a variational parameter $\lambda$
for a given Hamiltonian $H$. Then, the ground state energy is
estimated by minimizing the expectation value, $\langle
\Psi_{\lambda} | H | \Psi_{\lambda} \rangle$ against $\lambda$.

In the present formalism,
we also first choose a trial function $\Psi_{n}(\lambda)$, but we require
$ H_\lambda | \Psi_{n}(\lambda) \rangle = E_n (\lambda) | \Psi_{n}(\lambda) \rangle$,
where $H_\lambda$ is called a parent Hamiltonian. We will discuss implications of this requirement below.
The original Hamiltonian is now rewritten
\begin{eqnarray}
 H &=& H_{\lambda} + H - H_{\lambda} \nonumber \\
   &=& H_{\lambda} + H' \label{4.1.5} ,
\end{eqnarray}
where $H'= H - H_{\lambda}$ is the new renormalized perturbing
Hamiltonian. Clearly, success of the following perturbation
calculation depends how small $H'$ can be made. In order to obtain
an optimum $H_{\lambda}$, we first determine $\lambda$ through the
condition,
\begin{eqnarray}
  \delta\langle\Psi_{n}(\lambda) | H | \Psi_{n}(\lambda) \rangle = 0 .
\end{eqnarray}
Here, we note that this condition is not limited to the ground state only,
but is valid for all excited states. Therefore,
$\lambda$ will be generally dependent on the state number $n$, of which energy we want to estimate.

The standard Rayleigh-Schr\"{o}dinger perturbation expansion is given by [4]
\begin{eqnarray}
&&|n \rangle =\sum_{l=0}^{\infty} \left[Q_n {\frac {1}{H_0-E_n^{(0)}}}(E_n-E_n^{(0)}-H_I)\right]^l |n \snr , \label{4.1.7} \\
&&E_n =E_n^{(0)}+\sum_{l=0}^{\infty} \snl n| H_I
        \left[Q_n {\frac {1}{H_0-E_n^{(0)}}}(E_n-E_n^{(0)}-H_I) \right]^l
      |n\snr  \label{4.1.8}
\end{eqnarray}
where $Q_n = 1-P_n = 1- |n \snr \snl n|$.

In the present theoretical scheme, we use
$H_0 = H_\lambda$ and $H'= H - H_{\lambda}$. The energy expression up to the second order is given by
\begin{eqnarray}
  E_n =& & \langle\Psi_{n}(\lambda_{n}) | H_{\lambda_n} |
          \Psi_{n}(\lambda_{n}) \rangle \nonumber \\
       &+&\langle\Psi_{n}(\lambda_{n}) | H' |
          \Psi_{n}(\lambda_{n}) \rangle \nonumber \\
       &+& \sum_{k \neq n} \frac{| \langle\Psi_{k}(\lambda_{n}) | H' |
          \Psi_{n}(\lambda_{n}) \rangle |^{2}}{E_{n}^{(0)}(\lambda_{n})
          -E_{k}^{(0)}(\lambda_{n})}  \label{4.1.9} \nonumber \\
       &+& \cdot \cdot \cdot \ \ .
\end{eqnarray}
where $ H_{\lambda_n} | \Psi_{n}(\lambda_{n}) \rangle =
  E_{n}^{(0)}(\lambda_{n}) | \Psi_{n}(\lambda_{n}) \rangle$.
  We note that the first two terms correspond to the variational calculation result
$\langle\Psi_{n}(\lambda) | H |
          \Psi_{n}(\lambda) \rangle_{\lambda=\lambda_n}$ with $\lambda_n$ determined through minimization.
The third term is the second order perturbation term with a renormalized perturbation which provides a
systematic improvement over the simple variational ground state energy.
The perturbation expansion, Eq.(\ref{4.1.9}) differs from that obtained through conventional perturbation theory.
The basis function used in the perturbation are those obtained through a variational process.

As a simple example of the variational perturbation scheme, we consider the anharmonic oscillator problem.
We will see that seemingly simple example exhibits all the salient features of the variational perturbation
theory.

The Hamiltonian is given by
\begin{eqnarray}
  H = \frac{p^2}{2m} + \frac{m\omega^2}{2}x^{2} + bx^4 \ \ ,
\end{eqnarray}
where $b$ is positive.
An obvious choice for the trial function is given by
\begin{eqnarray}
  \Psi_{\Omega}(x) =\left( \frac{m \Omega}{\pi \hbar} \right)^{\frac{1}{4}}
      e^{-\frac{m \Omega}{2 \hbar}x^2} \label{4.1.11} \ \ ,
\end{eqnarray}
for the ground state, where $\Omega$ is the  variational
parameter. The corresponding parent Hamiltonian is given by
\begin{eqnarray}
  H_{\Omega} = \frac{p^2}{2m} + \frac{m \Omega^2}{2}x^2 \ \ .
\end{eqnarray}
Using this parent Hamiltonian, the original Hamiltonian is rewritten as
\begin{eqnarray}
  H &=& H_{\Omega} + H' \nonumber \\
    &=& \frac{p^2}{2m} + \frac{m \Omega^2}{2}x^2
       + \frac{m}{2}(\omega^2 - \Omega^2 )x^2 + b x^4  \label{4.1.13} \ \ .
\end{eqnarray}
We denote the $n$th eigenstate of $H_{\Omega}$ by $| n_\Omega \rangle$.
Then, the expectation value of
$\langle n_\Omega | H | n_\Omega \rangle$ is given by
\begin{eqnarray}
  && \langle n_\Omega | H | n_\Omega \rangle \nonumber \\
  &=& \langle n_\Omega | H_\Omega | n_\Omega \rangle +
   \langle n | H' | n \rangle  \nonumber \\
    &=& \langle n_\Omega | H_\Omega | n_\Omega \rangle -
    \frac{m(\Omega^2 -\omega^2 )}{2}
    \langle n_\Omega | x^2 | n_\Omega \rangle +
    b\langle n_\Omega | x^4 | n_\Omega \rangle \nonumber \\
   &=& \hbar \Omega \left(n+ \frac{1}{2}\right) - \frac{\hbar(\Omega^2 -\omega^2 )}
    {4 \Omega}(2n+1) + \frac{3b \hbar^2}{4m^2 \Omega^2}
    (2n^2 + 2n +1) \ \ .
\end{eqnarray}
Here, we used the standard quantum mechanical results [5]
\begin{eqnarray}
   \langle n_\Omega | x^2 | n_\Omega \rangle &=&
   \frac{\hbar}{2m \Omega}(2n+1) \ , \nonumber \\
   \langle n_\Omega | x^4 | n_\Omega \rangle &=&
   \left(\frac{\hbar}{2m \Omega}\right)^2 (6n^2 + 6n +3) \ \ .
\end{eqnarray}
Taking a variation on $\langle n_\Omega | H | n_\Omega \rangle $,
we obtain a relation which determines $\Omega_n$,
\begin{eqnarray}
  \Omega_{n}^{3} - \omega^2 \Omega_n - \frac{6b \hbar}{m^2}
  \frac{2n^2 +2n +1}{2n + 1} = 0 \label{4.1.16} \ \ .
\end{eqnarray}
\\
\begin{table}
\caption{Comparison of the ground state energies(in eV) obtained using
         different approximation schemes. The values in the parentheses
         are the ratios to the exact values.
         The values of $\frac{1}{2} m \Omega_{0}^2$ are also
         shown as references.}

\begin{tabular}{cccccccc}
 & & & b(eV$\AA^{-4}$) \\
\cline{3-8}
   & & 0.01     & & 0.05     & & 0.25   &  \\ 
 \hline
 & Perturbation        & 1.4318427 & & 1.5279252 & & does not converge.  & \vspace{-0cm} \\
  & Theory              & (99.935\%) & & (95.962\%) & &   & \vspace{0.1cm} \\ 

 & Variational         & 1.4333279 & & 1.5968858  & & 2.0664772 &  \vspace{-0cm} \\
 & Calculation         & (100.038\%) & & (100.293\%) & & (100.929\%) &  \vspace{0.1cm}  \\ 

 & Present             & 1.4327276   & & 1.5912088  & & 2.0412648  &  \vspace{-0cm} \\
 & Method              & (99.997\%)  & & (99.937\%)  & & (99.679\%)  &  \vspace{0.1cm}  \\  

 & Exact               & 1.4327725  & & 1.5922195  & & 2.0474629  &   \vspace{-0cm} \\
 & Energy Value        &           & &           & &         &   \vspace{0.1cm}  \\  
\hline
 & $\frac{1}{2} m \Omega_{0}^2 (eV \AA^{-2})$ & 0.5770839 & & 0.8227827 & & 1.6423320 & \\
\end{tabular}
\end{table}

Substituting these results into Eq.(\ref{4.1.9}), we obtain
\begin{eqnarray}
 E_n &=&  \frac{\hbar \Omega_n}{2}(2n + 1) - \frac{3b \hbar^2}{4m^2
          \Omega_{n}^2}(2n^2 + 2n +1)  \nonumber \\
      &+&\frac{1}{4\hbar \Omega_n}\left(\frac{b \hbar^2}{4m^2
       \Omega_{n}^2} \right)^{2} \frac{64 n^5 + 160 n^4 - 336 n^3 - 664 n^2
       - 280 n -24 }{(2n + 1 )^2} . \ \ \ \ \ \label{4.1.17}
\end{eqnarray}
\\
$<${\bf Problem}$>$ Show that $E_n$ to the second order is given by Eq.(\ref{4.1.17}). \\

In order to show the systematic improvement achieved by the variational perturbation theory
over the variational or the plain perturbation theory, we have carried out numerical calculations.
For this purpose, we choose $\frac{m \omega^2}{2} = 0.5 eV \AA^{-2}$
and carry out calculations for various values of $b$.
The results
for the ground state energy are given in Table 4.1 [Table I, Ref.1]. The result shows that
the present method is clearly superior to the conventional perturbation theory and provides a systematic
mean for improvement over the variational calculation. 
First of all, the present method gives highly
accurate values in the regime where the conventional perturbation theory is not applicable.
This is because the perturbing Hamiltonian has been renormalized through the variational process.

Before we discuss the convergence problem of the variational perturbation theory, we briefly mention
that the method can be applied problems other than anharmonic potentials such as helium atom and also
to excited states [1]. Also, any well-behaving potential can be expanded around local minima
\begin{eqnarray}
V(x)= V(x_0 ) + \sum_{n=2}^{\infty}\frac{1}{n!}
V^{(n)}(x)|_{x=x_0}(x-x_0 )^n .
\end{eqnarray}
Therefore, we observe that above discussion on anharmonic
oscillator is generally applicable to any well-behaving
potentials. In the next section, we will show that this
approximation corresponds to the variational perturbation process
based on the Gaussian basis.

Next, we discuss the convergence problem of the variational perturbation process. Recently, this problem
has been studied in detail and has been shown that the expansion does not converge at higher orders for
strong coupling constants [6]. In fact, this behavior is not totally unexpected. \\
In the early part of this section, we have shown that perturbation
theory is only asymtotically convergent and eventually divergent.
The variational perturbation is also basically a perturbation
expansion only with a renormalized perturbation. This process is
explained on Fig.IV.1. As indicated by the arrow, the variational
process renormalizes the the coupling constant first. Then,
perturbation calculation is carried out with the renormalized
coupling constant. As shown by the anharmonic example, this
renormalized perturbation generally allows one to obtain much
higher accuracy than allowed in the original perturbation
expansion, although the variational
perturbation theory will also eventually diverges at higher orders. \\ \\
\noindent
{\bf Optimized Perturbation Theory} \\

Optimized perturbation theory was proposed to overcome the
convergence problem in the perturbation theory[7,8]. It may look
formidable in the field theory, but, in quantum mechanics, it is a
rather simple and straightforward idea. In the optimized
perturbation expansion, the variational process is not made in the
beginning, but at the last stage of the perturbation calculation.
In the process, the Hamiltonian is divided as in the variational
perturbation method. And the perturbation calculation is carried
out using Eqs.(\ref{4.1.7}) and (\ref{4.1.8}). Variation is made
on $E^{(l)}(\lambda)$, the $l$th order perturbation result, and,
then the obtained $\lambda_n$ is substituted into
$E^{(l)}_n(\lambda)$. Therefore, the value of $\lambda_{n}^{(l)}$
is different on different $l$.

We believe the method is best understood through a simple example
of anharmonic oscillator[9]. \\
We use the same Hamiltonian, Eq.(\ref{4.1.13})
\begin{eqnarray}
  H &=& H_{\Omega} + H' \nonumber \\
    &=& \frac{p^2}{2m} + \frac{m \Omega^2}{2}x^2
       + \frac{m}{2}(\omega^2 - \Omega^2 )x^2 + b x^4 , \nonumber \ \
\end{eqnarray}
and the same trial function, Eq.(\ref{4.1.11})
\begin{eqnarray}
  \Psi_{\Omega}(x) =\left( \frac{m \Omega}{\pi \hbar} \right)^{\frac{1}{4}}
      e^{-\frac{m \Omega}{2 \hbar}x^2}. \nonumber
\end{eqnarray}
We carry out the perturbation calculation on $E_n$ using
Eq.(\ref{4.1.8}) but without the variational condition,
Eq(\ref{4.1.16}). Then, to the second order, we obtain
\begin{eqnarray}
  E_n  &=& \langle n_\Omega | H_\Omega | n_\Omega \rangle +
   \langle n_\Omega | H' | n_\Omega \rangle
+ \sum_{k \neq n} \frac{|\langle k_{\Omega}|H^{\prime}| n_{\Omega} \rangle |^2}{E^{(0)}_n - E^{(0)}_k} \nonumber \\
 &=& \frac{\hbar \Omega}{2} (2n+ 1) - \frac{\hbar(\Omega^2 -\omega^2 )}{4 \Omega}(2n+1)\nonumber \\
 &&+ \frac{3b \hbar^2}{4m^2 \Omega^2}(2n^2 + 2n +1) \nonumber \\
 &&+ \frac{3b \hbar^2}{4m^2 \Omega^4}(2n^2 + 2n +1) - \frac{\hbar(\Omega^2-\omega^2)^2}{16\Omega^3}(2n+1) \nonumber \\
 &&- \frac{b^2 \hbar^3}{8m^4 \Omega^5}(34n^3 +51n^2+59n +21). \label{4.1.19}
\end{eqnarray}

We find that, by taking a variation on $E_n$ with respect to
$\Omega$, $\partial E_n/ \partial \Omega =0$, $\Omega$ does not
any real roots. In order to cope with this problem, we examine the
Hamiltonian, Eq.(\ref{4.1.13}) and the perturbation expansion,
Eq.(\ref{4.1.8}). We note that, if the perturbation calculation is
to be carried out to infinite order, then, the final result will
be independent of $\Omega$, since the original Hamiltonian does
not have $\Omega$.  The $\Omega$ dependence in the energy values
comes from 'incomplete' perturbation calculation.

 Therefore, unlike in
the variational perturbation  where one looks for minimum, here,
we search for a point where the energy value is least sensitive to
the variational parameter $\Omega$. This is called 'the principle
of minimal sensitivity' [8,9]. In the principle of minimal
sensitivity, one takes the point which satisfies $\partial^2 E_n/
\partial \Omega^2 =0$, as the point of the least sensitivity, when there is no real value to satisfy
$\partial E_n/ \partial \Omega =0$. It will be shown that in the
odd order approximation , there exists at least one real value of
$\Omega$, whereas in the even order, one should choose an $\Omega$
value which satisfies $\partial ^2 E_n /
\partial \Omega^2 =0$. This comes from the fact that, at the
odd order, the highest order of $\Omega$ is even and odd for the
even order.

Therefore we take $\partial^2 E_n/ \partial \Omega^2 =0$ for
Eq.(\ref{4.1.19}) and, by numerically solving the equation, obtain
$\Omega_{ls}$. The value of $\Omega_{ls}$ is substituted into
Eq.(\ref{4.1.19}) to obtain the energy value $E_n$. Here, we note
that, in the optimized perturbation theory, it is not possible to
obtain an analytic expression for the energy unlike in the
variational perturbation theory, Eq.(\ref{4.1.17}).

In order to study the convergence problem of the three
perturbation theories, we have carried out the optimized
perturbation calculation up to the third order,
\begin{eqnarray}
  E_n  &=& \langle n_\Omega | H_\Omega | n_\Omega \rangle +
   \langle n | H' | n \rangle \nonumber \\
&&+ \sum_{k \neq n} \frac{|\langle k_{\Omega}|H^{\prime}|
n_{\Omega} \rangle |^2}{E^{(0)}_n - E^{(0)}_k} \nonumber \\ &&+
\sum_{l \neq n} \sum_{k \neq n} \frac{ \langle n_\Omega |H^\prime
| k_\Omega \rangle \langle k_\Omega | H^\prime |l_\Omega \rangle
\langle l_\Omega |H^\prime |n|\Omega \rangle}{(E^{(0)}_n
-E^{(0)}_l)(E_n^{(0)}-E^{(0)}_k)} \nonumber \\ && - \sum_{k \neq
n} \frac{ |\langle k_\Omega |H^\prime | n_\Omega \rangle|^2
\langle n_\Omega |H^\prime | n_\Omega \rangle}{(E^{(0)}_n -
E^{(0)}_k)^2} \nonumber \\ &=& \frac{\hbar\Omega}{2}(2n+1) -
\frac{\hbar(\Omega^2-\omega^2)}{4\Omega}(2n+1) \nonumber \\ && +
\frac{3b\hbar^2}{4m^2 \Omega^2}(2n^2+2n+1) \nonumber \\ &&
+\frac{3b\hbar^2(\Omega^2-\omega^2)}{4m^2\Omega^4}(2n^2+2n+1)-
\frac{\hbar(\Omega^2-\omega^2)^2}{16\Omega^3}(2n+1) \nonumber \\
&&- \frac{b^2\hbar^3}{8m^4\Omega^5}(34n^3+51n^2+59n+21) \nonumber
\\ &&- \frac{\hbar(\Omega^2-\omega^2)^3}{32\Omega^5}(2n+1) +
\frac{b\hbar^2(\Omega^2-\omega^2)^2}{4m^2\Omega^6}(6n^2+6n+3) \nonumber \\
&&- \frac{b^2\hbar^3(\Omega^2-\omega^2)}{16m^4\Omega^7}(170n^3+255n^2+292n+105) \nonumber \\
&& + \frac{b^3\hbar^4}{16m^6\Omega^8}(375n^4+750n^3+1416n^2+1041n+333). \label{4.1.20}
\end{eqnarray} \\
$<${\bf Problem}$>$ Prove  Eq.(\ref{4.1.19}) and (\ref{4.1.20}) .\\

 This equation has a real solution for $\partial E_n/ \partial \Omega =0$, which can be substituted into
the equation to obtain the energy value.\\ In Fig.IV.2, we compare
the three perturbation methods for the ground state($n=0$) with
parameters $m\omega^2/2=0.5eV\AA^{-2}$ and $b=0.01eV\AA^{-4}$.The
figure shows clearly that the optimized perturbation has the best
convergence. Now, we consider the relation between the optimized
perturbation theory and the conventional perturbation. \\ \\ \\ \\ \\
\hspace*{3cm} \epsfig{figure=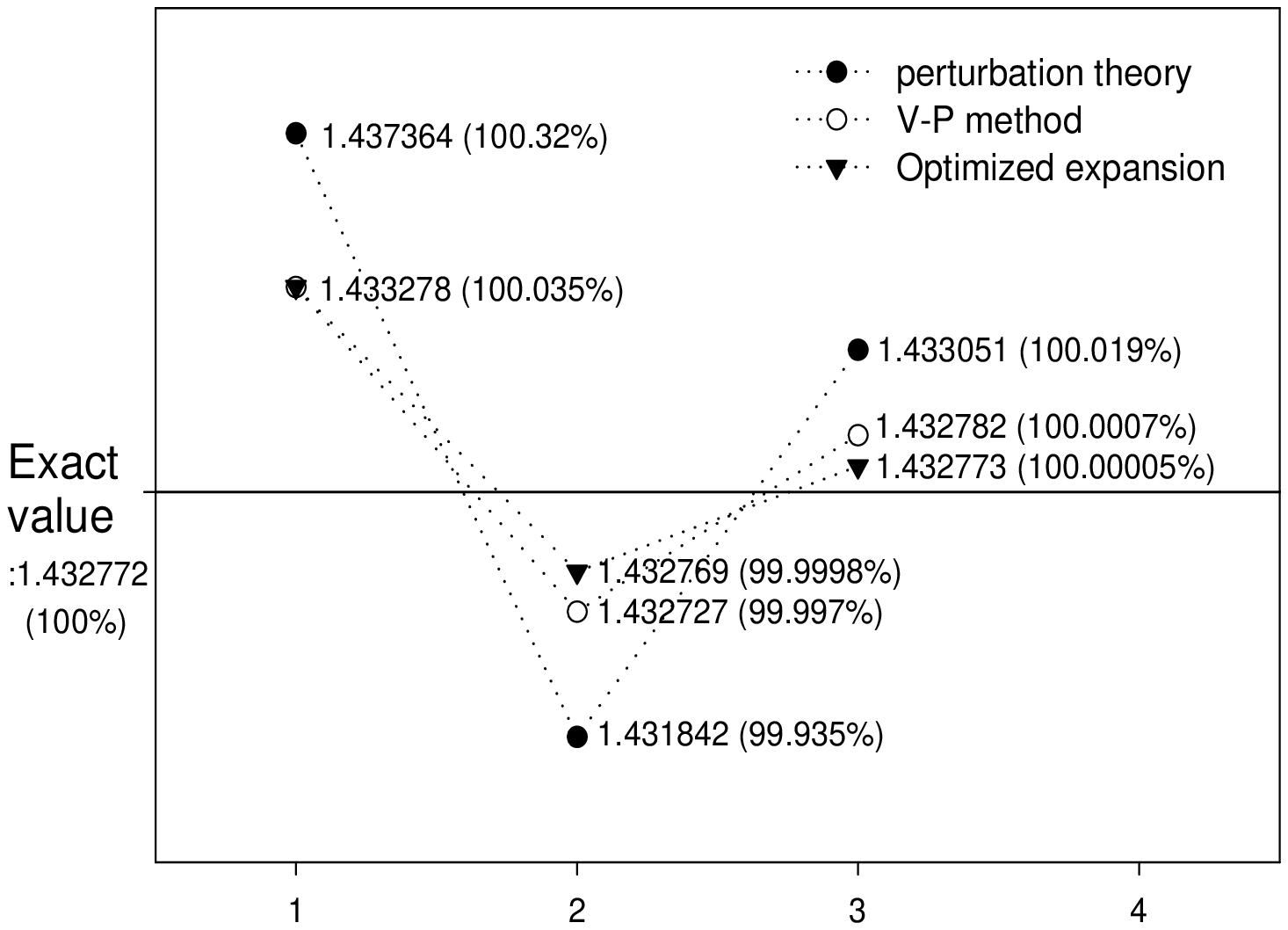,height=8cm,width=11cm}
{\small
\begin{itemize}
\item[] {\bf Fig. IV.2}   The three different perturbation theories
are compared for the parameters $m\omega^2/2=0.5eV\AA^{-2}$ and
$b=0.01eV\AA^{-4}$ up to the third order.
\end{itemize}}
\vspace{0.6cm}
The
optimized perturbation theory renormalizes the
perturbation(coupling constant) at each order and choose the
minimum(or near minimum) point for $E_n$. Therefore, the optimized
perturbation theory can be pictorially represented by solid line in Fig.IV.1.

Here, it should be mentioned that although the optimized
perturbation theory may solve the problem of convergence, its
practical use is rather limited due to the complicated nature of
the detailed calculation. Quite often, the one-time
renormalization scheme of the variational perturbation theory
allows enough accuracy required.

The variational and the optimized perturbation theories in quantum
mechanics via path integral have been intensively studied by
Kleinert and his colleagues and any higher terms can be readily
generated through computer[2].

In the remaining sections, the variation and the optimized
perturbation theories will be applied to bosonic field theories
using the $\lambda \phi^4$ model as a model system.

\subsection{Variational Perturbation Theory in The Functional Schr\"{o}dinger Picture}

\idn It has been shown that in Chapter III, Gaussian approximations
invariably lead into mean-field results. In order to go beyond
Gaussian results, various approximations schemes including
functional integrals and background field methods have been
proposed[1$\sim$3]. Among these methods, except several
variational calculations with correlated or non-Gaussian trial
functions, majority of the schemes are basically based on the
concepts of the variational or the optimized perturbation theory.
Therefore, here again the approaches going beyond the Gaussian
approximations can be classified by the methodologies employed,
namely, the operator(Green's function method), the functional
integrals, and the functional Schr\"{o}dinger picture methods. It
has been observed that the first two methods naturally lead into
Feynman diagrams again albeit renormalized and somewhat
simplified. Therefore, in order to go beyond the second order,
careful enumerations on inequivalent diagrams are still required.
However, functional Schr\"{o}dinger picture does not require such
counting processes. Being  a basic perturbtion theory but with a
Gaussian basis, it is enough to carry out the standard
Rayleigh-Schr\"{o}dinger perturbation calculation given by
Eqs.(\ref{4.1.7}) and (\ref{4.1.8}). The only modification is that
the basis are now functionals instead of functions.

Here, we present a bosonic Rayleigh-Schr\"{o}dinger perturbation
theory based on Gaussian wavefunctional. We consider a model with
the Lagrangian density
\begin{eqnarray}
{\cal L} = \frac{1}{2} \partial_\mu \phi_x \partial^\mu \phi_x -V(\phi_x), \label{4.2.1}
\end{eqnarray}
where $x=(x^1,x^2,\cdot\cdot\cdot,x^D)$ represents a position in
$D$-dimensional space and $\phi_x=\phi(\vec x)$ is the field at
$\vec x$. We note that this model can be readily reduced to
non-relativistic many-body theory by changing the kinetic part.
Also, the theory can be reduced to non-relativistic quantum
mechanics by requiring $D=0$. \\
For the variational perturbation process, we use the Gaussian trial functional
\begin{eqnarray}
|0 \rangle^{(0)} =N \exp \Big\{i \int_y {\cal P}_y \phi_y -
\frac{1}{2} \int_{x, \ y} (\phi_x-\varphi) f_{xy}(\phi_y-\varphi)
\Big\}, \label{4.2.2}
\end{eqnarray}
where $N$ is the normalized constant,$\int_{x, \ y} \equiv \int
d^Dx  \ d^Dy$, and $f_{xy}= \int d^Dp f(\varphi) e^{ip(x-y)}$ with
$p=(p^1,p^2,\cdot\cdot\cdot,p^D)$. The classical constant
$\varphi$ is equal to the Gaussian vacuum expectation value of
$\phi_x$:
\begin{eqnarray}
\varphi= \ ^{(0)} \langle 0| \phi_x | 0 \rangle^{(0)} . \label{4.2.3}
\end{eqnarray}
The Hamiltonian density corresponding to Eq.(\ref{4.2.1}) is given by
\begin{eqnarray}
{\cal H}&=&\pi_x\partial_t\phi_x -{\cal L} \nonumber \\
&=& \frac{1}{2} \pi^2_x + \frac{1}{2}(\partial_x \phi_x)^2 + V(\phi_x), \label{4.2.4}
\end{eqnarray}
where $\pi_x= \frac{1}{i} \frac{\delta}{\phi_x}$ as shown in
Eq.(190). Here, we set $c=\hbar=1$ following the field
theoretical notation. In order to calculate the ground energy
state density $\epsilon=\langle H \rangle$, it is necessary to
evaluate the following matrix elements[4],
\begin{eqnarray}
^{(0)}\langle 0 |(\partial_x \phi_x)^2 |0 \rangle^{(0)} &=&
(\partial_x \varphi)^2 -   \langle (\phi_x-\varphi) \partial_x^2 (\phi_x-\varphi) \rangle \nonumber \\
 &=& (\partial_x \varphi)^2 - \frac{1}{2}\int_y \delta(x-y) \partial_x^2 f_{xy}^{-1} \nonumber \\
&=& (\partial_x \varphi)^2 + \frac{1}{2} \int \frac{d^Dp}{(2\pi)^D} \frac{p^2}{f(p)}, \label{4.2.5}
\end{eqnarray}
where we utilized the Gaussian functional integrals, Eq.(154). We
also have
\begin{eqnarray}
^{(0)}\langle 0 |\pi_x^2 |0 \rangle^{(0)}&=& \Big\langle \Big( \frac{1}{i} \frac{\delta}{\delta \phi_x} \Big)^2 \Big\rangle \nonumber \\
&=& {\cal P}_x^2 + \frac{1}{2} \int \frac{d^Dp}{(2\pi)^D} \frac{p^2}{f(p)}, \label{4.2.6} \\ \nonumber \\
^{(0)}\langle 0 |V(\phi_x) |0 \rangle^{(0)} &=& \int^{\infty}_{-\infty} \frac{d\beta}{\sqrt{2\pi}}\tilde{V}(\beta) \langle e^{i\beta\phi_x} \rangle \nonumber \\
&=& \int^{\infty}_{-\infty} \frac{d\beta}{\sqrt{2\pi}}\tilde{V}(\beta)e^{-\frac{1}{4}\beta^2 f^{-1}_{xx}}e^{i\beta\varphi} \nonumber \\
&=& \int^{\infty}_{-\infty} \frac{d\beta}{\sqrt{2\pi}}\tilde{V}(\beta)
\int^{\infty}_{-\infty} \frac{d\alpha}{\sqrt{2\pi}} e^{-\alpha^2 /2}
 e^{\sqrt{2(-\frac{1}{4}\beta^2 f^{-1}_{xx})}\ \alpha}e^{i\beta\varphi} \nonumber \\
&=& \int^{\infty}_{-\infty} \frac{d\alpha}{\sqrt{2\pi}} e^{-\alpha^2 /2}
\int^{\infty}_{-\infty} \frac{d\beta}{\sqrt{2\pi}}\tilde{V}(\beta)e^{i\frac{\alpha}{\sqrt{2}}\beta\sqrt{f^{-1}_{xx}}+i\beta\varphi} \nonumber \\
&=& \int^{\infty}_{-\infty} \frac{d\alpha}{\sqrt{2\pi}} e^{-\alpha^2 /2} V\Big( \frac{\alpha}{\sqrt 2}\sqrt{f_{xx}^{-1}}+\varphi \Big) \nonumber \\
&=&\int^{\infty}_{-\infty} \frac{d\alpha}{\sqrt{\pi}}
e^{-\alpha^2} V\Big(\alpha\sqrt{f_{xx}^{-1}}+\varphi \Big).
\label{4.2.7}
\end{eqnarray}
\\
{\bf $<$Problem$>$} Prove Eqs.(\ref{4.2.5})$\sim$(\ref{4.2.7}). \\

Collecting the results, we obtain for the ground state energy
density, which is also called effective potential in the field
theory, for the Gaussian trial functional.
\begin{eqnarray}
\epsilon \left[\varphi,{\cal P}_x,f \right]& \equiv & \ ^{(0)} \langle 0| {\cal H}_x | 0 \rangle^{(0)} \nonumber \\
&=&\frac{1}{2}(\partial_x\varphi)^2+\frac{1}{2}{\cal P}_x^2+\frac{1}{4}\int \frac{d^Dp}{(2\pi)^D}f(p) \nonumber \\
&&+\frac{1}{4}\int \frac{d^Dp}{(2\pi)^D}\frac{p^2}{f(p)}
+\int^{\infty}_{-\infty} \frac{d\alpha}{\sqrt{\pi}} e^{-\alpha^2}
V\Big(\alpha\sqrt{f_{xx}^{-1}}+\varphi \Big) \label{4.2.8}
\end{eqnarray}
As the first process of the variational perturbation theory, we
take variations on $\epsilon$ with respect to ${\cal P}_x$ and
$f(p)$. We observe that $\delta\epsilon /\delta p=0$ simply gives
${\cal P}_x=0$ as expected. $\delta\epsilon /\delta f(p)=0$ gives
the crucial minimizing condition corresponding to
Eq.(\ref{4.1.16}) in quantum mechanics.
\begin{eqnarray}
\frac{\delta\epsilon}{\delta f(p)}=0&=&\frac{1}{4(2\pi)^D}+\frac{p^2}{4(2\pi)^D}\left(-\frac{1}{f^2(p)}\right) \nonumber \\
&+&\frac{1}{\sqrt{2\pi}}\int^{\infty}_{-\infty}
d\beta\tilde{V}(\beta)e^{i\beta\varphi}e^{-\frac{1}{4}\beta^2
f^{-1}_{xx}}
\left(-\frac{1}{4}\beta^2\right)\frac{1}{(2\pi^D)}\left(\frac{-1}{f^2(p)}\right) \nonumber\\
\label{4.2.9}
\end{eqnarray}
where we used the second relation in Eq.(\ref{4.2.7}). This
equation can be arranged as
\begin{eqnarray}
f^2(p)&=&p^2+\int^{\infty}_{-\infty} \frac{d\beta}{\sqrt{2\pi}}\tilde{V}(\beta)e^{i\beta\varphi}e^{-\frac{1}{4}\beta^2f^{-1}_{xx}}(-\beta^2) \nonumber \\
&=& p^2+ \int_{-\infty}^{\infty}\frac{d\alpha}{\sqrt{\pi}}e^{-\alpha^2}V^{(2)}(\alpha\sqrt{f^{-1}}+\varphi) \nonumber \\
&=&p^2+ \mu^2, \label{4.2.10}
\end{eqnarray}
where $V^{(n)}(z)=d^nV(z)/dz^n=\int_{-\infty}^{\infty}(dq/
\sqrt{2\pi})(iq)^n\tilde{V}(q)e^{iqz}$. \\
The renormalized mass or coupling constant $\mu^2$ is given by
\begin{eqnarray}
\mu^2(\varphi)=\int_{-\infty}^{\infty}\frac{d\alpha}{\sqrt{\pi}}e^{-\alpha^2}V^{(2)}(\alpha\sqrt{f^{-1}_{xx}}+\varphi).
\label{4.2.11}
\end{eqnarray}
Here, it is convenient to introduce the following notation
\begin{eqnarray}
I_n(Q^2)=\int \frac{d^Dp}{(2\pi)^D}\left[
\frac{\sqrt{p^2+Q^2}}{(p^2+Q^2)^n}\right]. \label{4.2.12}
\end{eqnarray}
Note that $f_{xx}^{-1}=I_1(\mu^2)$. \\ Substituting the above
relations, Eq(\ref{4.2.10})$\sim$(\ref{4.2.12}), into
Eq.(\ref{4.2.8}) and noting that $\partial\varphi/\partial x=0$,
the Gaussian effective potential becomes
\begin{eqnarray}
{\cal V}_G(\varphi)&=&\frac{1}{2}I_0(\mu^2)-\frac{\mu^2}{4}I_1(\mu^2) \nonumber \\
&&+
\int_{-\infty}^{\infty}\frac{d\alpha}{2\sqrt{\pi}}e^{-\alpha^2/4}
V \left(\frac{\alpha}{2}\sqrt{I_1(\mu^2)} \ +\varphi\right).
\label{4.2.13}
\end{eqnarray}
We note that when ${\cal V}_G(\varphi)$ has the absolute minimum
at $\varphi_0$, $\mu(\varphi_0)$ becomes the physical mass(or the
renormalized mass) and ${\cal V}_G(\varphi)$ represents the
vacuum(ground) state energy. The symmetry of the ground state can
be discussed using Eq.(\ref{4.2.13}).

For the second stage of the variational perturbation theory, it is
necessary to partition the Hamiltonian in two parts as we did in
Eq.(\ref{4.1.5}) and (\ref{4.1.13}). This can be best achieved
constructing annihilation and creation operators for Gaussian
particles (Gaussions), of which the Gaussian functions are exact solutions.\\
Following Ref.[4], we define
\begin{eqnarray}
A_f(p)&=& \left(\frac{1}{2(2\pi)^Df(p)}\right)^{1/2}\int_x  \ e^{-ipx}[f(p)(\phi_x-\varphi)+i\pi_x], \nonumber \\
A_f^{\dagger}(p)&=&\left(\frac{1}{2(2\pi)^Df(p)}\right)^{1/2}\int_x
\ e^{ipx}[f(p)(\phi_x-\varphi)-i\pi_x]. \label{4.2.14}
\end{eqnarray}
We can readily show that these operators satisfy
$[A_f(p),A_f^{\dagger}(p^\prime)]=\delta(p^\prime -p)$. Based on
these operators, one can construct the parent Hamiltonian
\begin{eqnarray}
H_0 &=& \int dp \ f(p)A_f^{\dagger}(p)A_f(p) \nonumber \\
    &=& \int_x \left[ \frac{1}{2}\pi^2_x+\frac{1}{2}(\partial_x \phi_x)^2+
\frac{1}{2}\mu^2(\phi_x-\varphi)^2-\frac{1}{2}I_0(\mu^2)   \right]
\label{4.2.15}
\end{eqnarray}
\\
$<${\bf Problem}$>$ Show that
$[A_f(p),A_f^{\dagger}(p^\prime)]=\delta(p^\prime -p)$ and prove
Eq.(\ref{4.2.15}). \\

The Gaussian ground state(vacuum),
Eq.(\ref{4.2.2}) is the ground state of $H_0$ with the zero energy
eigenvalue $E^{(0)}_0$.
The exited states of $H_0$ are
\begin{eqnarray}
|n \rangle^{(0)}=\frac{1}{\sqrt{n!}}\prod^n_{i=1}
A^{\dagger}_f(p_i) |0 \rangle^{(0)}, \ n=1,2,\cdot\cdot\cdot,
\label{4.2.16}
\end{eqnarray}
with the corresponding energy eigenvalues
\begin{eqnarray}
E^{(0)}_n=\sum^n_{i=1}f(p_i). \label{4.2.17}
\end{eqnarray}
$|n\rangle^{(0)}$ are normalized,
\begin{eqnarray}
 ^{(0)}\langle n|n\rangle^{(0)}=\frac{1}{n!}\sum_{P_{i}(n)}\prod^n_{k=1}\delta(p_k^\prime-p_{ik}), \label{4.2.18}
\end{eqnarray}
where the $P_i(n)$ represents a permutation of the set
\{$i_k$\}=\{$1,2,\cdot\cdot\cdot,n$\} and the summation is over
all $P_i(n)$s. $|n\rangle^{(0)}$ describes a n-particle state with
the continuous momenta $p_1,p_2,\cdot\cdot\cdot,p_n$.
$|0\rangle^{(0)}$ and $|n\rangle^{(0)}$ with
$n=1,2,\cdot\cdot\cdot,\infty$  constitute the complete set for
$H_0$. \\ We note that the above Gaussian formalism is quite
general and convenient for any system with a well behaving
potential $V(\phi_x)$. Now, the Hamiltonian is rewritten
$H=H_0+H_I=H_0+(H-H_0)$ with
\begin{eqnarray}
H_I=\int_x
\left[-\frac{1}{2}\mu^2(\phi_x-\varphi)^2+\frac{1}{2}I_0(\mu^2)+V(\phi_x)
\right]. \label{4.2.19}
\end{eqnarray}
Following the Rayleigh-Schr\"{o}dinger perturbation procedures,
Eq.(\ref{4.1.7}) and (\ref{4.1.8}),we write
\begin{eqnarray}
|n\rangle=\sum_{l=0}^{\infty}\left[
Q_n\frac{1}{H_0-E_n^{(0)}}(E_n-E_n^{(0)}-H_I)
\right]^l|n\rangle^{(0)}, \label{4.2.20}
\end{eqnarray}
and
\begin{eqnarray}
E_n(\varphi)=E_n^{(0)}+\sum_{l=0}^\infty \  ^{(0)}\langle n|H_I\left[
 Q_n\frac{1}{H_0-E_n^{(0)}}(E_n-E_n^{(0)}-H_I) \right]^l|n\rangle^{(0)},\label{4.2.21}
\end{eqnarray}
where $Q_n=\sum_{j \neq n}^\infty \int d^Dp_1d^Dp_2\cdot\cdot\cdot
d^Dp_j |j\rangle^{(0)}{^{(0)}\langle j|}$. \\ For the case, $n=0$
and $l=0$, Eq.(\ref{4.2.21}) gives the vacuum(ground state) energy
up the first order $E_0^1=E_0^{(0)}+E_0^{(1)}=\int_x{\cal
V}_G(\varphi)$,which is just the product of the Gaussian
effective potential and the space volume. \\

\noindent{\bf Application to the $\lambda \phi^4$ Field Theory.} \\

As an example, we consider the potential,
$V(\phi_x)=\frac{1}{2}m^2\phi_x^2+(\lambda/4!)\phi_x^4$, which was
widely studied in connection with the Gaussian approximation[5].
Thus, this model allows direct comparisons with existing results
and can be readily reduced to non-relativistic many-body theory
and quantum mechanics. \\
With the above potential, Eq.(\ref{4.2.11}) becomes
\begin{eqnarray}
\mu^2(\varphi)&=&\int_{-\infty}^{\infty} \frac{d\alpha}{\sqrt \pi}e^{-\alpha^2}V^{(2)}(\alpha\sqrt{I_1}+\varphi) \nonumber \\
&=&m^2 +\frac{1}{2}\lambda\varphi^2 +\frac{1}{4}\lambda
I_1(\mu^2),  \label{4.2.22}
\end{eqnarray}
where we used
$\int_{-\infty}^{\infty}\frac{d\alpha}{\sqrt\pi}\alpha^{2n+1}e^{-\alpha^2}=0
$ and $\int_{-\infty}^{\infty}\frac{d\alpha}{\sqrt \pi}
\alpha^{2n}e^{-\alpha^2}=2^{-n}\cdot1\cdot3\cdot5\cdot\cdot\cdot(2n-1)$.
Also, Eq.(\ref{4.2.13}) becomes
\begin{eqnarray}
{\cal V}_G(\varphi)=\frac{1}{2}m^2\varphi^2
+\frac{1}{4!}\lambda\varphi^4+\frac{1}{2}I_0(\mu)-\frac{1}{32}\lambda
I_1^2(\mu^2), \label{4.2.23}
\end{eqnarray}
which is the standard Gaussian effective potential for the $\lambda\phi^4$ model. \\ \\
$<${\bf Problem}$>$ Prove Eqs.(\ref{4.2.22}) and (\ref{4.2.23}).
\\

To obtain the effective potential of the $\lambda\phi^4$
 model up to a given order using Eqs.(\ref{4.2.20}) and
(\ref{4.2.21}), we need the following matrix elements:
\begin{eqnarray}
\snl n|H_I|n\snr &=&\snl 0|H_I|0\snr \snl n|n\snr + {\frac {1}{n!}}{\frac
    {\lambda}{4(2\pi)^D}}\sum_{\{P_{ij}(n-2)\}}
     \prod_{k=1}^{n-2}\Delta(p_{i_k}-p'_{j_k}) \nonumber  \\
     &&\times  \delta(p'_{j_{(n-1)}}+p'_{j_n}
      -p_{i_{(n-1)}}-p_{i_n}) \nonumber \\
       &&\times[f(p'_{j_{(n-1)}})f(p'_{j_n})
      f(p_{i_{(n-1)}})f(p_{i_n})]^{-{\frac {1}{2}}} \;, \label{4.2.24} \\
\snl n|H_I|n-1\snr &=&{\frac {1}{\sqrt{n!(n-1)!}}}\biggl\{
   {\frac {\lambda\varphi}{2\sqrt{2(2\pi)^D}}}
     \sum_{P_{ij}(n-2)}\prod_{k=1}^{n-2}
      \Delta(p_{i_k}-p'_{j_k})  \nonumber  \\
 &&\times\delta(p'_{j_{(n-1)}}-p_{i_{(n-1)}}-p_{i_n})[f(p'_{j_{(n-1)}})
      f(p_{i_{(n-1)}})f(p_{i_n})]^{-{\frac {1}{2}}}   \nonumber   \\
      &&  +\sqrt{\frac {(2\pi)^D}{2}}
         (\mu^2-{\frac {\lambda}{3}}\varphi^2)\varphi \nonumber \\
       &&\times  \sum_{P_{ij}(n-1)}\prod_{k=1}^{n-1}
      \Delta(p_{i_k}-p'_{j_k})\delta(p_{i_n})
      [f(p_{i_n})]^{-{\frac {1}{2}}} \biggr\} \;, \label{4.2.25} \\
\snl n|H_I|n-2\snr &=& {\frac {1}{\sqrt{n!(n-2)!}}}{\frac {\lambda}{4(2\pi)^D}}
     \sum_{P_{ij}(n-3)}\prod_{k=1}^{n-3}\Delta(p_{i_k}-p'_{j_k})  \nonumber  \\
   && \times\delta(p'_{j_{(n-2)}}-p_{i_{n-2}}
      -p_{i_{(n-1)}}-p_{i_n})    \nonumber \\
       &&\times[f(p'_{j_{(n-2)}})f(p_{i_{n-2}})
      f(p_{i_{(n-1)}})f(p_{i_n})]^{-{\frac {1}{2}}}  \; , \label{4.2.26} \\
\snl n|H_I|n-3\snr &=&
   {\frac {1}{\sqrt{n!(n-3)!}}}{\frac {\lambda\varphi}{2\sqrt{2(2\pi)^D}}}
     \sum_{P_{ij}(n-3)}\prod_{k=1}^{n-3}\Delta(p_{i_k}-p'_{j_k})
      \nonumber  \\  && \times\delta(p_{2_{(n-2)}}+p_{i_{(n-1)}}+p_{i_n}) \nonumber \\
&&\times [f(p_{i_{(n-2)}})f(p_{i_{(n-1)}})f(p_{i_n})]^{-{\frac {1}{2}}} \label{4.2.27} \\
\snl n|H_I|n-4\snr &=&
   {\frac {1}{\sqrt{n!(n-4)!}}}{\frac {\lambda}{4(2\pi)^D}}
     \sum_{P_{ij}(n-4)}\prod_{k=1}^{n-4}\Delta(p_{i_k}-p'_{j_k}) \nonumber  \\
&& \times\delta(p_{i_{(n-3)}}+p_{i_{n-2}}+p_{i_{(n-1)}}+p_{i_n}) \nonumber  \\
&&\times
[f(p_{i_{(n-3)}})f(p_{i_{(n-2)}})f(p_{i_{(n-1)}})f(p_{i_n})]^{-{\frac
{1}{2}}}  \;, \label{4.2.28}
\end{eqnarray}
with
\begin{eqnarray}
   \Delta(p_{i_k}-p'_{j_k})=\left \{
    \begin{array}{ll}
     0,& \ \ \ \ \  {\rm  for} \ \ \ \ \ k<0 \\
     1,& \ \ \ \ \  {\rm  for} \ \ \ \ \ k=0 \\
     \delta(p_{i_k}-p'_{j_k}), & \ \ \ \ \  {\rm  for} \ \ \ \ \ k>0
    \end{array} \;.  \right.   \nonumber
\end{eqnarray}
Here, the index $i_k\in \{1,2,\cdots,n\}$ with $k=1,2,\cdots,n$ corresponds
to $|n\snr$ and $j_k\in\{1,2,\cdots,n'\}$ with $k=1,2,\cdots, n'$ to
$\snl n'|$. $P_{ij}(l)$ represents a given permutation of $l$ momenta
$p_{i_1},p_{i_2},\cdots,p_{i_l}$ paired respectively with $p'_{j_1},p'_{j_2},
\cdots,p'_{j_l}$, and $\sum_{P_{ij}(l)}$ is over all different $P_{ij}(l)s$.
For any $P_{ij}(l)$, $i_1, i_2, \cdots, i_l$ are different from one another,
and so are $j_1, j_2, \cdots, j_l$. \\ \\
$<${\bf Problem}$>$ Prove Eqs.(\ref{4.2.24})$\sim$(\ref{4.2.28}).
For the calculation, it is much more convenient to carry out the
normal ordering process beforehand[See Appendix C.].\\

Employing the above matrix elements, a straightforward, yet
lengthy calculation according to Eq.(\ref{4.2.21}) gives the
effective potential of the $\lambda\phi^4$ field theory up to the
third order as
\begin{eqnarray}
{\cal V}^{iii} (\varphi) &\equiv& {\frac {E_0^{iii}}{(2\pi)^D\delta(0)}} \nonumber \\
    &=&{\cal V}_G (\varphi)   -{\frac {1}{2}}{\frac {1}{\mu^2}}\varphi^2
    (\mu^2-{\frac {\lambda}{3}}\varphi^2)^2
      -{\frac {A}{48}}{\frac {\lambda^2}{\mu^{4-2D}}}\varphi^2
       -{\frac {B}{384}}{\frac {\lambda^2}{\mu^{5-3D}}}   \nonumber  \\
     && +{\frac {A+A_1+A_2}{48}}{\frac {\lambda^2}{\mu^{6-2D}}}\varphi^2
           (\mu^2-{\frac {\lambda}{3}}\varphi^2) \nonumber \\
          && +{\frac {2 B_1+B_2}{128}}{\frac {\lambda^3}{\mu^{7-3D}}}\varphi^2
           +{\frac {C}{512}}{\frac {\lambda^3}{\mu^{8-4D}}}   \label{4.2.29}
\end{eqnarray}
with
\[
 \begin{array}{l}
  A=\int {\frac {dx}{(2\pi)^D}} {\frac {dy}{(2\pi)^D}}
   [f_1(x)f_1(y)f_1(x+y)]^{-1}[f_1(x)+f_1(y)+f_1(x+y)]^{-1} \;, \\
  A_1=\int {\frac {dx}{(2\pi)^D}} {\frac {dy}{(2\pi)^D}}
   [f_1(x)f_1(y)f_1(x+y)]^{-1}[1+f_1(x)+f_1(y)+f_1(x+y)]^{-1}  \; ,  \\
  A_2=\int {\frac {dx}{(2\pi)^D}} {\frac {dy}{(2\pi)^D}}
   [f_1(x)f_1(y)f_1(x+y)]^{-1}[f_1(x)+f_1(y)+f_1(x+y)]^{-1} \\
   \hspace{1cm}  \times  [1+f_1(x)+f_1(y)+f_1(x+y)]^{-1}  \; ,  \\
  B=\int {\frac {dx}{(2\pi)^D}} {\frac {dy}{(2\pi)^D}} {\frac {dz}{(2\pi)^D}}
   [f_1(x)f_1(y)f_1(z)f_1(x+y+z)]^{-1}  \\
   \hspace{1cm} \times [f_1(x)+f_1(y)+f_1(z)+f_1(x+y+z)]^{-1} \; , \\
  B_1=\int {\frac {dx}{(2\pi)^D}}{\frac {dy}{(2\pi)^D}}
      {\frac {dz}{(2\pi)^D}}
   [f_1(x)f_1(y)f_1(z)f_1(x+y)f_1(x+y+z)]^{-1}  \\
   \hspace{1cm}  \times  [f_1(x)+f_1(y)+f_1(x+y)]^{-1}
   [f_1(x)+f_1(y)+f_1(z)+f_1(x+y+z)]^{-1}  \\
  B_2=\int {\frac {dx}{(2\pi)^D}}{\frac {dy}{(2\pi)^D}}
      {\frac {dz}{(2\pi)^D}}
   [f_1(x)f_1(y)f_1(z)f_1(x+y)f_1(x+z)]^{-1}  \\
   \hspace{1cm}  \times  [f_1(x)+f_1(y)+f_1(x+y)]^{-1}
   [f_1(x)+f_1(z)+f_1(x+z)]^{-1}  \\
  C=\int {\frac {dx}{(2\pi)^D}}{\frac {dy}{(2\pi)^D}}
      {\frac {dz}{(2\pi)^D}} {\frac {d\omega}{(2\pi)^D}}
   [f_1(x)f_1(y)f_1(z)f_1(\omega)f_1(x+y+z)f_1(x+y+\omega)]^{-1}  \\
   \hspace{1cm}  \times  [f_1(x)+f_1(y)+f_1(z)+f_1(x+y+z)]^{-1}
   [f_1(x)+f_1(y)+f_1(\omega)+f_1(x+y+\omega)]^{-1}
  \end{array}
\]

and $f_1(w)=\sqrt{1+w^2}$. In Eq.(\ref{4.2.29}), the second,
third, and fourth terms are the second order corrections and the
last three terms represent the third order corrections to the
Gaussian
effective potential. \\ \\
$<${\bf Problem.}$>$ Prove Eq.(\ref{4.2.29}). \\

Quite surprisingly, we note that the second order result of
Eq.(\ref{4.2.29}) has an additional term,
$-\frac{1}{2}(1/\mu^2)\varphi^2[\mu^2-(\lambda/3)\varphi^2]^2$
which does not appear in other references[6]. In order to see the
origin of this term, we note that the normal ordered form of $H_I$
is given by [See Eq.(513).]
\begin{eqnarray}
H_I&=&\int_x \Big\{ {\cal
V}_G(\varphi)+\varphi\Big(\mu^2-\frac{\lambda}{3}\varphi^2\Big):(\phi_x-\varphi): \nonumber \\
   &&+\frac{\lambda}{3!}\varphi :(\phi_x-\varphi)^{3}:+\frac{\lambda}{4!}:(\phi^x-\varphi)^4:\Big\},\label{4.2.30}
\end{eqnarray}
where : : denotes normal ordering. The additional term arises from
the second term, which disappears if we choose $<\phi_x>=\varphi$.
However, we note that this result is from the simple Gaussian
calculation and does not include higher order contribution.
therefore, when higher order contribution is included
self-consistently, it is expected that this term produces nonvanishing contribution.

It should be noted that in obtaining the above results, no
diagrammic considerations are made. Although the calculation maybe
lengthy, it is rather straightforward and conceptually simple. We
believe that the complexity in selecting relevant diagrams and
corresponding symmetry factors have, so far, prevented obtaining
the third order results using the operator and the function
integral method based on the Heisenberg picture.

Before concluding this section, we compare the present results
with the ones in the quantum mechanics, Eqs.(\ref{4.1.17}) and
(\ref{4.1.20}). Quantum mechanics requires $D=0 \ ; \
\varphi(x,y)\to x(t), \ d^Dp=1$ and $(2\pi)^D=1$. Also, we observe
$\varphi=0$ for anharmonic oscillator. Then, Eq.(\ref{4.2.29}) is
reduced to become
\begin{eqnarray}
{\cal V}_G(0)&=&\frac{1}{2}I_0(\mu^2)-\frac{1}{32}\lambda I_1^2(\mu^2) \nonumber \\
&&-\frac{B}{384}\frac{\lambda^2}{\mu^5}+\frac{C}{512}\frac{\lambda^3}{\mu^8} \nonumber \\
&=&
\frac{1}{2}\mu-\frac{3}{4}\frac{b}{\mu^2}-\frac{3b^2}{8\mu^5}+\frac{27}{16}\frac{b^3}{\mu^8}.
\label{4.2.31}
\end{eqnarray}
Here, we used $I_0=\mu,
I_1=\frac{1}{\mu},B=\frac{1}{4},C=\frac{1}{16}$ and
$b=\lambda/46$.  Eq.(\ref{4.2.31}) can be directly compared with
Eq.(\ref{4.1.17}) and (\ref{4.1.20}) and confirms the correctness
of the present calculation.

The discussion in this section has been confined to the bosonic
case. In order that the variational perturbation theory can be
applied to condensed matter systems, it is necessary to develop
Rayleigh-Schr\"{o}dinger perturbation theory for fermionic
functionals, which is a future subject.

\subsection{Optimized Pertubation Theory in The Functional Schr\"{o}dinger Picture}

In this section, we present an optimized perturbation theory in
the functional Schr\"{o}dinger picture with an external source[1].
First, a free field theory with an external source will be
constructed. Based on this free-field basis, a general expression
on the optimized effective potential for a class of model
potentials will be given[1,2]. Then, the theory will be applied to
the $\lambda\phi^4$ model.

The Hamiltonian with an external source, $J_x\equiv J(\vec x)$, is given by
\begin{eqnarray}
H_0^{J, \ \mu}&=&\int_x \Big[ \frac{1}{2}\pi_x^2+\frac{1}{2}(\partial_x\phi_x)^2 +\frac{1}{2}\mu^2\phi_x^2  \nonumber \\
  &&-J_x\phi_x-\frac{1}{2}f_{xx}+\frac{1}{2}\int_y J_x h_{xy}^{-1}J_y    \Big], \label{4.3.1}
\end{eqnarray}
where $f_{xy}\equiv (\sqrt{-\partial_x^2+\mu^2})\delta(x-y)$ with
$\int_z f_{xz}f^{-1}_{zy}=\delta(x-y)$  and  $h_{xy}\equiv
(-\partial_x^2+\mu^2)\delta(x-y)$ with $\int_z
h_{xz}h_{zy}^{-1}=\delta(x-y)$. The last term is introduced to
make $E_0^{0}[J]=0$ in the presence of the external source term
$-J_x\phi_x$(Compare with Eq.(\ref{4.2.13}).). The functional
Schr\"{o}dinger equation is readily solved using the
power-counting method. The ground state wavefunctional is a
Gaussian-type functional
\begin{eqnarray}
|0;J\rangle^{(0)}=N \exp\Big\{-\frac{1}{2}\int_{x, \ y}\left(\phi_x-\int_y h^{-1}_{xz}J_z\right)
 f_{xy}\left(\phi_y-\int_z h_{yz}^{-1}J_z\right)   \Big\}, \label{4.3.2}
\end{eqnarray}
where $N$ is the normalization constant. It should be noted that,
here, $ ^{(0)}\langle J;0|\phi_x|0;J\rangle^{(0)}=\int_z
h_{xz}^{-1}J_z$ which, unlike in Section IV.B, is dependent on $x$.
Then, accordingly, annihilation and creation operators can be
constructed as
\begin{eqnarray}
A_f(p;J)=\left(\frac{1}{2(2\pi)^Df(p)}\right)^{1/2}\int_x
e^{-ipx}[f(p)(\phi_x -\int_z h_{xz}^{-1}J_z)+i
\pi_x],\label{4.3.3}
\end{eqnarray}
and
\begin{eqnarray}
A^{\dagger}_f(p;J)=\left(\frac{1}{2(2\pi)^Df(p)}\right)^{1/2}\int_x
e^{ipx}[f(p)(\phi_x -\int_z h_{xz}^{-1}J_z)-i \pi_x],\label{4.3.4}
\end{eqnarray}
with $[A_f(p;J),A^{\dagger}_f(p\prime;J)]=\delta(p\prime-p)$ and
$A_f(p;J)|0\rangle^{(0)}=0$. \\ \\ $<${\bf Problem.}$>$ Show that
$H_0^{J, \ \mu}=\int d^Dp \ f(p)A_f^{\dagger}(p;J)A_f(p;J)$, where
$f(p)=\sqrt{p^2+\mu^2}$. \\ \\
The eigenwavefunctionals for exited states can be written as
\begin{eqnarray}
|n;J\rangle^{(0)}=\frac{1}{\sqrt{n!}}\prod^n_{i=1}A_f^{\dagger}(p_i;J)|0;J\rangle^{(0)}, n=1,2,\cdot\cdot\cdot,\infty \label{4.3.5}
\end{eqnarray}
and the corresponding eingenergies are
\begin{eqnarray}
E_n^{(0)}[J]=\sum_{i=1}^n f(p_i). \label{4.3.6}
\end{eqnarray}

The eigenwavefunctionals $|n;J\rangle^{(0)}$ and
$|0;J\rangle^{(0)}$ are orthogonal and normalized, $ ^{(0)}\langle
J;m|n;J\rangle^{(0)}=\delta_{mn}\frac{1}{n!}\sum_{P_i(n)}\prod^n_{k=1}\delta(p^{\prime}_k-p_{ik})$.
Here $P_i(n)$ represents a permutation of the set
$\{i_k\}=\{1,2,\cdot\cdot\cdot,n\}$ and the summation is over all
$P_i(n)$'s. $|n;J\rangle^{(0)}$ describe a n-particle state with
the continuous momenta $p_1,p_2,\cdot\cdot\cdot,p_n$.
$|0;J\rangle^{(0)}$ and $|n;J\rangle^{(0)}$ with
$n=1,2,\cdot\cdot\cdot,\infty$ constitute the complete set for
$H_0^{J, \ \mu}$ and satisfy the closure
$|0;J\rangle^{(0)}{^{(0)}}\langle J;0|+\sum^{\infty}_{n=1}\int
d^Dp_1d^Dp_2\cdot\cdot\cdot d^Dp_n|n;J\rangle^{(0)}{^{(0)}}\langle
J;n|=1$.

We now introduce the model potential through the following scalar Lagrangian density
\begin{eqnarray}
{\cal L}=\frac{1}{2}\partial_\mu\phi_x\partial^\mu\phi_x-V(\phi_x). \label{4.3.7}
\end{eqnarray}
We assume that $V(\phi_x)=\int\frac{d\Omega}{\sqrt{2\pi}}\tilde
V(\Omega)e^{i\Omega\phi_x}$, at least, in a sense of tempered
distributions[3]. In fact, this assumption also applies to the
previous section. For the system, Eq.(\ref{4.3.7}). the
time-independent functional Schr\"{o}dinger equation in the
presence of an external source of $J_x$ is
\begin{eqnarray}
\left(H-\int_x J_x\phi_x \right)|\Psi_n\rangle=E_n[J]|\Psi_n\rangle \label{4.3.8}
\end{eqnarray}
with the Hamiltonian
$H=\int_x[\frac{1}{2}\pi_x^2+\frac{1}{2}(\partial_x\phi_x^2)+V(\phi_x)]$.
\\ Here, the eigenvalue $E_n[J]$ is a functional of $J_x$. For
convenience in the calculation, we shift $\phi_x \to
\phi_x+\Phi$($\Phi$ is a constant). Thus, the above equation is
modified
\begin{eqnarray}
\left[ H(\phi_x+\Phi)-\int_x J_x(\phi_x+\Phi)\right]\Psi_n[\phi_x+\Phi, J] \nonumber \\
=E_n[J;\Phi]\Psi_n[\phi_x+\Phi,J]. \label{4.3.9}
\end{eqnarray}
This shift is made in the spirit of the background field
method[4]. However, we will not go into the background field
method.\\ Therefore, the above shift can be regarded as a just
mathematical convenience. Normal ordering the Hamiltonian with
respect to a normal-ordering mass $M$(See Appendix C.) and
inserting a vanishing term
$\int_x[\frac{1}{2}\mu^2\phi_x^2-\frac{1}{2}\mu^2\phi_x^2]$ with
$\mu$ an arbitrary mass parameter into the Hamiltonian, we have
\begin{eqnarray}
N_M\left[H(\phi_x+\Phi)-\int_xJ_x(\phi_x+\Phi)\right] \nonumber \\
=H_0^{J, \mu}+H^{\mu, \Phi}_I-C \label{4.3.10}
\end{eqnarray}
with
\begin{eqnarray}
H^{J, \mu}_I=\int_x
\left\{-\frac{1}{2}\mu^2\phi_x^2+N_M\left[V(\phi_x+\Phi)\right]\right\}
\label{4.3.11}
\end{eqnarray}
and
\begin{eqnarray}
C=\int_x\left[-\frac{1}{2}f_{xx}+\frac{1}{2}\int_y
J_xh^{-1}_{xy}J_y+\frac{1}{2}I_0(M^2)-\frac{M^2}{4}I_1(M^2)+J_x\Phi\right].\label{4.3.12}
\end{eqnarray}
Here, the normal ordering of the shifted potential is given by[5]
\begin{eqnarray}
N_M[V(\phi_x+\Phi)]=\int \frac{d\Omega}{\sqrt{2\pi}}\tilde
V(\Omega)e^{i\Omega(\phi_x+\Phi)+\frac{\Omega^2}{4}I_1(M^2)}.\label{4.3.13}
\end{eqnarray}
For $I_n(M^2)$, see Eq.(\ref{4.2.10}). Since C is a constant, we
can now carry out the Rayleigh-Schr\"{o}dinger perturbation
calculation only for $H^{\mu, \Phi}_I$. To check the order of the
perturbation calculation, an index factor $\delta$ will be
attached to $H^{\mu, \Phi}_I$, so that Eq.(\ref{4.3.9}) is
modified to become
\begin{eqnarray}
[H_0^{J, \mu}+\delta H^{\mu,\Phi}_I]\Psi_n[\phi_x+\Phi,J;\delta] \nonumber \\
=(E_n[J;\Phi,\delta]+C)\Psi_n[\phi_x+\Phi,J;\delta]. \label{4.3.14}
\end{eqnarray}
Now, we carry out the optimized perturbation expansion as
explained in Section IV.A. Obviously, the zeroth-order
approximation to $E_0[J;\Phi,\delta],E_0^{(0)}[J;\Phi]$
satisfies
\begin{eqnarray}
E_0^{(0)}[J,\Phi]+C=E_0^{(0)}=0. \label{4.3.15}
\end{eqnarray}
To the $n$th order of $\delta$,
\begin{eqnarray}
E_0^{(0)}[J;\Phi]= {^{(0)}}\langle J;0|H_I^{\mu,\Phi}
\left[Q_0\frac{1}{H_0^{J,\mu}-E_0^{(0)}[J]}\Big(E_0^{(1)}[J;\Phi]-H_I^{\mu,\Phi}\Big)\right]^{n-1}|0;J\rangle^{(0)},\nonumber \\
\label{4.3.16}
\end{eqnarray}
with $Q_n=\sum_{j \neq n}^\infty \int d^Dp_1d^Dp_2\cdot\cdot\cdot d^Dp_j|j;J\rangle^{(0)}{^{(0)}}\langle J;j|$. \\
Thus
$E_0[J;\Phi,\delta]=E^{(0)}_0[J;\Phi]+\sum^{\infty}_{n=1}\delta^nE_0^{(n)}[J;\Phi]$.

Before carrying out the perturbation calculation, we note that a
Legendre transformation of $E_0[J;\Phi,\delta]$ yields the static effective action[6]
\begin{eqnarray}
\Gamma_s[\varphi;\Phi,\delta]=-E_0[J;\Phi,\delta]-\int_x J_x \varphi_x. \label{4.3.17}
\end{eqnarray}
To calculate the effective potential, one can conveniently take
$\phi_x=\Phi$ to fix the arbitrary shifted parameter $\Phi$[4].
Then, the effective potential is given by
\begin{eqnarray}
{\cal V}(\Phi)\equiv -\frac{\Gamma_s[\varphi;\Phi,\delta]}{\int_x}\Biggl |_{\varphi=\Phi,\delta=1}. \label{4.3.18}
\end{eqnarray}
When truncated at a given order of $\delta$, ${\cal V}(\Phi)$ will
depend on $\mu$. To obtain an approximated effective potential,
one determines $\mu$ according to the principle of minimal sensitivity explained in Section IV.A.

The matrix elements, which appear in Eq.(\ref{4.3.16}), involve
only Gaussian integrals except commutators of creation and
annihilation operators and, thus, can be readily calculated as
follows,
\begin{eqnarray}
&&\snl n|{\cal N}_M[V(\phi_x+\Phi)]|m\snr \nonumber \\    && \ \ \
={\frac{1}{\sqrt{n!m!}}}
     \sum_{i=0}^n C_n^i C_m^{m-n+i} (n-i)!
     (2(2\pi)^{D})^{-{\frac {m-n+2i}{2}}}     \nonumber \\   &&\ \ \
    \times (\prod_{j=n-i+1}^{n}f(p_j)
     \prod_{k=n-i+1}^{m}f(p'_k))^{-{\frac {1}{2}}}
     e^{i(\sum_{k=n-i+1}^{m}p'_k-\sum_{j=n-i+1}^{n}p_j)x}
    \prod_{l=1}^{n-i}\delta(p'_l-p_l) \nonumber \\    && \ \ \
    \times \int_{-\infty}^{\infty}{\frac {d \alpha}{\sqrt{\pi}}} e^{-\alpha^2}
    V^{(m-n+2i)}\Big(\alpha \sqrt{f_{xx}^{-1}-I_1(M^2)}+\Phi+\int_z
        h_{xz}^{-1}J_z\Big) \label{4.3.19}
\end{eqnarray}
with $n\leq m$. In Eq.(\ref{4.3.19}), $V^{k}(z)\equiv {\frac {d^k
V(z)}{(dz)^k}}$. For simplicity, in getting the above results, we
have employed the permutation symmetry of momenta in
Eq.(\ref{4.3.16}) for various products of $\delta$ functions. Note
that matrix elements of $\phi_x^2$ are special cases of
Eq.(\ref{4.3.19}). Substituting the above matrix elements into
Eq.(\ref{4.3.16}), one can obtain the first- and the second-order
corrections to $E_0[J;\phi]$ as
\begin{eqnarray}
E_0^{(1)}[J;\phi]&=&
    \int_x\biggl\{-{\frac {\mu^2}{2}}[(\int_z
       h_{xz}^{-1}J_z)^2+{\frac {1}{2}}f_{xx}^{-1}]         \nonumber  \\
       &&+ \int_{-\infty}^{\infty}{\frac {d \alpha}{\sqrt{\pi}}} e^{-\alpha^2}
    V\Big(\alpha \sqrt{f_{xx}^{-1}-I_1(M^2)}+\Phi \nonumber \\
    &&+\int_z h_{xz}^{-1}J_z\Big)\biggl\}   \label{4.3.20}
\end{eqnarray}
and
\begin{eqnarray}
E_0^{(2)}[J;\phi]&=&
   -{\frac {\mu^4}{2}}\int {\frac {d^D p}{(2\pi)^D}}{\frac {1}{f^2(p)}}\biggl
   |\int_{xz} e^{ipx} h_{xz}^{-1}J_z\biggl |^2 -
   {\frac {\mu^4}{16}}\int {\frac {d^D p}{(2\pi)^D}}{\frac {1}{f^3(p)}}\int_x
               \nonumber \\
  &&+\mu^2\int {\frac {d^D p}{(2\pi)^D}}{\frac {1}{f^2(p)}}
   \int_{x_1 z} e^{-ipx_1} h_{x_1 z}^{-1}J_z\int_{x_2} e^{ipx_2}
   \int_{-\infty}^{\infty}{\frac {d \alpha}{\sqrt{\pi}}} e^{-\alpha^2}
               \nonumber \\
&&\times V^{(1)}\Big(\alpha \sqrt{f_{xx}^{-1}-I_1(M^2)}+\Phi+\int_z h_{xz}^{-1}J_z\Big)  \nonumber \\
   &&+{\frac {\mu^2}{8}}\int {\frac {d^D p}{(2\pi)^D}}{\frac {1}{f^3(p)}}
   \int_x \int_{-\infty}^{\infty}{\frac {d \alpha}{\sqrt{\pi}}} e^{-\alpha^2}
    V^{(2)}\Big(\alpha \sqrt{f_{xx}^{-1}-I_1(M^2)}+\Phi+\int_z
        h_{xz}^{-1}J_z\Big)
          \nonumber \\
&& -\sum_{j\not=0}^{\infty} {\frac {1}{j!2^j}}
    \int {\frac {\prod_{k=1}^j d^D p_k}{(2\pi)^{jD}}}
    {\frac {1}{\prod_{k=1}^j f(p_k) (\sum_{k=1}^j f(p_k))}}
               \nonumber \\
&& \times \biggl |\int_x e^{ix\sum_{k=1}^j p_j}
      \int_{-\infty}^{\infty}{\frac {d \alpha}{\sqrt{\pi}}} e^{-\alpha^2}
    V^{(j)}\Big(\alpha \sqrt{f_{xx}^{-1}-I_1(M^2)}+\Phi+\int_z
        h_{xz}^{-1}J_z\Big)\biggl |^2   \;, \nonumber \\
      \label{4.3.21}
\end{eqnarray}
respectively. Here, $``|\cdots |$'' represents the absolute
value.

Next, we extract the approximated effective potential for the system order by order.

At the zeroth order, $E_0^{(0)}[J;\phi]=-C$ from Eq.(\ref{4.3.15}), and so, taking
$-{\frac {\delta E_0^{(0)}[J;\phi]}{\delta J_x}}=\int_y
h_{xy}^{-1}J_y+\Phi=\varphi_x^{(0)}$ as $\Phi$, one has $J^{(0)}=0$.
Consequently, the effective potential at the zeroth-order of $\delta$ is
\begin{equation}
{\cal V}^{(0)}(\Phi) =-{\frac {\Gamma_s^{(0)}[\varphi;\Phi,\delta]}{\int_x}}
                    \biggl |_{\varphi_x= \Phi} =
                    {\frac {1}{2}} f_{xx}
  -{\frac {1}{2}}I_0(M^2)
  +{\frac {M^2}{4}}I_1(M^2)  \;. \label{4.3.22}
\end{equation}

Up to the first order (Hereafter, any Greek-number superscript,
such as $``I", ``II"$, means ``up to the order whose number is
consistent with the Greek number".),
\begin{equation}
E_0^{I}[\Phi,\delta;J]=E_0^{0}[J;\phi]+\delta E_0^{(1)}[J;\phi] \label{4.3.23}
\end{equation}
and $-{\frac {\delta E_0^{I}[\Phi,\delta;J]}{\delta J_x}}=
\varphi_x^{I}=\Phi$ yields
\begin{eqnarray}
&&\int_y h_{xy}^{-1}J_y +\delta \int_y h_{xy}^{-1}\biggl\{\mu^2\int_z h_{yz}^{-1}J_z
   \nonumber  \\  &&  \ \
   -\int_{-\infty}^{\infty}{\frac {d \alpha}{\sqrt{\pi}}} e^{-\alpha^2}
    V^{(1)}\Big(\alpha \sqrt{f_{xx}^{-1}-I_1(M^2)}+\Phi+\int_z
        h_{xz}^{-1}J_z\Big)\biggl\}=0     \;. \label{4.3.24}
\end{eqnarray}
When extracting the effective potential up to first order,
Eqs.(\ref{4.3.17}), (\ref{4.3.18}) and (\ref{4.3.23}) imply that
only the $J_x$ up to the first order, $J^{I}$, is necessary. Owing
to $J^{(0)}=0$, it is enough to take $J_x=0$ for the last term in
the left hand of Eq.(\ref{4.3.24}). Thus, $J^{I}$ can be solved
from Eq.(\ref{4.3.24}) as
\begin{equation}
J^{I}=\delta \int_{-\infty}^{\infty}{\frac {d \alpha}{\sqrt{\pi}}}
    e^{-\alpha^2} V^{(1)}\Big(\alpha \sqrt{f_{xx}^{-1}-I_1(M^2)}+\Phi\Big) \;.\label{4.3.25}
\end{equation}
Even $J^I$ will not be needed to get the effective potential up to
the first order of $\delta$, because there exists no linear, but
the quadratic term of $J_x$ in the zeroth-order term of
$(E_0^{I}[J;\Phi,\delta]-\int_x J_x \Phi)$ as shown in
Eq.(\ref{4.3.12}). In fact, to obtain the effective potential up
to the $n$th order, one need the approximated $J$ only up to the
$(n-1)$th order. Now, one can write down the effective potential
up to the first order
\begin{eqnarray}
{\cal V}^{I}(\Phi,\delta) &=&
                {\frac {1}{2}} [f_{xx} -I_0(M^2)] +{\frac {1}{4}}M^2I_1(M^2)
                    -\delta {\frac {1}{4}}\mu^2f_{xx}^{-1}
                    \nonumber \\  &\ \ \ &
     +\delta \int_{-\infty}^{\infty}{\frac {d \alpha}{\sqrt{\pi}}}
    e^{-\alpha^2} V\Big(\alpha \sqrt{f_{xx}^{-1}-I_1(M^2)}+\Phi\Big) \;.\label{4.3.26}
\end{eqnarray}
Obviously, this result will yield nothing but the Gaussian
effective potential[5].

Finally, we consider the second order. $\varphi_x=\varphi_x^{II}=-{\frac
{\delta E_0^{II}[J;\Phi,\delta]}{\delta J_x}}=\Phi$ can be solved for $J^{II}$.
In the present case, however, it is enough to use only $J^{I}$ for the effective potential.
Substituting $J^I$ into Eq.(\ref{4.3.18}), we obtain the effective potential for the system, Eq.(\ref{4.3.7}), up
to the second order as
\begin{eqnarray}
{\cal V}^{II}(\Phi,\delta)&=&
              {\frac {1}{2}} [f_{xx} -I_0(M^2)]+{\frac {1}{4}}M^2I_1(M^2)
                    -\delta {\frac {1}{4}}\mu^2f_{xx}^{-1}
                    \nonumber \\  &\ &
       +\delta \int_{-\infty}^{\infty}{\frac {d \alpha}{\sqrt{\pi}}}
    e^{-\alpha^2} V\Big(\alpha \sqrt{f_{xx}^{-1}-I_1(M^2)}+\Phi\Big)
        \nonumber \\  &\ &
 -\delta^2 {\frac {\mu^2}{16}}\int {\frac {d^D p}{(2\pi)^D}}{\frac {1}{f^3(p)}}
  \biggl[\mu^2-2\int_{-\infty}^{\infty}{\frac {d \alpha}{\sqrt{\pi}}} e^{-\alpha^2}
    V^{(2)}\Big(\alpha \sqrt{f_{xx}^{-1}-I_1(M^2)}+\Phi\Big)\biggl]
        \nonumber \\  &\ &
    - \delta^2\sum_{j=2}^{\infty} {\frac {1}{j!2^j}}
    \int {\frac {\prod_{k=1}^{j-1} d^D p_k}{(2\pi)^{(j-1)D}}}
    {\frac {1}{f(\sum_{k=1}^{j-1} p_k)\prod_{k=1}^{j-1} f(p_k)}}
        \nonumber \\  &\ &
  \times  {\frac {1}{\Big(f(\sum_{k=1}^{j-1} p_k)+\sum_{k=1}^{j-1} f(p_k)\Big)}}
 \biggl [\int_{-\infty}^{\infty}{\frac {d \alpha}{\sqrt{\pi}}} e^{-\alpha^2}
    V^{(j)}\Big(\alpha \sqrt{f_{xx}^{-1}-I_1(M^2)}+\Phi\Big)\biggl ]^2 \;,\nonumber \\
  \label{4.3.27}
\end{eqnarray}
where, one should take $\delta=1$ after renormalizing ${\cal V}^{II}
(\Phi,\delta)$, and $\mu$ is determined from the stationary condition
\begin{equation}
{\frac {\partial {\cal V}^{II} (\Phi)} {\partial \mu}} =0 \;.\label{4.3.28}
\end{equation}
Here, ${\cal V}^{II} (\Phi)$ is the effective potential after
${\cal V}^{II} (\Phi,\delta)$ is renormalized. If
Eq.(\ref{4.3.28}) has no real solutions, $\mu$ can be fixed by
${\frac {\partial^2 {\cal V}^{II} (\Phi)} {(\partial \mu)^2}}=0$
as explained in Section IV.A. Note that in (1+1) dimensions,
$\{{\frac {1}{2}} [f_{xx} -I_0(M^2)] +{\frac {1}{4}}M^2I_1(M^2) -
{\frac {1}{4}}\mu^2f_{xx}^{-1}\}$ and $[f_{xx}^{-1} -I_1(M^2)]$ in
Eq.(\ref{4.3.27}) with $\delta=1$ is finite and, thus, for any
(1+1)-dimensional theories which make the series in
Eq.(\ref{4.3.27}) finite, no renormalization procedure is needed.
Similarly, employing Eq.(\ref{4.3.19}), one can obtain higher
order corrections to the Gaussian effective potential from
Eq.(\ref{4.3.16}).

As an example of the above optimized perturbation theory, we
consider the potential,
\begin{equation}
V(\phi_x)={\frac {1}{2}}m^2\phi_x^2+\lambda\phi_x^4  \;, \label{4.3.29}
\end{equation}
which was also studied with the variational perturbation
technique.

Substituting Eq.(\ref{4.3.29}) into Eq.(\ref{4.3.27}), and noting
that $\int_{-\infty}^{\infty} \alpha^{2n+1}e^{-\alpha^2}{\frac {d
\alpha}{\sqrt{\pi}}}=0$ for $n=1,2,\cdots$ and
$\int_{-\infty}^{\infty} \alpha^{2n} e^{-\alpha^2}{\frac {d
\alpha} {\sqrt{\pi}}}=2^{-n}\cdot 1\cdot3\cdots (2n-1)$ for
$n=0,1,2,\cdots$, one can easily obtain the effective potential
for the system up to the second order
\begin{eqnarray}
&&{\cal V}^{II}(\Phi,\delta)= {\frac {1}{2}} (f_{xx} -I_0(M^2))
             +{\frac {1}{4}}M^2I_1(M^2)+
         \delta({\frac {1}{2}}m^2\Phi^2+\lambda\Phi^4
         -{\frac {1}{4}}\mu^2f_{xx}^{-1}) \nonumber \\  & & \ \ \
        +\delta{\frac {1}{4}}(f_{xx}^{-1}-I_1(M^2))
        [m^2+12\lambda\Phi^2+3\lambda (f_{xx}^{-1}-I_1(M^2))]
        \nonumber \\  && \ \ \
   -\delta^2{\frac {1}{16}}\int {\frac {d^D p}{(2\pi)^D}}{\frac {1}{f^3(p)}}
   [m^2-\mu^2+12\lambda\Phi^2+6\lambda (f_{xx}^{-1}-I_1(M^2))]^2
        \nonumber \\  && \ \ \
   -\delta^2 12\lambda^2\Phi^2\int {\frac {d^D p_1 d^D p_2}{(2\pi)^{2D}}}
    {\frac {1}{f(p_1)+f(p_2)+f(p_1+p_2)}}{\frac {1}{f(p_1)f(p_2)f(p_1+p_2)}}
        \nonumber \\  &&  \ \ \
   -\delta^2{\frac {3}{2}} \lambda^2\int {\frac {d^D p_1 d^D p_2 d^D p_3}
   {(2\pi)^{3D}}}{\frac {1}{f(\sum_{k=1}^{3} p_k)+\sum_{k=1}^{3} f(p_k)}}
   \times{\frac {1} {f(\sum_{k=1}^{3} p_k)\prod_{k=1}^{3} f(p_k)}} \;.  \nonumber \\
\label{4.3.30}
\end{eqnarray}
Discarding terms with $I_n(M^2) (n=0,1)$, the above result becomes
identical to Eq.(2.36) in Ref.7(1990) in Section IV.A. This can be
verified by carrying out integrations of $I_n(\Omega) (n=0,1)$ and
$I^{(n)}(\Omega) (n=2,3,4)$ in the above reference over one
component of each Euclidean momentum. Using $\mu$ fixed at the
Gaussian approximation result for each order will simultaneously
imply that up to each order, the vacuum expectation value of the
field operator $\phi_x$ is identical to that in the Gaussian
approximation. If one chooses to do so, Eq.(\ref{4.3.30}) will
yield the second order result of the variational perturbation
theory in Eq.(\ref{4.2.27}).

Finally, we point out that taking $\mu=m$ will lead to the conventional
perturbation result on the effective potential.

\section{Summary and Future Works}

\idn In this lecture note, we have presented a theoretical schemes
for study of many-particle systems based on the functional
Schr\"{o}dinger picture. In Chapter I, the basic conventional
tools for many-body theory including the second quantization and
the Green's function method are briefly reviewed. In Chapter II,
basic ingredients for functional Schr\"{o}dinger picture are
treated. They include functional calculus, Grassmann algebra, and
the basic functional Sch\"{o}dinger picture formalism. In Chapter
III, functional Schr\"{o}dinger picture formalism is applied to
various condensed matter problems including electron gas, the
Hubbard model, the BCS super conductivity, and dilute bose gas
with the Bose-Einstein condensation. However, the formalism in the
Chapter III is limited to the variational approximation and shown
that it only leads to mean-field(Gaussian) results.

In Chapter IV, recent efforts to go beyond the Gaussian
approximation are presented. Two methods, the variational and the
optimized perturbation methods are first introduced with quantum
mechanical languages. Then, they are discussed in depth using
scalar field theory. As results of the study, the effective
potential(the ground state energy) are given, up to the third
order in the variational perturbation method and up to the second
order in the optimized perturbation theory.

However, it should be noted that no concrete condensed matter
models are discussed using above two methods. This is because the
theory presented in Chapter IV is limited to bosonic case and,
thus, cannot be applied to electronic systems such as electron gas
and the Hubbard model. The work extending the variational and the
optimized perturbation theories to fermionic system is currently
under study. Only when this extension is successful, the
functional Schr\"{o}dinger picture approach to many-particle
systems can be claimed truly useful.

\section{Appendices}

\subsection{Anticommutation Relations for Fermionic Field Theory}
In order to obtain anticommutation relation for fermion field, it is necessary to start from the spinor field
theory [1-3]. In general, the fermionic field theory can be studied through either non-hermitian or
hermitian field operators. Here, we start with a non-hermitian field operators.
It is an easy matter to see that the Dirac equation
\begin{eqnarray}
(i \gamma^\mu \partial_\mu -m) \psi =0
\end{eqnarray}
follows from the Euler-Lagrange equation
\begin{eqnarray}
\frac{\partial {\cal L}}{\partial \bar{\psi}} - \partial_\mu
\left( \frac{\partial {\cal L}} {\partial(\partial_\mu
\bar{\psi})} \right) =0 ,
\end{eqnarray}
if we choose
\begin{eqnarray}
{\cal L} &=& i \bar{\psi} \gamma^\mu \tensor{{\partial}}_\mu \psi - m \bar{\psi} \psi  \nonumber \\
&=& \frac{i}{2} [\bar{\psi} \gamma^\mu ( \partial_\mu \psi) - (\partial_\mu \bar{\psi} )\gamma_\mu  \psi]
-m \bar{\psi} \psi ,
\end{eqnarray}
where
\begin{eqnarray}
\bar{\psi} = \psi^\dagger \gamma^0
\end{eqnarray}
and
\begin{eqnarray}
A \tensor{\partial}^\mu B = \frac{1}{2} [A \partial^\mu B - (\partial^\mu A ) B] .
\end{eqnarray}
The $4 \times 4$ Dirac matrices are defined by the relation
\begin{eqnarray}
\gamma^\mu \gamma^\nu + \gamma^\nu \gamma^\mu = 2 g^{\mu \nu} ,
\end{eqnarray}
where the metric tensor $ g^{\mu \nu} $ is given by
\begin{eqnarray}
g^{\mu \nu} =g_{\mu \nu} = \left(
\begin{array}{cccc}
1&0&0&0\\
0&-1&0&0\\
0&0&-1&0\\
0&0&0&-1
\end{array} \right) ,
\end{eqnarray}
in Minkowski space.

From ${\cal L}$, we find the canonical momentum field $\pi(x)$
\begin{eqnarray}
\pi(x) = \frac{\partial {\cal L}}{\partial \dot{\psi}(x)} = i \psi^{\dagger}(x).
\end{eqnarray}
This relation naturally leads into the anticommutation relations
[1],
\begin{eqnarray}
&&\{ \psi_i (\vec{x},t), \psi_j^{\dagger} (\vec{x}',t)  \} = \delta^3 (\vec{x}-\vec{x}')
\delta_{ij} ,  \nonumber \\
&&\{ \psi_i (\vec{x},t), \psi_j (\vec{x}',t)  \} = 0 = \{
\psi_i^{\dagger} (\vec{x},t), \psi_j^{\dagger} (\vec{x}',t)  \}.
\end{eqnarray}
Note that the above non-hermitian field operators $\psi$ and $\psi^{\dagger}$ are
eigenstates of the charge operator.

Fermionic field theory can also be studied using hermitian field operators [2,3],
$\psi^{\dagger}(\vec{x})=\psi(\vec{x})$. Non-hermitian field operator can be
constructed from the hermitian operators as follows
\begin{eqnarray}
\psi_{\rm charge \  eigenstate} = \psi^{1}_{\rm hermitian}+i \psi^{2}_{\rm hermitian} .
\end{eqnarray}
In such a case, the Lagrangian density is expressed as
\begin{eqnarray}
{\cal L} = \frac{i}{2} \psi \alpha^\mu \partial_\mu \psi - \frac{1}{2} m \psi \beta \psi ,
\end{eqnarray}
where $\{ \alpha^i , \alpha^j  \} = 2 \delta^{ij}$ , $\{ \alpha, \beta \}=0,
\beta^2 =1, \alpha^0 =1$. \\
In this case, we can readily show that
\begin{eqnarray}
\{ \psi_a (x), \psi_b (x') \}_{x^0 = x^{0'}}=\delta_{ab} \delta(\vec{x}-\vec{x}') .
\end{eqnarray}
However, we note that the second half of Eq.(486) is not satisfied
in this case.
\\ \\
\subsection*{References}
\begin{itemize}
\item[[1]] L. H. Ryder, Quantum Field Theory, Cambridge University
Press (London) 1985. \item[[2]] P. Ramond, Field Theory; A Modern
Primer, 2nd ed., Addison-Wesley 1989. \item[[3]] J. Schwinger, MIT
Lecture Note 1960. \item[[4]] C. Itzykson and J. E. Zuber, Quantum
Field Theory, McGraw Hill 1980.
\end{itemize}

\subsection{Dual for Gaussian Functionals}

Here, we prove that the dual to a Gaussian functional is given by
another Gaussian functional. Specifically we prove
Eqs.(226)$\sim$(230).

For a real $\theta(\vec{x})$ basis, Eq.(223) show that the dual
$\bar{\Psi}[\theta]$ of ${\Psi}[\theta]$ is given by
\begin{eqnarray}
\bar{\Psi} [\theta] = \int D \theta' \Psi^{\dagger}[\theta'] e^{\theta'\theta} ,
\end{eqnarray}
where ${\Psi} [\theta] = e^{\frac{1}{2} \theta M \theta}$ . First,
the kernel $M$ should have the antisymmetric property to have
nonvanishing contribution,
\begin{eqnarray}
M^t = -M .
\end{eqnarray}
This property can be proved easily as follows,
\begin{eqnarray}
\theta_i M_{ij} \theta_j &=& \theta_i (M_{ji})^{t} \theta_j
= -\theta_j (M_{ji})^{t} \theta_i \nonumber \\
&=& -\theta_j (M^{t})_{ji} \theta_i = -\theta_i (M^{t})_{ij} \theta_j . \nonumber
\end{eqnarray}
Thus, we have $M^t = -M$ .

Upon substituting ${\Psi} [\theta] = e^{\frac{1}{2} \theta M \theta}$ into
Eq.(490), we have
\begin{eqnarray}
\int D \theta'  e^{\frac{1}{2} \theta' M^{\dagger} \theta'}e^{\theta'\theta} .
\end{eqnarray}
Let $\theta_{i}' = u_i + K_{ij} \theta_j$, then
\begin{eqnarray}
\theta'\theta = ( u_i + K_{ij} \theta_j ) \theta_i = u\theta -
\theta {K} \theta
\end{eqnarray}
\begin{eqnarray}
&&\frac{1}{2} \theta' M \theta' = \frac{1}{2} (u_i + K_{ij} \theta_j ) M^{\dagger}_{ik}
(u_k + K_{kl} \theta_l )  \nonumber \\
&=& \frac{1}{2}  u M^{\dagger} u + \frac{1}{2} ( u_i
M^{\dagger}_{ik} K_{kl} \theta_l + K_{ij} \theta_j
M^{\dagger}_{ik} u_k ) + \frac{1}{2} K_{ij} \theta_j
M^{\dagger}_{ik}
K_{kl} \theta_l \nonumber \\
&=&  \frac{1}{2}  u M^{\dagger} u + \frac{1}{2} (u M^\dagger K \theta + \theta \tilde{K}
M^\dagger u) + \frac{1}{2} \theta \tilde{K}
M^\dagger K \theta \nonumber \\
&=& \frac{1}{2}  u M^{\dagger} u + \frac{1}{2} (u M^\dagger K \theta - u \tilde{K}
M^\dagger \theta ) +\frac{1}{2} \theta \tilde{K}
M^\dagger K \theta  . \nonumber
\end{eqnarray}
Thus,
\begin{eqnarray}
&&\frac{1}{2} \theta' M^{\dagger} \theta' + \theta'\theta \nonumber \\
&&=  \frac{1}{2}  u M^{\dagger} u + \frac{1}{2}  u (M^\dagger K -\tilde{M}^\dagger K + 2) \theta
+ \theta \tilde{K} \theta + \frac{1}{2} \theta \tilde{K} M^\dagger K \theta .
\end{eqnarray}
Here, we require $M^\dagger K -\tilde{M}^\dagger K + 2 =0$. \\
Using $\tilde{M} =-M$, we have
\begin{eqnarray}
2 M^\dagger K = -2 , \nonumber
\end{eqnarray}
which gives
\begin{eqnarray}
K= - (M^\dagger )^{-1} ,
\end{eqnarray}
or
\begin{eqnarray}
\tilde{K}= (M^\dagger )^{-1} .
\end{eqnarray}
Here, we used $\tilde{K}= -K$, which is required for $K$ as a new
kernel.

Finally, we have
\begin{eqnarray}
\frac{1}{2} \theta' M^{\dagger} \theta' + \theta'\theta
= \frac{1}{2}  u M^{\dagger} u +  \frac{1}{2} \theta \tilde{K} \theta .
\end{eqnarray}
Therefore
\begin{eqnarray}
&&\int D \theta'  e^{\frac{1}{2} \theta' M^{\dagger} \theta'}e^{\theta'\theta}
=\int D u e^{\frac{1}{2} u M^{\dagger} u + \frac{1}{2} \theta \tilde{K} \theta} \nonumber \\
&&= {\rm det} (M)^{1/2}e^{\frac{1}{2} \theta (M^{\dagger})^{-1} \theta},
\end{eqnarray}
which is Eq.(226). Here, we used Eq.(179).

Now, we continue to prove Eq.(228), where complex Grassmann variables are used.

For complex variables,
\begin{eqnarray}
u= \frac{1}{\sqrt{2}} (u_1 + i u_2 ) , \ \ u^\dagger = \frac{1}{\sqrt{2}} (u_1 - i u_2 ) ,
\end{eqnarray}
and
\begin{eqnarray}
u_1 = \frac{1}{\sqrt{2}} (u + u^\dagger ) , \ \ u_2 = \frac{1}{\sqrt{2} i} (u - u^\dagger ) .
\end{eqnarray}
The dual vector for $\Psi(u_1 , u_2 )= \langle u_1 , u_2 | \Psi \rangle$ is given by
\begin{eqnarray}
\bar{\Psi}[u_1 , u_2 ] = \int D \bar{u}_1 D \bar{u}_2 e^{\bar{u}_1  {u}_1 } e^{\bar{u}_2  {u}_2 }
\Psi^\dagger[\bar{u}_1 ,\bar{u}_2 ] .
\end{eqnarray}
Now,
\begin{eqnarray}
&& \bar{u}_1  {u}_1 + \bar{u}_2  {u}_2  \nonumber \\
&=& \frac{1}{2} [ ( \bar{u} + \bar{u}^\dagger )( {u} + {u}^\dagger )-
( \bar{u} - \bar{u}^\dagger )( {u} - {u}^\dagger )]  \nonumber \\
&=& \bar{u} u^\dagger + \bar{u}^\dagger u = \bar{u}^\dagger u -
u^\dagger \bar{u}.
\end{eqnarray}
Thus,
\begin{eqnarray}
\bar{\Psi}[u , u^\dagger ] = \int D \bar{u} D \bar{u}^\dagger \langle \Psi |  \bar{u}, \bar{u}^\dagger \rangle
e^{\bar{u}^\dagger u - u^\dagger \bar{u}}
\end{eqnarray}
which is Eq.(228). \\
When $\Psi[u , u^\dagger ]= e^{u^\dagger G u} $ , $\Psi^{\dagger}[u , u^\dagger ]= e^{u^\dagger G^\dagger u} $ . \\
Thus, we have
\begin{eqnarray}
&&\bar{\Psi}[u , u^\dagger ] = \int D \bar{u} D \bar{u}^\dagger
e^{\bar{u}^\dagger G^\dagger \bar{u}+ \bar{u}^\dagger {u} - u^\dagger \bar{u}} \nonumber \\
&&= e^{u^\dagger (G^{\dagger})^{-1} u} \int D \bar{u} D \bar{u}^\dagger
e^{(\bar{u}^\dagger  -u^\dagger (G^{\dagger})^{-1} )G^{\dagger}(\bar{u} +(G^{\dagger})^{-1} u )} \nonumber \\
&&= C e^{u^\dagger \bar{G} u} ,
\end{eqnarray}
which is Eq.(230).

\subsection{Normal-Ordering of $H_I$}
A normal-ordered product of field operators is very convenient for
calculation of the ground state energy(effective potential),
because its expectation value vanishes identically. Therefore
before carrying out calculations on matrix elements,
Eqs.(\ref{4.2.22})$\sim$(\ref{4.2.26}), it is necessary to normal
order $H_I$. $\phi_x$ can be rewritten using Eq.(\ref{4.2.12}),
\begin{eqnarray}
\phi_x&=& \varphi+\int \frac{d^Dp}{(2\pi)^{D/2}} \ \frac{1}{(2f(p))^{1/2}}[A_f(p)e^{ipx}+A_f^{\dagger}(p)e^{-ipx}] \nonumber \\
&=&\varphi+\varphi^- +\varphi^+, \label{c.1}
\end{eqnarray}
where
\begin{eqnarray}
\varphi^- &=&\int\frac{d^Dp}{(2\pi)^{D/2}} \ \frac{1}{(2f(p))^{1/2}}A_f(p)e^{ipx} ,\nonumber \\
\varphi^+&=&\int\frac{d^Dp}{(2\pi)^{D/2}} \ \frac{1}{(2f(p))^{1/2}}A_f^{\dagger}(p)e^{-ipx} .\label{c.2}
\end{eqnarray}
Using the above expressions, we readily obtain
\begin{eqnarray}
[\varphi^-,\varphi^+]&=&\int \frac{d^Dp}{(2\pi)^D}\frac{1}{2f(p)}=\frac{1}{2}I_1(\mu^2), \label{c.3} \\
\phi_x-\varphi&=&:(\phi_x-\varphi):, \label{c.4} \\
(\phi_x-\varphi)^2 &=& :(\phi_x-\varphi)^2:+\frac{1}{2}I_1(\mu^2), \label{c.5} \\
(\phi_x-\varphi)^3 &=& :(\phi_x-\varphi)^3:+3[\varphi^-,\varphi^+](\phi_x-\varphi), \label{c.6} \\
(\phi_x-\varphi)^4 &=& :(\phi_x-\varphi)^4:+6[\varphi^-,\varphi^+]:(\phi_x-\varphi):+3[\varphi^-,\varphi^+]^2 \label{c.7}
\end{eqnarray}
$H_I$ is now rewritten
\begin{eqnarray}
H_I&=&\int_x \Big[\frac{1}{2}m^2\phi_x^2-\frac{1}{2}\mu^2(\phi_x-\varphi)^2+\frac{\lambda}{4!}\phi_x^4\Big] \nonumber \\
&& \ \ \ +\frac{1}{2}\int_xI_0(\mu^2) \nonumber \\
&=& \int_x\biggl\{\frac{1}{2}m^2\varphi^2+\frac{\lambda}{4!}+\frac{1}{2}I_0(\mu^2) \nonumber \\
&& \ \ \ +(\phi_x-\varphi)\varphi\Big[\frac{\lambda}{6}\varphi^2+m^2\Big] \nonumber \\
&& \ \ \ +(\phi_x-\varphi)^2\Big[\frac{1}{2}m^2-\frac{1}{2}\mu^2+\frac{\lambda}{4}\varphi^2\Big] \nonumber \\
&& \ \ \ +\frac{\lambda}{6}\varphi(\phi_x-\varphi)^3+\frac{\lambda}{4!}(\phi_x-\varphi)^4\biggl\}. \label{c.8}
\end{eqnarray}
Substituting Eqs.(\ref{c.3})$\sim$(\ref{c.7}) into (\ref{c.8}), we
obtain
\begin{eqnarray}
H_I&=&\int_x \Big\{{\cal V}_G(\varphi)+(\phi_x-\varphi)\varphi(\mu^2-\frac{\lambda}{3}\varphi^2) \nonumber \\
&& \ \ \ +\frac{\lambda}{6}\varphi:(\phi_x-\varphi)^3:+\frac{\lambda}{4!}:(\phi_x-\varphi)^4:\Big\}, \label{c.9}
\end{eqnarray}
which is Eq.(\ref{4.2.30}).

\section*{Acknowledgements} 

 This work was partly supported by
Korea Science and Engineering Foundation(KOSEF) through SRC
program, Center for Strongly Correlated Materials Research at
Seoul National University. C. K. K. acknowledges support from
KOSEF for the visit to Austria.  He also acknowledges warm
hospitality and support by Prof. E. Krotscheck while this lecture
was given at Johannes Kepler Universit\"{a}t-Linz, Austria.

\section*{Bibliography} 

 The general reference are text books and
lecture notes on many-body theory and the Schr\"{o}dinger picture
method. References relevant to each chapter are listed separately. \\

\noindent{\bf General References}
\begin{itemize}
\item[1.] B. Hatfield, {\it Quantum Field Theory of Point
Particles and Strings} (Addison-Wesley, 1992); This book is the
only textbook which has a detailed treatise on the Schr\"{o}dinger
picture representation of the field theory to the authors'
knowledge. \item[2.] J. W. Negele and H. Orland, {\it Quantum
Many-Particle Systems} (Addison-Wesley,1998); This many-body
textbook
 gives a detailed treatment on the functional integral approach to many-particle systems.
\item[3.] A. L. Fetter and J. D. Walecka, {\it Quantum Theory of
Many-Particle Systems} (McGraw-Hill, 1971); This is a standard
textbook on the many-particle theory based on the Green's function
formalism. \item[4.] H. Haken, {\it Quantum Field Theory of
Solids-An Introduction} (North-Holland, 1796); An introductory
level treatment on the second quantization and its applications is
given. \item[5.] J. H. Yee, {\it "Schr\"{o}dinger Picture
Representation of Quantum Field Theory", in Recent Developments in
Field Theory}, ed. by J. E. Kim (Min Eum Sa, Seoul, 1991) p.210;
This lecture note gives a good introduction on bosonic and
fermionic Schr\"{o}dinger picture field theory. \item[6.] J. H.
Yee, {\it "Variational Approach to Quantum Field Theory: Gaussian
Approximation and the Perturbative Expansion around It"},
hep-th/9707234; This lecture note provides an introduction for
variational approach in the functional Schr\"{o}dinger picture
method and the background field method. \item[7.] C. Kiefer, {\it
"Functional Schr\"{o}dinger Equation for Fermions in External
Gauge Fields"}, Ann. Phys. {\bf 236}, 241-285 (1994).
\end{itemize}
{ \bf Section II.B}
\begin{itemize}
\item[1.] R. Floreanini and R. Jackiw, Phys. Rev. D {\bf 37}, 2206 (1988).
\item[2.] A. Duncan, H. Meyer-Ortmanns, and R. Roskies, Phys. Rev. D {\bf 36}, 3788 (1987).
\end{itemize}
{ \bf Section III.A}
\begin{itemize}
\item[1.] H. S. Noh, C. K. Kim and K. Nahm, Phys. Lett. A {\bf
204}, 162 (1995). \item[2.] S. K. Kim, J. Yang, K. S. Soh, and J.
H. Yee, Phys. Rev. D {\bf 40}, 2647 (1989). \item[3.] E. Fradkin,
{\it Field Theories of Condensed Matter Systems} (Addison-Wesley,
1991). \item[4.] M. Rasseti, {\it The Hubbard Model-Recent
Results} (World Scientific, 1991). \item[5.] E. Daggoto, Rev. Mod.
Phys. {\bf 66}, 763 (1994). \item[6.] S. -A. Ahn, I. E. Dikstein,
and S. -H. S. Salk, Physica C {\bf 282-287}, 1707 (1997); S. -H.
S. Salk,
           Physica B {\bf 259-261}, 926 (1999); S. -A. Ahn, K. Park, and S. -H. S. Salk, cond-mat/9903159.
\end{itemize}
{ \bf Section III.B}
\begin{itemize}
\item[1.] H. S. Noh, C. K. Kim, and K. Nahm, Phys. lett. A {\bf
210}, 317 (1996). \item[2.] J. S. Song, S. Hyun, and C. K. Kim, J.
Korean Phys. Soc. {\bf 29}, 821 (1996). \item[3.] J. R.
Schrieffer, {\it Theory of Superconductivity} (Benjamin, 1964)
p.41.
\end{itemize}
{ \bf Section III.C}
\begin{itemize}
\item[1.] S. -H. Kim, H. S. Noh, D. K. Kim, C. K. Kim, and K.
Nahm, Ann. Phys.(Leipzig) {\bf 9}, 579 (2000). \item[2.] E.
Braaten and A. Nieto, Phys. Rev. B {\bf 55}, 8090 (1997).
\item[3.] E. Braaten and A. Nieto, Eur. Phys. J. B {\bf 11}, 143
(1999); E. Braaten, H. -W. Hammer, and S. Hermans, Phys. Rev. A
{\bf 63}, 063609 (2001).
\end{itemize}
{ \bf Section III.D}
\begin{itemize}
\item[1.] H. S. Noh, S. K. You, and C. K. Kim, Int. J. Mod. Phys. B {\bf 11}, 1829 (1997).
\item[2.] J. Lee, K. Na, and J. H. Yee, Phys. Rev. D {\bf 51}, 3125 (1995).
\item[3.] R. P. Feynman, {\it Statistical Mechanics} (Benjamin,1972), p.46.
\item[4.] D. K. Kim and C. K. Kim, J. Phys. A: Math. Gen. {\bf 31}, 6029 (1998).
\item[5.] H. S. Noh, M. D. Kim, C. K. Kim, K. Nahm, and C. -M. Ryu, J. Phys. A: Math. Gen. {\bf 34}, 9793 (2001).
\end{itemize}
{ \bf Section IV.A}
\begin{itemize}
\item[1.] S. K. You, K. J. Jeon, C. K. Kim, and K. Nahm, Eur. J.
Phys. {\bf 19}, 179 (1998). \item[2.] H. Kleinert, {\it Path
Integral in Quantum Mechanics, Statistics and Polymer Physics},
2nd ed.(World Scientific, 1995). \item[3.] W. -F. Lu, C. K. Kim,
J. H. Yee, and K. Nahm, Phys. Rev. D {\bf 64}, 025006 (2001); {\it
ibid} {\bf 66},  069901(E) (2002) and references therein.
\item[4.] A. Z. Capri, {\it Nonrelativistic Quantum Mechanics}
(Benjamin/Cummings, 1985). \item[5.] S. Fl\"{u}gge, {\it Practical
Quantum Mechanics} (Springer, 1971) p.80. \item[6.] W. -F. Lu, S.
K. You, J. Bak, C. K. Kim, and K. Nahm, J. Phys. A: Math. Gen.
{\bf 35}, 21 (2002). \item[7.] I. Stancu and P. M. Stevenson,
Phys. Rev. D {\bf 42}, 2710 (1990); P. M. Stevenson, Phys. Rev. D
{\bf 23}, 2916 (1981). \item[8.] A. Okopinska, Phys. Rev. D {\bf
35}, 1835 (1987). \item[9.] B. Y. Lee, S. U. Cha, K. Nahm, S. N.
Park, S. K. You, and C. K. Kim, to be published.
\end{itemize}
{ \bf Section IV.B}
\begin{itemize}
\item[1.] S. K. You, C. K. Kim, K. Nahm, and H. S. Noh, Phys. Rev.
C {\bf 62}, 045503 (2000). \item[2.] G. H. Lee, T. H. Lee, and J.
H. Yee, Phys. Rev. D {\bf 58}, 125001 (1998). \item[3.] Ref.3 in
Section IV.A and references therein. \item[4.] W. -F. Lu, S. -Q.
Chan, and G. -J. Ni, J. Phys. A: Math. Gen. {\bf 28}, 7233 (1995).
\item[5.] See for example, H. Kleinert and V. Schulte-Frohlinde,
{\it Critical Properties of $\phi^4$-theories} (World Scientific,
2001). \item[6.] P. Cea, Phys. Lett. B {\bf 236}, 191 (1990); P.
Cea and L. Tedesco, Phys. Rev. D {\bf 55}, 4967 (1997).
\end{itemize}
{ \bf Section IV.C}
\begin{itemize}
\item[1.] W. -F. Lu, C. K. Kim, and K. Nahm, Phys. Lett. B {\bf
540}, 309 (2002). \item[2.] W. -F. Lu, C. K. Kim, and K. Nahm,
Phys. Lett. B {\bf 546}, 177 (2002). \item[3.] N. Boccara, {\it
Functional Analysis - An Introduction for Physicists} (Academic
Press, 1990) Chapt.4. \item[4.] G. H. Lee and J. H. Yee, Phys.
Rev. D {\bf 56}, 6573 (1997). \item[5.] W. -F. Lu and C. K. Kim,
J. Phys. A: Math. Gen. {\bf 35}, 393 (2002). \item[6.] See
Chapt.18 in Gen. Ref.1.
\end{itemize}

\end{document}